\documentclass[aps,floatfix,onecolumn,tightenlines,superscriptaddress]{revtex4}
\usepackage{graphicx}
\usepackage{amsmath}
\usepackage{latexsym}
\usepackage{graphicx}
\usepackage[pdftitle={Macroscopic quantum resonators},
            pdftex
            ]{hyperref}

\newcommand{\Karolyhazy}{K\'a{}rolyh\'a{}zy}
\newcommand{\bra}[1]{\langle #1 \vert}
\newcommand{\ket}[1]{\vert #1 \rangle}
\newcommand{\ci}{\mathrm{i}}
\begin{document}

\title{Macroscopic quantum resonators (MAQRO)}

\author{Rainer Kaltenbaek}
\email[Corresponding author - ]{rainer.kaltenbaek@univie.ac.at}
\affiliation{Vienna Center for Quantum Science and Technology, Faculty of Physics, University of Vienna, Vienna, Austria}
\author{Gerald Hechenblaikner}
\affiliation{EADS Astrium Friedrichshafen, Immenstaad, Germany}
\author{Nikolai Kiesel}
\affiliation{Vienna Center for Quantum Science and Technology, Faculty of Physics, University of Vienna, Vienna, Austria}
\author{Oriol Romero-Isart}
\affiliation{Max-Planck-Institut f\"ur Quantenoptik, Garching, Germany}
\author{Keith C. Schwab}
\affiliation{Applied Physics, California Institute of Technology, Pasadena, CA 91125, USA}
\author{Ulrich Johann}
\affiliation{EADS Astrium Friedrichshafen, Immenstaad, Germany}
\author{Markus Aspelmeyer}
\affiliation{Vienna Center for Quantum Science and Technology, Faculty of Physics, University of Vienna, Vienna, Austria}

\begin{abstract}
Quantum physics challenges our understanding of the nature of physical reality and of space-time and suggests the necessity of radical revisions of their underlying concepts. Experimental tests of quantum phenomena involving massive macroscopic objects would provide novel insights into these fundamental questions. Making use of the unique environment provided by space, MAQRO aims at investigating this largely unexplored realm of macroscopic quantum physics. MAQRO has originally been proposed as a medium-sized fundamental-science space mission for the 2010 call of Cosmic Vision.
MAQRO unites two experiments: DECIDE (DECoherence In Double-Slit Experiments) and CASE (Comparative Acceleration Sensing Experiment). The main scientific objective of MAQRO, which is addressed by the experiment DECIDE, is to test the predictions of quantum theory for quantum superpositions of macroscopic objects containing more than $10^{8}$ atoms. Under these conditions, deviations due to various suggested alternative models to quantum theory would become visible. These models have been suggested to harmonize the paradoxical quantum phenomena both with the classical macroscopic world and with our notion of Minkowski space-time.
The second scientific objective of MAQRO, which is addressed by the experiment CASE, is to demonstrate the performance of a novel type of inertial sensor based on optically trapped microspheres. CASE is a technology demonstrator that shows how the modular design of DECIDE allows to easily incorporate it with other missions that have compatible requirements in terms of spacecraft and orbit. CASE can, at the same time, serve as a test bench for the weak equivalence principle, i.e., the universality of free fall with test-masses differing in their mass by 7 orders of magnitude.
\end{abstract}

\maketitle

\section{Introduction}
\label{sec::intro}
Testing the predictions of quantum theory on macroscopic scales is one of today's outstanding challenges of modern physics and addresses fundamental questions on our understanding of the world. Specifically: will the counterintuitive phenomena of quantum theory prevail on the scale of macroscopic objects? This is at the heart of the so-called ``quantum measurement problem'', also known as Schr\"odinger's cat paradox. Another question is whether quantum superposition states of massive macroscopic objects are consistent with our notion of space-time or whether quantum theory will break down in such situations. Investigating quantum superpositions of massive objects might also open up a new route for experimental investigations of quantum gravity. Questions of this kind, i.e., at the interface between quantum laws and the macroscopic world and gravity, address the basic building blocks of our world view and cannot be answered given presently available experimental results.

MAQRO is a proposal for a medium-size space mission that carries two, largely independent experiments: DECIDE and CASE. DECIDE is designed to test the limits of quantum theory using quantum optomechanics. It will make use of a novel combination of a thermal shield and an extra-spacecraft platform in order to achieve the low temperatures and high vacuum needed for the proposed experiments. CASE implements a new type of inertial sensor based on optically trapped microspheres that can be used to test the weak equivalence principle with mass ratios of more than $10^7$. The experiments are planned to be hosted on an LTP-type platform (LTP: LISA Technology Package of the LISA Pathfinder mission) \cite{Armano2009}. An ideal orbit for DECIDE would be a halo orbit around Lagrange point L1 or L2, and a test of the equivalence principle using CASE would ideally be done in a low Earth orbit (LEO). The reason for that is is that the gradient of the gravitational field in a LEO configuration is typically around two orders of magnitude stronger than at L1 or L2. The requirements of both experiments could be met in a mission that uses a highly elliptical orbit (HEO). The design of both experiments is modular and light-weight, allowing for a combination with other experiments that have similar orbit and spacecraft requirements. 
Given that the main scientific objective of MAQRO is the experiment DECIDE, it would be preferable to use a halo orbit around L1 or L2. While CASE can still be performed in this case, it would yield a significantly lower sensitivity for the test of the equivalence principle due to the weak gravitational-field gradient. It might then be preferable to combine DECIDE with some other mission or to investigate the possibility whether CASE could be operated already during an LPF-like spiral-up transfer orbit from a LEO to L1 (see figure \ref{figure_alternative_orbits} and Ref. \cite{Armano2009}).

\section{Scientific Objectives \& Requirements}
\label{sec::obj}
The main objective of the mission is implemented by DECIDE (DECoherence In a Double-slit Experiment). DECIDE will be able to test the predictions of quantum theory against alternative theories that predict a transition from quantum to classical behaviour for the massive objects investigated. The second experiment, CASE (Comparative Acceleration Sensing Experiment), demonstrates a novel, optomechanical inertial sensor and compares it to existing inertial-sensor architectures, allowing to perform a test of the weak equivalence principle.

\subsection{Can we observe interference of massive objects?}
\label{subsec::interference}
Quantum theory is one of the most successful theories known today. It has not only been confirmed in countless experiments but also lies at the heart of important technologies like semiconductors, computer memories, superconductors and lasers, to name a few. Important elements of quantum physics, however, challenge our conceptual understanding of the world. In particular, the counterintuitive nature of quantum superposition and entanglement \cite{Schroedinger1935a,Nielsen2000a} has given rise to both scientific and philosophical discussions ever since the times of Einstein and Bohr \cite{Bohr1949a}. A physical system is said to be in a quantum superposition if it is \textit{in principle} impossible to distinguish whether the system is in one or another of multiple possible states. An example is the famous double-slit experiment where a particle can go through either of two slits. If one cannot possibly tell which of the two slits the particle went through, and if one repeatedly performs position-sensitive measurements behind the double slit, then the distribution of measured particle positions will follow an interference pattern. It is important to note that \textit{each single} particle interferes with itself, i.e., the experiment can be designed such that one and only one particle passes through the setup and is detected before the next particle follows \cite{Zeilinger1981a,Tsuchiya1986a,Tonomura1989a}. The experiment has to be repeated for many single particles in order to acquire enough data points to clearly resolve the interference pattern. Quantum entanglement \cite{Einstein1935a,Schroedinger1935a,Bell1964a} is a direct consequence of quantum superposition if one considers the evolution of composite systems where at least one of the subsystems initially is in a quantum superposition.

Schr\"odinger showed \cite{Schroedinger1935a} that one cannot simply ignore the weirdness of quantum concepts as being restricted to abstract things that happen on very small scales only. He devised a gedankenexperiment that is now known as the paradox of Schr\"odinger's cat. In essence, he considers a cat that is confined within a box the contents of which are principally inaccessible from the outside. In the same box, there are a radioactive atom and a Geiger counter. If the atom decays, the counter clicks and breaks a bottle of poison, killing the cat. From outside the box, the state of the atom cannot be known in principle, and so it has to be described as a superposition between being decayed and not decayed. The atom's state will be entangled with that of the Geiger counter clicking or not clicking, i.e., the bottle of poison being broken or not. According to quantum theory, because no information is, in principle, available about the state of the contents of the box, one concludes that the cat will be in a superposition of being dead and alive, its state entangled with that of the atom and the Geiger counter. Only once the box is opened, will the fate of the cat (and the atom) be decided.

Instead of a cat, we can use any simpler but still macroscopic system that can be prepared in different, macroscopically distinct states. Can we isolate such a system well enough from the environment to bring it into a superposition of those clearly distinct states, i.e., in a superposition of ``dead'' and ``alive''? Typically, the unavoidable coupling of a physical system to its environment leads to decoherence, i.e., it is possible, in principle, by looking at the environment to determine the state of the physical system. With increasing size and complexity of an object, decoupling it from its environment becomes increasingly difficult.

There are various methods to investigate massive objects in the quantum regime.  Matter-wave interference has been observed with a variety of objects, with electrons \cite{Marton1953a,Marton1954a}, neutrons \cite{Rauch1974a,Zeilinger1981a,Rauch2000a}, atoms \cite{Cronin2009a}, and with increasingly large molecules (see, e.g., \cite{Arndt1999a,Hackermueller2004a,Gerlich2011a}). In the case of matter waves with atoms, entangled states with a high number of atoms have been realized \cite{Julsgaard2001a}, and it is even possible to realize atom lasers \cite{Hagley1999a,Bloch1999a}. However, in these cases, the question can be raised whether the states of the atoms considered are indeed macroscopically distinguishable. An intuitive notion for macroscopically distinct states would be for a human observer to be able to distinguish such states with their senses. Because of these considerations, we will only consider superpositions of states that are distinct in the center-of-mass position of a macroscopic object. Mechanical resonators, in principle, allow for the preparation of such states and provide a new route for studying quantum theory with massive objects; in particular, in combination with quantum-optical control techniques. These devices allow for studying the collective center-of-mass motion of massive objects that span the size range from hundreds of nanometers in the case of nano electromechanical or nano optomechanical systems (NEMS/NOMS) to tens of centimeters in the case of gravitational wave antennae. The quickly developing field of quantum optomechanics \cite{Schwab2005a,Kippenberg2008a,Aspelmeyer2010a} opens - aside from numerous novel sensing and actuation technologies at and beyond the quantum limit - a unique opportunity to generate superposition states of massive objects, thereby revealing new opportunities for macroscopic quantum experiments.

DECIDE uses the techniques of quantum optomechanics to realize quantum states with objects on a macroscopic scale. This is achieved by using the following procedure, parts of which have been inspired by related work \cite{Chang2009a,RomeroIsart2010a,RomeroIsart2011a,RomeroIsart2011b}:
\begin{enumerate}
 \item dielectric nanospheres are loaded into an optical trap within a high-finesse cavity
 \item all nanospheres except one are ejected from the cavity mode via radiation pressure excerted by a focused UV laser
 \item using quantum back-action cooling, the center-of-mass motion is cooled to the mechanical ground-state of motion (see, e.g., \cite{WilsonRae2008a,Groeblacher2009a,OConnell2010a}).
 \item once the ground state is reached, the optical trap is switched off, and the wavefunction expands for a time $t_1$
 \item a tightly focused UV laser pulse is shot through the center of the expanded wavefunction. If light is scattered off the nanosphere, the nanosphere will be localized, the superposition is destroyed, and the whole procedure has to be repeated. If no light is scattered, the wavefunction will resemble a Schr\"odinger-cat-like state \footnote{While, here, we have used the notion of post-selection of cases where no UV light is scattered, it can be shown that no post-selection is necessary for an interference pattern to form \cite{Kaltenbaek2012a}.}.
 \item the wavefunction expands for a time $t_2$ long enough for the two parts of the wavefunction to overlap and interfere
 \item the cavity field is switched back on in order to read out the particle position with high accuracy
\end{enumerate}
As long as the nanosphere is not accidentially lost, the steps 3 to 7 can be repeated without the need for loading new particles. If this procedure is repeated many times under the same initial conditions, quantum theory predicts that the distribution of measured positions will follow an interference pattern. Its visibility will depend on the amount of decoherence resulting from, e.g., the scattering of gas particles and blackbody radiation as well as the absorption and emission of blackbody radiation. These effects will be discussed in section \ref{sub::DECqm}. 

The goal of DECIDE is to demonstrate and to investigate the interference of massive objects using dielectric nanospheres as well as to look for possible deviations from the predictions of quantum theory like they are suggested by alternative, macrorealistic theories. Those spheres have a diameter of up to $\sim 100\,$nm and will be used in a double-slit-like experiment as described above. The experiment will be repeated with varying particle sizes and masses in order to study the parameter dependence of the underlying decoherence mechanisms. DECIDE will provide experimental conditions that are impossible to fulfill on Earth (see section \ref{subsec::CaseForSpace}). 

\subsection{What are the limits of quantum theory?}
\label{subsec::limits}
According to quantum theory it is, in principle, possible to observe interference with arbitrarily large and complex objects as long as they are isolated well enough from the environment. Since Schr\"odinger's famous cat \cite{Schroedinger1935a} and the insight of its remarkable consequences for our world view, physicists have been considering the question whether there is a limit to quantum theory, i.e., a parameter regime where objects will behave classical no matter how well they are isolated from their environment. Up to date, all experiments are in agreement with the predictions of quantum theory. Yet, various theoretical models have been suggested that introduce additional physical mechanisms leading to a transition between the quantum and the classical regime in order to explain the classical nature of our world at the large scale. Such models are called macrorealistic. In the folowing, we will describe the models we will concentrate on in this proposal:

\begin{itemize}
 \item \textbf{The CSL model:}\\
       The continuous-spontaneous-localization (CSL) model \cite{Ghirardi1990a,Collett2003a} is based on the work of Ghirardi, Rimini and Weber (GRW) \cite{Ghirardi1986a} and of Pearle \cite{Pearle1976a,Pearle1989a}. Related work was published by Gisin \cite{Gisin1989a}. In this model, all microscopic particles are continuously localized with a rate $\lambda$ and a spatial accuracy $r_c\equiv\alpha^{-1/2}$. The CSL model is very general and does not assume a particular physical mechanism causing the localization. While the localization rate is negligible for elementary particles, the effect increases with the number of particles in an object, and, for macroscopic objects, localization occurs nearly immediately.
 \item \textbf{The Quantum Gravity (QG) model:}\\
       The QG model, has been introduced by Ellis, Mohanty, Mavromatos, and Nanopoulos \cite{Ellis1989a,Ellis1992a}. Essentially, it is assumed that any future theory of quantum gravity will allow for the occurrence of wormholes on a microscopic scale in an otherwise flat spacetime. Particles become entangled with degrees of freedom in these wormholes. Because these degrees of freedom are inaccessible, any initially pure state will become mixed over time.
 \item \textbf{The model of \Karolyhazy{}:}\\
       In 1966, \Karolyhazy{} presented the first model that predicted the decoherence of massive superposition states due to gravitation. In particular, he assumed that the spacetime metric fluctuates, leading to the dephasing of superpositions of massive systems involving large spatial separations \cite{Karolyhazy1966a}. We refer to his model as the K model.
 \item \textbf{The Di\'osi and Penrose models:}\\
       For several decades, Di\'o{}si and Penrose have independently proposed models that predict a gravitationally induced collapse of superposition states involving massive objects (see, e.g.,  \cite{Diosi2005a,Penrose1996a}). While the physical mechanisms causing decoherence are fundamentally different in the two models, the resulting predictions for the decoherence rate are basically identical \cite{Diosi2005a}. We will refer to this model as the Di\'osi-Penrose (DP) model.
\end{itemize}

So far, no experiment has been performed to cleary confirm or rule out any of these models but the experiments of the Arndt group on matter-wave interference with large molecules should soon be able to test Adler's version of the CSL model \cite{Adler2007a,Adler2009a,Nimmrichter2011a}. But this still leaves a lot of leeway for the CSL model by varying the parameters $\lambda$ or $\alpha$, and all other models still remain far out of range of current state-of-the-art experiments.

\subsection{Decoherence according to quantum theory}
\label{sub::DECqm}
In quantum theory, the evolution of a closed physical system is always unitary and can be described by the Schr\"odinger equation. Because such a system is completely isolated from its environment, it does not experience decoherence. For all practical purposes, however, a physical system is never completely decoupled from its environment. If we take this into account, we deal not with a closed but with an open quantum system. For all standard quantum decoherence mechanisms and also for all macrorealistic models we will consider, the time evolution can be described by a master equation of the form \cite{Gallis1990a}:
\begin{equation}
 \dot{\rho} = \frac{1}{\ci\hbar} \left[\mathcal{H}, \rho\right] - \Lambda \left[\hat{x},\left[\hat{x}, \rho\right]\right],
\end{equation}
where the dot represents the derivative with respect to time, $\mathcal{H}$ is the Hamiltonian, $\rho$ is the density matrix, and $\hat{x}$ denotes the position operator. The first term on the right-hand is simply the unitary evolution due to quantum theory while the second term leads to the decay of off-diagonal terms of the density matrix, i.e., to decoherence. If we take the position representation of the density matrix, $\rho(x_1, x_2) = \bra{x_1}\rho\ket{x_2}$, and if we only consider the decoherence term of the equation above, then we can describe the evolution of the elements of the density matrix by \cite{Ghirardi1986a}:
\begin{equation}
 \frac{\mathrm{d}}{\mathrm{d}t} \rho(x_1, x_2) = -\Lambda (x_1 - x_2)^2 \rho(x_1, x_2) = -F(x_1,x_2) \rho(x_1, x_2).\label{eq::decterm}
\end{equation}
The function $F(x_1,x_2)$ has the dimension of a frequency and can be interpreted as the decay rate of the coherent, off-diagonal elements of the density matrix. In the following, we will discuss the main decoherence mechanisms:
\begin{itemize}
 \item \textbf{scattering of background gas}:\\
  For all cases, we are interested in here, the decoherence rate $F(x_1,x_2)$ can be assumed to be constant because the de-Broglie wavelength of the gas molecules is much shorter than the dimension of the nanosphere and $\vert x_1-x_2\vert$ \cite{Schlosshauer2007a}:
  \begin{equation} 
  F_{gas} = \frac{2 \sqrt{6 \pi} r^2 p}{m_a v_a}.
  \end{equation}
This yields an upper bound on the decoherence due to gas scattering. $m_a$ and $v_a$ are the average mass and velocity of the gas particles, $p$ is the gas pressure, and $r$ is the radius of the sphere. We design DECIDE such that the vacuum level is low enough to render the decoherence due to gas scattering negligible (see section \ref{subsub::vacuum}).
 \item \textbf{scattering of blackbody radiation}:\\
 In contrast to the scattering of gas particles, the wavelength of blackbody radiation is typically much larger than the dimensions of the nanosphere and the displacements in the superpositions we consider. We describe the corresponding decoherence due to the scattering of blackbody radiation via a decoherence parameter \cite{Schlosshauer2007a}:
   \begin{equation} 
   \Lambda_{bb,sca} = \frac{8\times8! r^6  c \zeta\left( 9\right)}{9 \pi} \left(\frac{k_B T}{c \hbar}\right)^9   \mathrm{Re}\left(\frac{\epsilon_{bb} - 1}{\epsilon_{bb} + 2}\right)^2,
   \end{equation}
  where $\zeta(9)$ is Riemann's $\zeta$ function, and $T$ is the temperature of the environment. Here and below, we assume that the relative permittivity, $\epsilon_{bb}$, is constant over the spectrum of the blackbody radiation \cite{Chang2009a}.
 \item \textbf{absorption of blackbody radiation}:\\
 The decoherence parameter due to the absorption of blackbody-radiation photons can be described via the same formula as the decoherence parameter for the emission of blacbody-radiation (see below) except that we have to use the internal temperature instead of the temperature of the environment:
   \begin{equation} 
   \Lambda_{bb,abs} = \frac{16 \pi^5 r^3  c }{189} \left(\frac{k_B T}{c \hbar}\right)^6 \mathrm{Im}\left(\frac{\epsilon_{bb} - 1}{\epsilon_{bb} + 2}\right),
   \end{equation}
   where  $T$ is the temperature of the environment.
 \item \textbf{emission of blackbody radiation}:\\
From Bohren and Huffman \cite{Bohren1998a} one can calculate the rate of photons absorbed by a nanosphere in a mode $k=\frac{2\pi}{\lambda}$ if the temperature of the environment is $T$. If we replace that temperature instead with the internal temperature $T_i$ of the sphere, the same relation gives the rate of emitted photons in the mode $k$:
\begin{equation}
R(k) = \frac{3 V k^3 c}{\pi} \frac{1}{\exp\left(\frac{\hbar c k}{k_B T_i}\right)-1} \mathrm{Im}\left(\frac{\epsilon_{bb}-1}{\epsilon_{bb}+2}\right).
\end{equation}
The overall rate of emitted photons is $R_{tot}=\int dk R(k)$. We can use these expressions to calculate the decoherence rate in the master equation due to emission of a single photon \cite{Hackermueller2004a}:
\begin{equation}
\frac{d}{dt} \langle \mathbf{r}_1\vert \rho(t) \vert \mathbf{r}_2\rangle = F(\Delta r)\langle \mathbf{r}_1\vert \rho(t) \vert \mathbf{r}_2\rangle,
\end{equation}
where $\Delta r = \vert \mathbf{r}_2 - \mathbf{r}_1\vert$, and
\begin{equation}
F(\Delta r) = \frac{1}{R_{tot}} \int dk R(k) \frac{\sin(k \Delta r)}{k \Delta r}.
\end{equation}
If one assumes that $\Delta r$ is much smaller than any thermal wavelengths, one can Taylor expand the sinc function in the integral to get:
\begin{equation}
F(\Delta r) \approx 1 - \frac{\Delta r^2}{6 R_{tot}} \int dk R(k) k^2 \approx \exp(-\frac{\Delta r^2}{6 R_{tot}} \int dk R(k) k^2).
\end{equation}
In order to get the decoherence rate due to the emission of $n=t R_{tot}$ photons during some time $t$, one has to take $F(\Delta r)$ to the power of $n$. One can then show that the decoherence parameter due to the emission of blackbody radiation is:
   \begin{equation} 
   \Lambda_{bb,em} = \frac{1}{6} \int dk R(k) k^2 = \frac{16 \pi^5 r^3  c }{189} \left(\frac{k_B T_i}{c \hbar}\right)^6 \mathrm{Im}\left(\frac{\epsilon_{bb} - 1}{\epsilon_{bb} + 2}\right).
   \end{equation}
\end{itemize}

In order to compare the decoherence rates according to quantum theory with those predicted by macrorealistic models, we introduce a new parameter, which we denote as the \textit{coherent expansion distance} (CED) and will define in the following. In the course of a typical experimental run of DECIDE, the wavefunction of the nanosphere will expand for a time $\tau$ to a width $\sigma(\tau) = \sqrt{x^2_0 + v^2_m t^2}$, where $x_0$ is the ground-state extension of the optically trapped nanosphere, and $v_m$ is the expansion velocity of the wave packet once the harmonic potential is switched off. At any given moment, the decoherence rate experienced by the matrix element $\langle-\sigma(t)\vert\rho\vert\sigma(t)\rangle$ is $F(t) = \Lambda \left(2 \sigma(t)\right)^2$, see equation (\ref{eq::decterm}). We then assume that the overall decoherence that is experienced by that matrix element $\langle-\sigma(t)\vert\rho\vert\sigma(t)\rangle$ up to some time $\tau$ can be calculated by integrating the decoherence rate over the time the wave packet expands:
\begin{equation}
 \Gamma(\tau) \equiv \int^\tau_0{}\mathrm{d}t\,F(t) \approx 4 \Lambda \int^\tau_0{}\mathrm{d}t\,\sigma(t)^2 \approx 4 x^2_0 \Lambda \tau + \frac{4}{3} v^2_m \Lambda \tau^3.\label{eq::GammaCET}
\end{equation}
After this expansion, a UV pulse is used to probabilistically split the wave packet into two parts separated by the distance $\sigma(\tau)$. We then let the wave packet evolve again freely for a time $\tau$ such that the two parts of the wavefunction have time to overlap and interfere \footnote{As we have described in section \ref{sub::DECcomp}, this is oversimplified. A more detailed analysis will be given in a separate study \cite{Kaltenbaek2012a}.}. The off-diagonal terms that lead to interference in the center of the wave packet will be reduced by a factor $\exp\left[-2  \Gamma(\tau)\right]$. The visibility of the interference pattern should then be proportional to the square of that factor, i.e., $\exp\left[-4  \Gamma(\tau)\right]$. We then introduce the coherent expansion distance as $\mathrm{CED} = v_m \mathrm{CET}$, where the coherent expansion time (CET) is defined as the time it takes for the expected decoherence visibility to decohere by a factor $1/e$, i.e., $4\Gamma\left(\mathrm{CET}\right) \equiv 1$. We have to take a slightly different approach for the decoherence due to gas scattering because $F(x_1,x_2)=F_{gas}$ is constant and does not depend on $\sigma(t)$. The overall  decoherence is then simply given by: $\Gamma_{gas}(\tau) = \tau F_{gas}$, and we can simply add it to the right-hand side of equation (\ref{eq::GammaCET}).

\subsection{Decoherence according to macrorealistic models}
\label{sub::DECmrm}
Similar to decoherence in quantum theory, we can ascribe decoherence rates to the various macrorealistic models. For a desicisve experiment comparing the predictions of these models with the predictions of quantum theory, it is necessary to find a parameter regime where the macrorealistic models predict significantly stronger decoherence rates than quantum theory. Refs.~\cite{RomeroIsart2011b,RomeroIsart2011c} also provide an overview of the decoherence parameters of the macrorealistic models we discuss.

The CSL model depends on two parameters, $\alpha$ and $\lambda_0$, which were originally chosen to be $\alpha = r^{-2}_c= 10^{14}\,\mathrm{m}^{-2}$ and $\lambda_0 = 10^{-16}\,\mathrm{s}^{-1}$. Using these parameters, the decoherence rate for a nanosphere is given by (see section 3 and appendix A of Ref.~\cite{Collett2003a}):
\begin{equation}
F_{CSL} = \Lambda_{CSL} \Delta{}x^2 = \frac{m^2 \lambda_0 \alpha  f_{CSL}(\sqrt{\alpha} r)}{4 m^2_0} \Delta{}x^2,
\end{equation}
where $m$ and $r$ are mass and radius of the nanosphere, $m_0$ is the mass of a nucleon, $\Delta{}x = \vert x_1-x_2\vert$ the wave-packet separation, and
\begin{equation}
 f_{CSL}(\frac{r}{a}) = 6 \frac{a^4}{r^4} \left[ 1 - 2 \frac{a^2}{r^2} + (1+2 \frac{a^2}{r^2}) \mathrm{e}^\frac{r}{a}\right].
\end{equation}
Recently, Adler argued that the value of $\lambda_0$ should be chosen to be significantly larger than the standard one such that decoherence already occurs after the time it takes for the formation of a latent image in an analogue camera \cite{Adler2007a}. He suggests a value of $\lambda_{\mathrm{Adler}} = 10^{8\pm 2}\,\mathrm{s}^{-1}$. 

In the QG model, each consituent of matter experiences decoherence. Following Ellis and his collaborators \cite{Ellis1989a,Ellis1992a}, the corresponding decoherence parameter for a proton in natural units is $\Lambda_{\mathrm{micro},\mathrm{QG}} = \frac{m^6_0}{m^3_P}$, where $m_0 \approx 0.94\,\mathrm{GeV}$ is the mass of a proton, and $m_P \approx 1.22\times 10^{19}\,\mathrm{GeV}$ is the Planck mass. In SI units, we have $\Lambda_{\mathrm{micro},\mathrm{QG}} = \frac{c^4 m^6_0}{\hbar^3 m^3_P}$, with $m_0 \approx 1.673\times 10^{-27}\,\mathrm{kg}$, and $m_P \approx 2.18\times 10^{-8}\,\mathrm{kg}$. We can calculate the decoherence parameter for a body of mass $m$ approximately by assuming that the number of consituent microscopic particles is $N=\frac{m}{m_0}$ and multiplying it with the microscopic decoherence parameter:
\begin{equation}
\Lambda_\mathrm{QG} = N \Lambda_{\mathrm{micro},\mathrm{QG}} = m \frac{c^4 m^5_0}{\hbar^3 m^3_P}.
\end{equation}

For the K model, one gets $\gamma = \alpha \lambda = 1/(2 a^2_c \tau_c)=\hbar/(2 a^4_c m)$ \cite{Frenkel1990a}, where 
\begin{equation}
a_c = \left\{ 
\begin{array}{cl}
 \left(\frac{r}{\Lambda_p}\right)^{\frac{2}{3}} L &\mathrm{ if } \;r > a_c \\
 \left(\frac{L}{\Lambda_p}\right)^2 L &\mathrm{ if } \;r > a_c
\end{array}
\right.\label{ac}
\end{equation}
Here, $m$, $r$, and $L=\frac{\hbar}{m\,c}$ are the mass, the radius, and the Compton wavelength of our particle, respectively. $\Lambda_p = \left(\frac{G \hbar}{c^3}\right)^{\frac{1}{2}} \approx 10^{-35}\,\mbox{m}$ is the Planck length. Because $\Lambda \equiv \gamma/4$, the decoherence rate according to the K model then is:
\begin{equation}
F_K = \Lambda_K \Delta{}x^2 = \frac{\hbar}{8 m a^4_c} \Delta{}x^2,
\end{equation}

The Di\'osi-Penrose model predicts the decoherence rate \cite{Penrose1996a,Diosi2005a}
\begin{equation}
F_{DP} = \frac{E_G}{\hbar},
\end{equation}
where $E_G$ is the gravitational self energy of the difference between the mass distributions belonging to the different states in the superposition. If we assume a spherical, continuous mass distribution, this becomes \cite{RomeroIsart2011b,RomeroIsart2011c}:
\begin{equation}
F_{DP} = \left\{ 
\begin{array}{cl}
 \frac{20\,G \rho^2 r^3 \Delta{}x^2}{\hbar} &\mathrm{ if } \;\Delta{}x \ll r \\
 \frac{20\,G \rho^2 r^5}{\hbar} &\mathrm{ if } \;\Delta{}x > r 
\end{array}
\right.,
\end{equation}
where $\rho$ is the mass density of the sphere, $r$ is the radius, $\Delta{}x$ is the displacement, and $G$ is Newton's constant.

\subsection{Comparison of quantum theory and macrorealistic models}
\label{sub::DECcomp}
\begin{figure}[t]
 \begin{center}
  \includegraphics[width=0.55\linewidth]{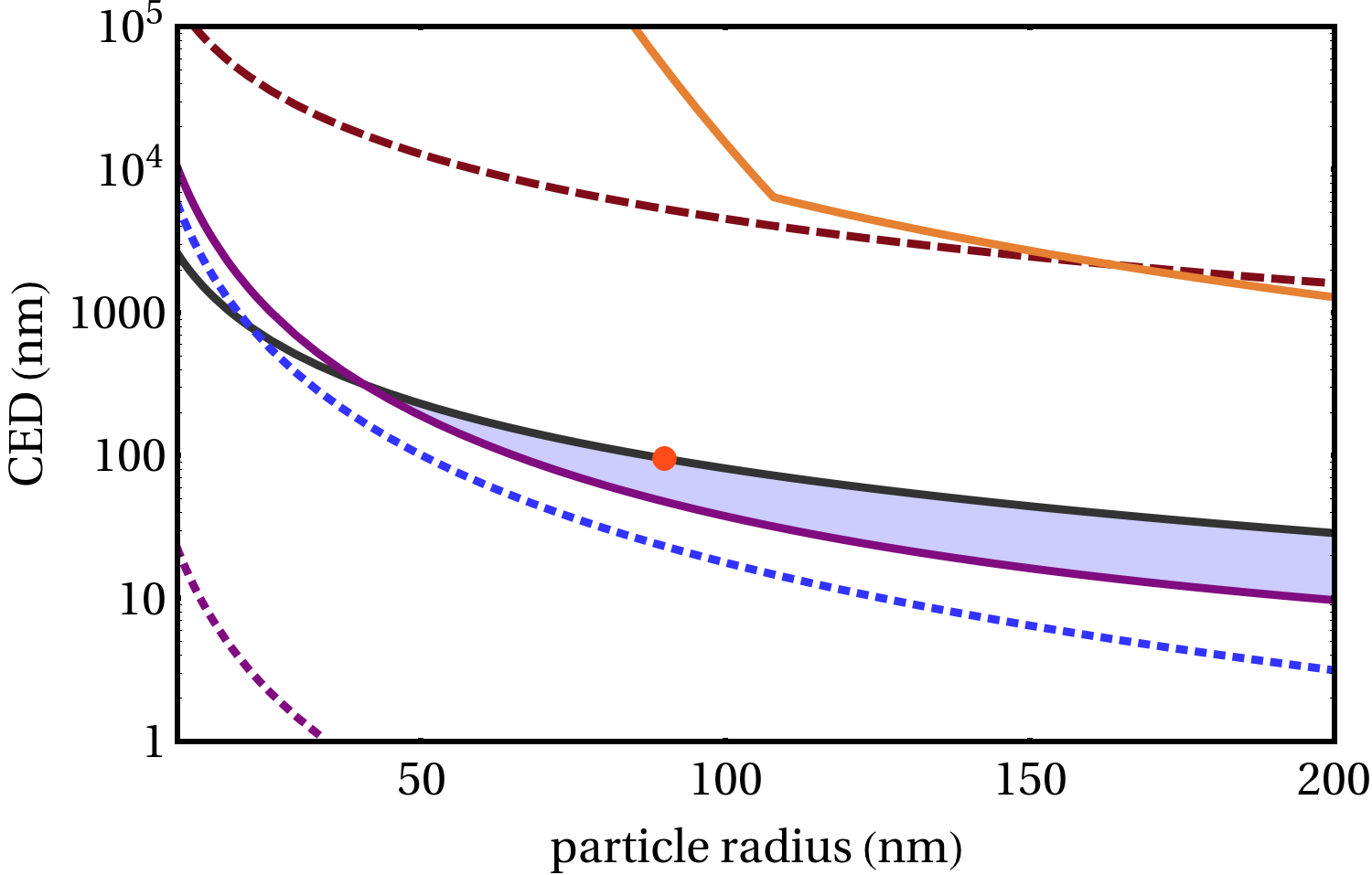}
  \caption{\textbf{Quantum theory vs. macrorealism - state of the art.} We compare the CED as predicted by quantum theory (dark-gray, solid) against the CED predicted by the original CSL model ($\lambda_{\mathrm{CSL}} = 10^{-16}\,\mathrm{s}^{-1}$, magenta, solid), Adler's modified CSL model ($\lambda_\mathrm{Adler}=10^{-8}\,\mathrm{s}^{-1}$, magenta, dotted), the QG model (blue, dotted), the DP model (red, dashed), and the K model (orange, solid). We have assumed the baseline configuration of DECIDE, i.e., the nanospheres consist of fused silica with a mass density of $2201\,\mathrm{kg}\mathrm{m}^{-3}$, and a relative permittivity of $\epsilon_\mathrm{la} = 2.1 + \ci 2.5\times 10^{-10}$ at the wavelength, $\lambda = 1064\,$nm, of the trapping laser. The environment temperature is $T_e = 32\,$K, the gas pressure $P \le 10^{-12}\,$Pa, and the internal temperature of the sphere is $T_i=98\,$K, resulting from a trapping field with a power of $0.1\,$W and a waist of $10\,\mu$m. We assume that the permittivity for blackbody radiation is constant with the value $\epsilon_{bb} = 2.1 + \ci 0.57$\cite{Chang2009a}. The blue-shaded region indicates where a decisive test of quantum theory against any CSL model with $\lambda \ge 10^{-16}\,\mathrm{s}^{-1}$ and also against the QG model is possible. The orange dot highlights the CED for a particle with a radius of $90\,$nm. \label{fig::CED}}
 \end{center}
\end{figure}

Using the expressions we gave above for the various decoherence rates, it is possible to compare the CED as predicted by quantum theory with that predicted by the CSL model, the QG model, the K model and the DP model. In Fig.\ref{fig::CED}, we plot the CED for the spacecraft design proposed here and for state-of-the-art material properties.

The blue-shaded region indicates the range of particle radii where all CSL models with $\lambda \ge \lambda_\mathrm{CSL} = 10^{-16}\,\mathrm{s^{-1}}$ as well as the QG model predict a smaller CED than quantum theory. That means, for these radii, quantum theory predicts a violation of those models because both of them predict a ``collapse'' of the wavefunction while quantum theory does not. The baseline configuration of the proposed mission is indicated with an orange dot.

\begin{figure}[t]
 \begin{center}
  \includegraphics[width=0.49\linewidth]{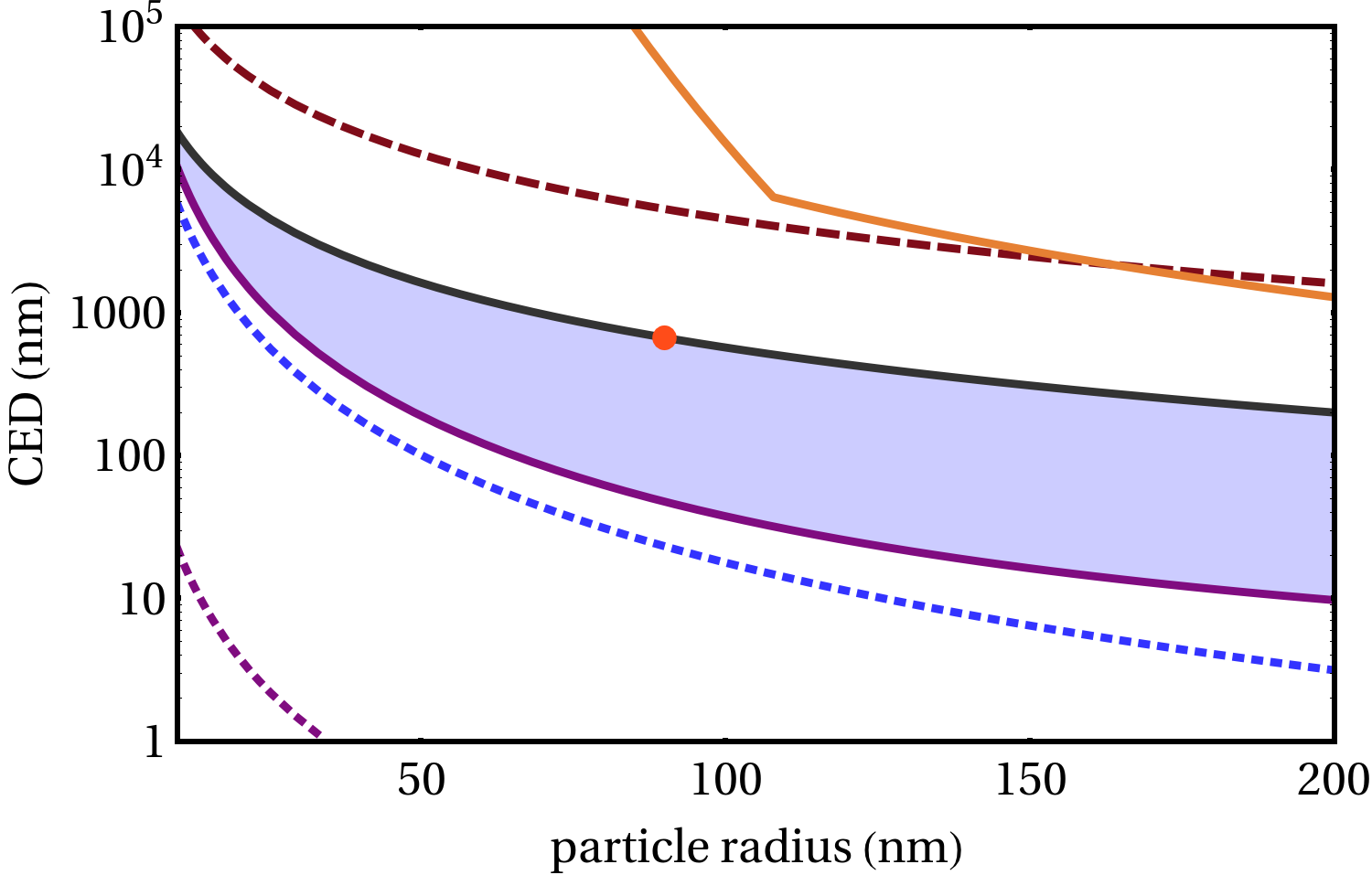}
  \includegraphics[width=0.49\linewidth]{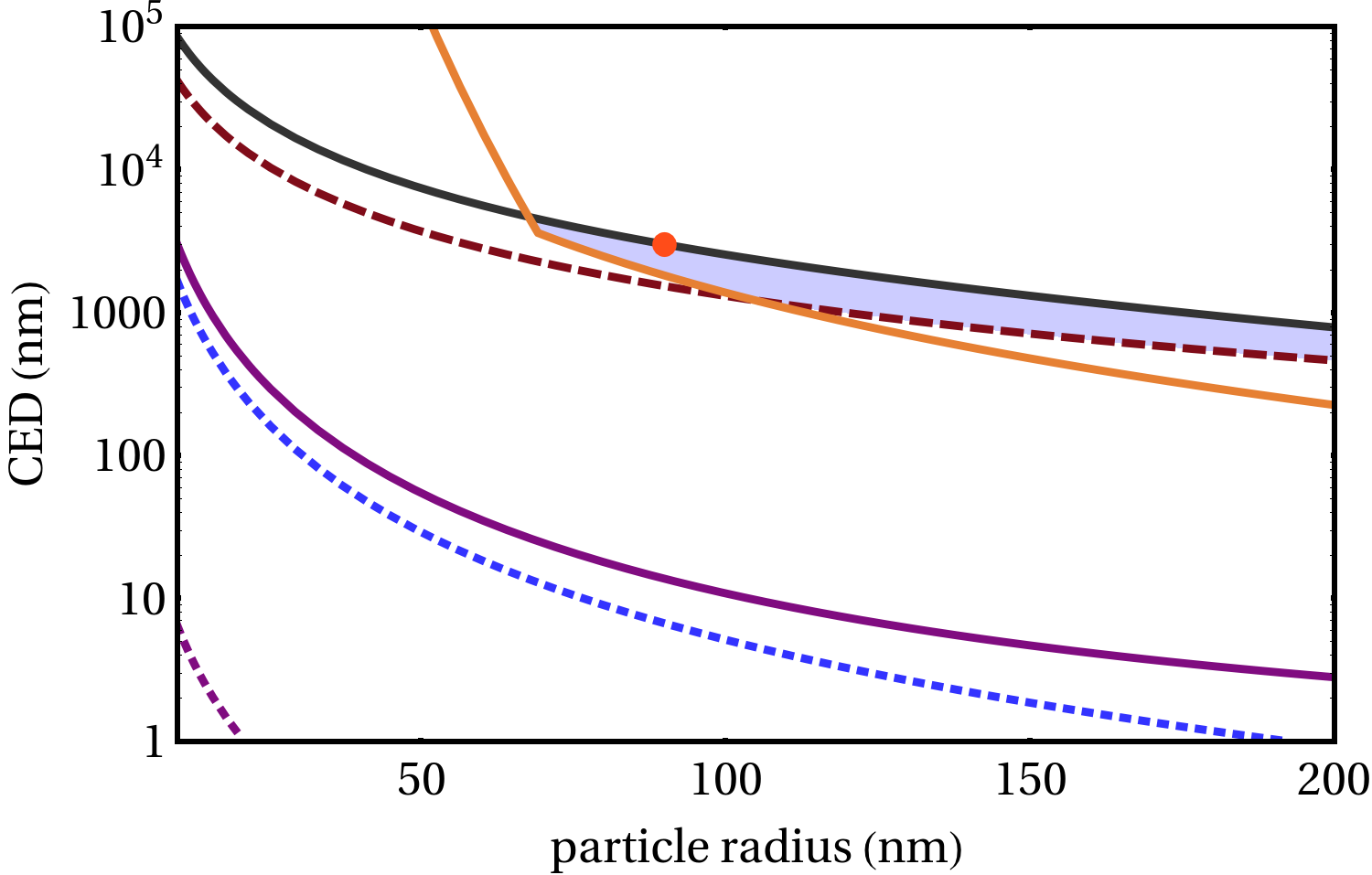}
  \caption{\textbf{Quantum theory vs. macrorealism - for envisioned future parameters.} Using state-of-the-art material parameters and the design of the spacecraft proposed here, only tests of quantum theory against the predictions of the CSL model and the QG model are possible, see fig. \ref{fig::CED}. In order to also test the K model and/or the DP model, significant improvements have to be made. The two plots here show the CED for two sets of improved parameters that we will aim to achieve in the future. \textbf{(left)} Here, we have assumed the same parameters as in figure \ref{fig::CED} but with lower absorption at the wavelength of the trapping laser, i.e., $\epsilon_\mathrm{la} = 2.1 + \ci 2.5\times 10^{-13}$. These parameters still will not allow to test the K model or the DP model. \textbf{(right)} In order to allow for such a test, further improvements have to be made. Here, we assume even lower absorption, i.e., $\epsilon_\mathrm{la} = 2.1 + \ci 2.5\times 10^{-15}$, as well as a higher mass density of $9680\,\mathrm{kg}\mathrm{m}^{-3}$ and a lower environment temperature, $T_e=12\,$K. The blue-shaded region indicates where quantum theory predicts a violation of all macrorealistc models considered here.\label{fig::CEDfuture}}
 \end{center}
\end{figure}

For a focused Gaussian UV beam, the minimum waist achievable is $w_0 \ge \lambda_{UV}/2$. In order to allow for some off-diagonal elements of the density matrix to ``survive'' the preparation of the double slit, the CED must be much larger than the UV waist. Considering the blue-shaded region in figure \ref{fig::CED}, we see that, for the currently proposed spacecraft design and material parameters, this condition is not fulfilled. Possible solutions to this problem are:
\begin{itemize}
\item \textbf{use a shorter UV wavelength of around $200\,$nm}\\
  An advantage to this approach is that no significant changes have to be made to the proposed setup. The problem is that for wavelengths much shorter than $350\,$nm, the radiation damages and/or charges UV-fused silica, see \cite{Arai1988a} and, e.g.,  \cite{Tsai1994a,Bagratashvili1996a}. These issues will have to be further investigated.
\item \textbf{use an even shorter UV wavelength of $\le 50\,$nm}\\
For even shorter wavelengths, fused silica becomes transparent again. Lasers at this wavelength are available \cite{Heinbuch2005a}, and we are confident that the need of the semi-conductor industry for short-wavelength lasers will lead to a fast increase of the TRL of these devices.
\item \textbf{improve the material and/or environment parameters}\\
Figure \ref{fig::CEDfuture} shows how various changes to the absorption or mass-density of the material of the nanosphere and/or a lower temperature of the environment can lead to significantly larger values of the CED.
\end{itemize} 

Even if we forget about this specific problem, the way to go seems to be given by the last point in the list. If we compare figures \ref{fig::CED} and \ref{fig::CEDfuture}, we see that only a significant improvement of the material parameters of the nanospheres used as well as improvements on the environment temperature will allow to test all the macrorealistic models considered here. While the improvement of material parameters can be pursued independently of the overall design of DECIDE, a further reduction of the environment temperature will require an adaptation of the design of the thermal shield and possibly of the whole spacecraft. The design of the thermal shield as it is presented here only allows for a minimum temperature of $30-40\,$K.

Most of the changes that would be needed for a violation of the K model and the Di\'osi-Penrose model aim at reducing the decoherence rates due to the scattering and the emission of blackbody radiation. These are the main decoherence mechanisms for an ultra-high-vacuum and low-temperature environment. In addition to these changes, it would help to use materials with higher mass density (see right-hand plot of figure \ref{fig::CEDfuture}) because that reduces the CED predicted by all macrorealistic models.

It should be noted that while CET and CED are very useful tools for roughly estimating parameters for which the predictions of quantum theory violate the predictions of macrorealistic models, recent results \cite{Kaltenbaek2012a} show that a central assumption we have made here is too simple. In particular, we have assumed that $t_1 = t_2$. This assumption was based on the approximation that the double slit has infinitely sharp edges. This can be compared to a standard double slit experiment, where the two slits are very narrow compared to their distance. In a more detailed analysis that takes into account that the edges of the double-slit are smooth (the slope is determined by the UV wavelength), it turns out that we must have $t_2 \gg t_1$ in order for the two parts of the wave packet to recombine \cite{Kaltenbaek2012a}. In a very similar context, the condition that $t_2 \gg t_1$ also occurs in Refs.~\cite{RomeroIsart2011b,RomeroIsart2011c}. These new results require slight improvements of the overall design of the mission (lower environment temperature, lower absorption materials) but the central concepts developed here as well as our main conclusions remain the same.

\subsection{Using optically trapped microspheres for an all-optical inertial sensor}
\label{sec::cat}
For many space applications, it is of imminent importance to accurately measure accelerations. In many recent and planned missions, capacitive inertial sensors are used for this purpose, e.g., the sensor ASTRE was used in various space-shuttle missions, the accelerometer STAR was used for the Earth-observation mission CHAMP. Further examples are the accelerometer used in GOCE \cite{Onera2008} and the one that is going to be used in MICROSCOPE\cite{Touboul2001a}). The sensitivity of the intertial sensors in GOCE and MICROSCOPE are around $10^{-12}\,\mathrm{m s^{-2}}$. Many of the inertial sensors for these missions are made by the French company ONERA \cite{Onera2008}, and we will, therefore, sometimes refer to capacitive inertial sensor as ONERA sensors. 

Such sensors typically have a small dynamical range. Because of their size, they have a large cross section for cosmic radiation that charges the test masses. Thin wires are used to discharge the test masses but these wires further restrict the dynamical range as well as the sensitivity of the devices. 

The novel design of an inertial sensor proposed here is based on the use of an optically trapped microsphere as a mechanical resonator. Its mean position can then be used to determine the acceleration of the spacecraft. While DECIDE is based on optically trapped nanospheres, it is also possible to optically trap significantly larger spheres \cite{Ashkin1971a} and to optically read out changes in their position (see, e.g., \cite{Volpe2007a}). While the goal of CASE is to achieve acceleration sensitivities similar to state-of-the-art capacitive sensors ($\sim 10^{-12}\,\mathrm{m s^{-2}}$ as in GOCE \cite{Onera2008} or MICROSCOPE \cite{Touboul2001a}), CASE in its currently suggested form exhibits several limitations due to the heating of the center-of-mass motion by the trapping laser and the sensitivity of the read-out mechanism. Yet, promising, alternative designs are currently under investigation, and CASE, in its current or a slightly modified form, is interesting due to several reasons. CASE promises a larger dynamic range and a significantly smaller cross section for cosmic radiation as well as a comparatively large distance between the microsphere and any surrounding elements that might also experience charging due to radiation. CASE also allows for an easy comparison of gravitational acceleration for different materials by loading microspheres of various materials into the optical trap. Finally, CASE allows for a test of the universality of free fall, i.e., the weak equivalence principle, with vastly different masses in the tradition of Galilei's original experiments. While the test mass of the capacitive sensor is around $0.1\,$kg, the microspheres can have masses between $10^{-14}\,$kg and $10^{-8}\,$kg if we assume them to be fused silica spheres with radii between $1\,\mu$m and $100\,\mu$m.

Theories that aim for a unification of the standard model of physics with general relativity in general predict deviations from the equivalence principle \cite{Damour1996a}. This has triggered a renewed interest in tests of the equivalence principle (see, e.g., \cite{Touboul2001a,Adelberger2009a}). CASE implements such a test of the equivalence principle. In contrast to other experiments, it allows to compare the free fall of test masses with significantly different masses, and it allows for the flexibility to perform tests with nanospheres of different materials, providing an interesting test bed for the search of possible violations of the equivalence principle due to modification of the standard model of physics \cite{Damour1996a}.

The sensitivity of the test of the universality of free fall is higher for strong gravitational field gradients. Depending on whether MAQRO is operated either in a HEO or in an orbit around L1 or L2, the gravitational acceleration is either up to $\sim 0.4\,g$ or $\sim 0.06\,g$. Here, $g=9.81\,\mathrm{m s^{-2}}$ is the gravitational acceleration on the surface of the Earth. If a HEO is chosen, CASE could be performed when the spacecraft is close to the perigee of the orbit, while DECIDE is ideally performed as far as possible from Earth.

\section{Proposed Payload Instrument Requirements and Design}
\subsection{Overview over all elements}
The MAQRO mission comprises two independent experiments named
\begin{enumerate}
  \item DECIDE (DECoherence In a Double-slit Experiment) and
  \item CASE (Comparative Acceleration Sensing Experiment).
\end{enumerate}
Although DECIDE and CASE are not cleanly split into two instruments (they share the laser and data-management unit), an alternative cold redundancy concept (i.e., a separate laser and data-management unit for CASE) would easily allow to define two separate instruments. The two experiments comprise the principle subsystems and units shown in table \ref{table_experimental_components}.

\begin{table}[t]
 \begin{center}
  \begin{tabular}{lll}
   \textbf{Experiment} & \textbf{subsystem} & \textbf{component}\\
   \hline
   DECIDE && \\
   & Nano-sphere trap & Optical bench (exterior) \\
   && CCD chip \& electronics \\
   && IR Laser Assembly \\
   && UV Laser Assembly \\
   && Cryo-Harness (optical \& electrical)\\
   & Thermal control subsystem & Heat shield and struts \\
   && Launch Lock mechanism \\
   & Data Management Unit & Processor \\
   && Software \\
   \hline
   CASE && \\
   & Microsphere accelerometer & Optical bench (interior)\\
   && Phase-meter \\
   && IR Laser Assembly \\
   && Venting ducts \\
   & Gravitational reference sensor & Sensor Unit \\
   && Interface \& Control Unit \\
   & Data-Management Unit & Processor \\
   && Software \\
  \end{tabular}
  \caption{Overview over the experimental components used in DECIDE and CASE. \label{table_experimental_components}}
 \end{center}
\end{table}

\begin{figure}[th]
 \begin{center}
 \includegraphics[width=0.49 \linewidth]{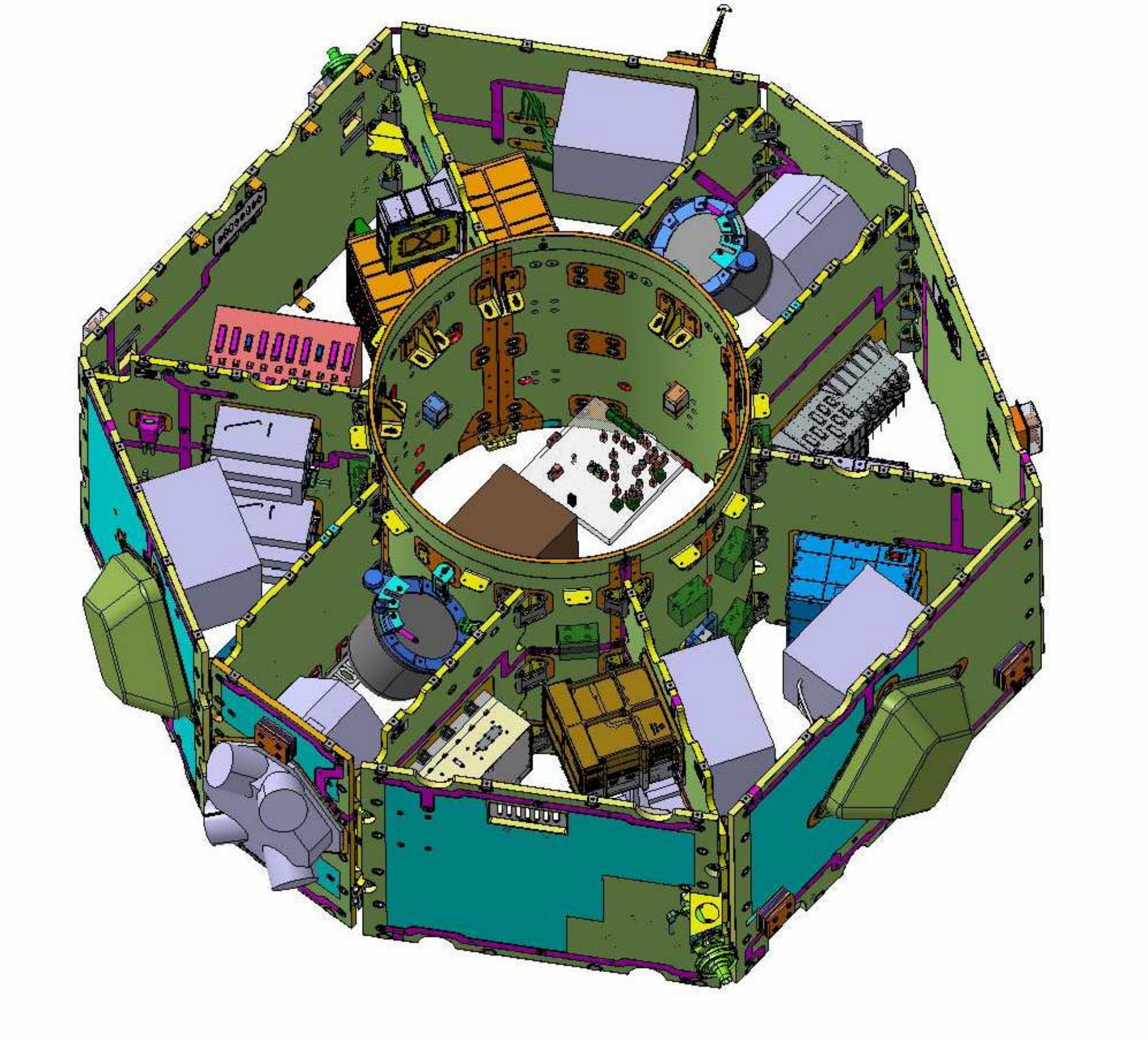}
 \includegraphics[width=0.49 \linewidth]{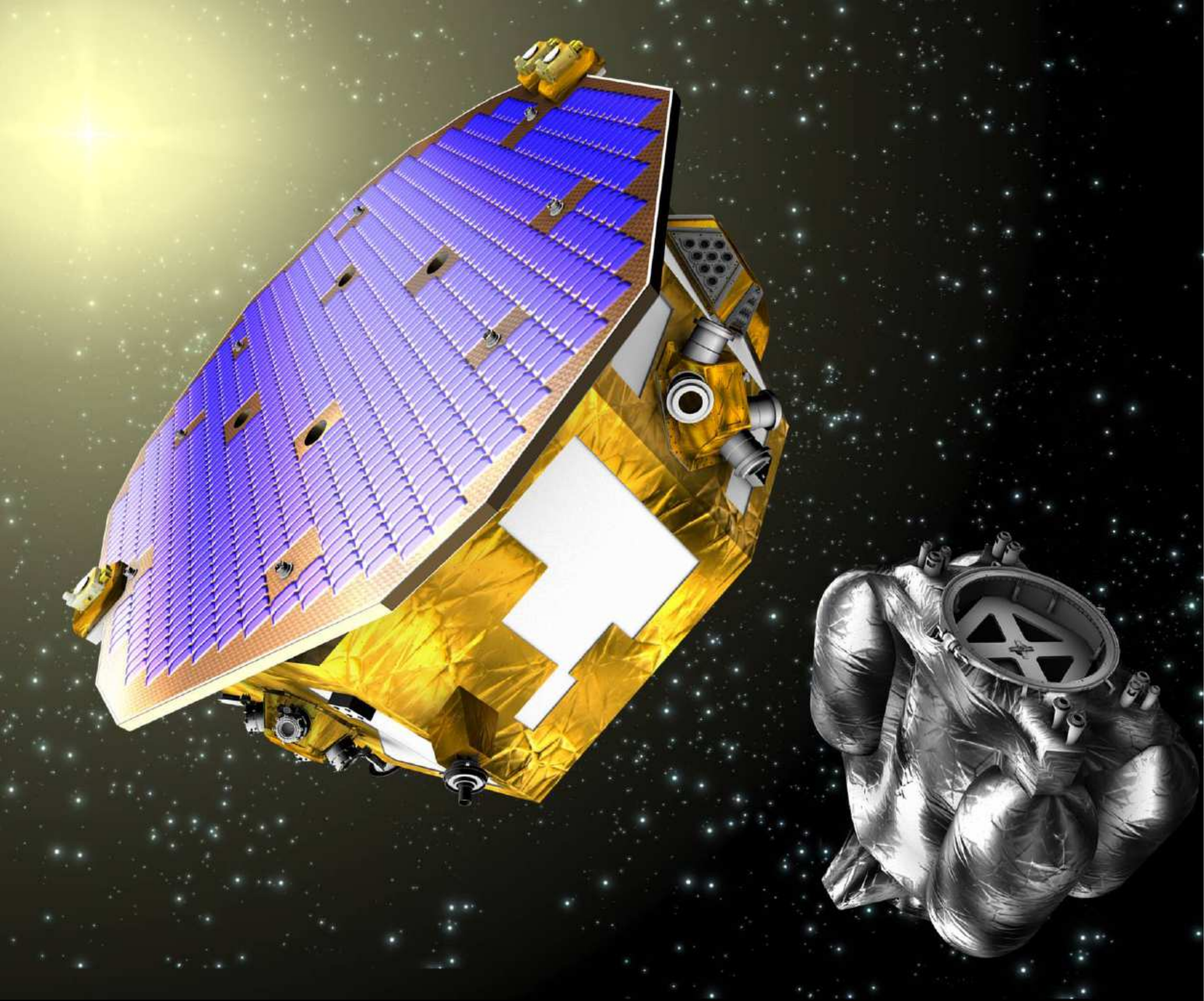}
 \end{center}
\caption{left: The octagonal structure of the LPF science craft right with the inner cylinder containing the optical bench and reference sensor for CASE. The compartments around the cylinder contain the various units and equipment. right: The LPF science-craft with the solar array on top is separated from the propulsion module. Image source: ESA, \cite{LPFgraphics}.}
 \label{figure_LPF_schematic}
 \end{figure}

The instrument package is largely based on technologies developed for ESA's LISA Pathfinder (LPF) mission payload, the LISA Technology Package (LTP), and could also take advantage of the LPF science-craft with MAQRO-specific adaptations and the attached propulsion module. This is shown in Figure \ref{figure_LPF_schematic} right.  The LPF science-craft architecture consists of an asymmetric octagonal carbon-fiber-reinforced-plastic (CFRP) structure with shear walls and an inner cylinder of $\approx0.8$ m diameter and various units and equipment attached to the panels (see figure \ref{figure_LPF_schematic}). For illustration, the LPF central assembly has been replaced by MAQRO instrument envelopes to indicate location and fit: in the inner cylinder one can see the optical bench for CASE and a cube indicating the reference acceleration sensor (200 mm edge length).

\subsection{CASE Design}
In CASE, two acceleration sensors are used: The microsphere-trap inertial sensor and the reference accelerometer. The sensitive cavity axis of the former defines the x-axis and is aligned with the respective x-axis of the reference sensor (ONERA accelerometer). The Drag-Free Attitude and Control System (DFACS) takes the input from the reference sensor to control the micro-propulsion thrusters of the spacecraft. As soon as the reference test-mass (and the microsphere as well) moves away from its nominal initial position, the  DFACS commands the thrusters such that the spacecraft remains centered on it. Unlike LPF \cite{Fichter2005a}, MAQRO only uses drag-free control referenced to a single test-mass (the reference sensor) and only along one degree of freedom (the x-axis), which greatly simplifies the control and the propulsion system requirements. Note that in default operating mode the position of the second test-mass (microsphere sensor) does not feed-back to the DFACS but is coupled to the spacecraft via laser metrology which is referenced itself to the S/C structure. A schematic of the MAQRO DFACS in default mode is shown in Figure \ref{figure_CASE_setup} below.
\begin{figure}[th]
 \begin{center}
 \includegraphics[width=0.79 \linewidth]{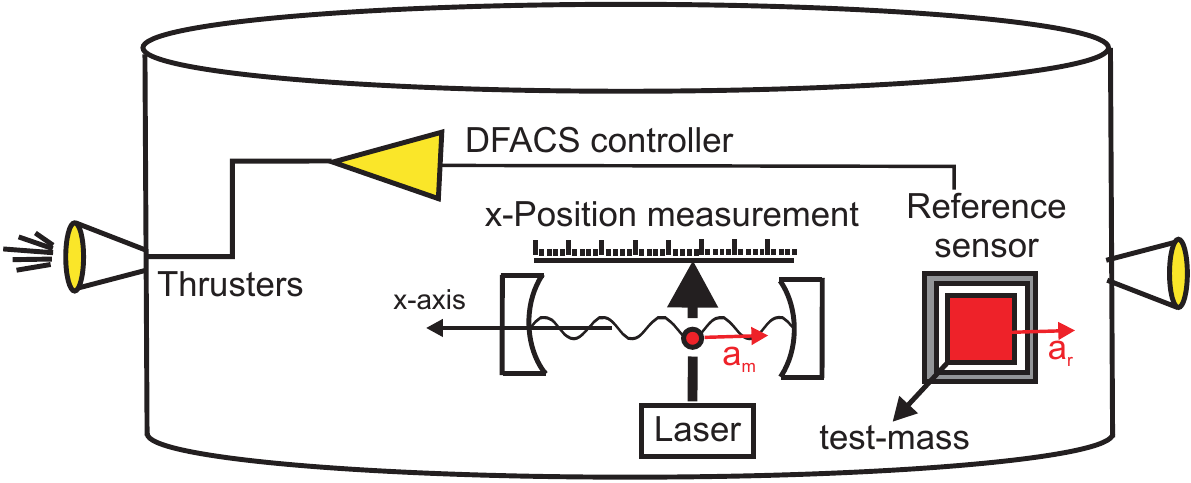}
 \caption{The drag-free attitude and control system of CASE. The spacecraft is symbolized by the cylinder with thrusters attached to the sides. The microsphere (red dot) is manipulated and its position sensed by the two laser beams. The DFACS reference is provided by an ONERA accelerometer.}
 \label{figure_CASE_setup}
 \end{center}
\end{figure}
The control laws and sensitivities are specified such as to meet the central science requirement for CASE, namely to measure the differential acceleration between the microsphere and the reference sensor with an accuracy better than ${\rm 10^{-12}ms^{-2}/\sqrt{Hz}}$ in the measurement band between 1 mHz and 1 Hz.\\
One must keep in mind that the static and dynamic gravity-field gradients inside the spacecraft cannot be completely nulled (e.g. for LPF there is an acceleration of ${\rm 10^{-9}ms^{-2}}$ of each test-mass due to remaining gravitational field gradients) and that the remaining gravitational field gradient is not known better than ${\rm 10^{-11}ms^{-2}}$ (typical error for LPF). The error on the remaining gravitational field gradient is determined by the finite accuracy of the spacecraft mass model which documents the exact position and mass distribution of all units and the spacecraft structure. This gradient error makes it impossible to discern whether any measured acceleration differences originate from the gradient error itself or from a violation of the weak equivalence principle. 
To push the measurement accuracy further, either the accuracy of the gravity model has to be improved, or re-calibration is used to determine the uncertainty in the interior gravitational field gradient.

One -- rather sophisticated -- way to perform such a calibration, could be through a movable compensation mass between the two test-masses. Another possibility would be to calibrate the gradient through rotation of the measurement axis by 180 degrees:

Assuming that the  measurement axis (x-axis) is orthogonal to the spacecraft cylinder axis (as depicted in figure \ref{figure_CASE_setup}) and that the spacecraft flies a highly inclined HEO, the measurement axis is approximately aligned with the direction of the earth gravitational field gradient during the first passage of the perigee. If, during the second passage, the spacecraft cylinder is rotated by 180 degrees with respect to the previous orientation, the sign of the spacecraft gradient on the two respective test-masses is similarly reversed with respect to the initial orientation, from which it is possible to infer the remainder gravitational field gradient.
Most of the thruster noise is removed when the two measured accelerations are subtracted from one another. However, imperfect common mode cancellation sets an upper limit on the allowed thruster noise and - assuming the common mode noise rejection is identical to LPF - requires the thruster noise for MAQRO to be less than ${\rm 2\times10^{-10}ms^{-2}/\sqrt{Hz}}$ .

\subsection{DECIDE Design}
\label{sub::benchDECIDE}
Figure \ref{fig::benchDECIDE} (left) shows a top-down view of the optical bench (20$\times$20 cm) for DECIDE, which is attached outside of the spacecraft, as is illustrated in the description of the thermal shield in subsection \ref{sub::thermal}. Figure \ref{fig::benchDECIDE} (right) shows a corresponding 3D illustration of the setup where it is possible to see the three struts connecting the setup to the spacecraft inner cylinder. The central elements of the experimental setup as illustrated are the following:

\begin{figure}[t]
 \begin{center}
  \includegraphics[width=0.54\linewidth]{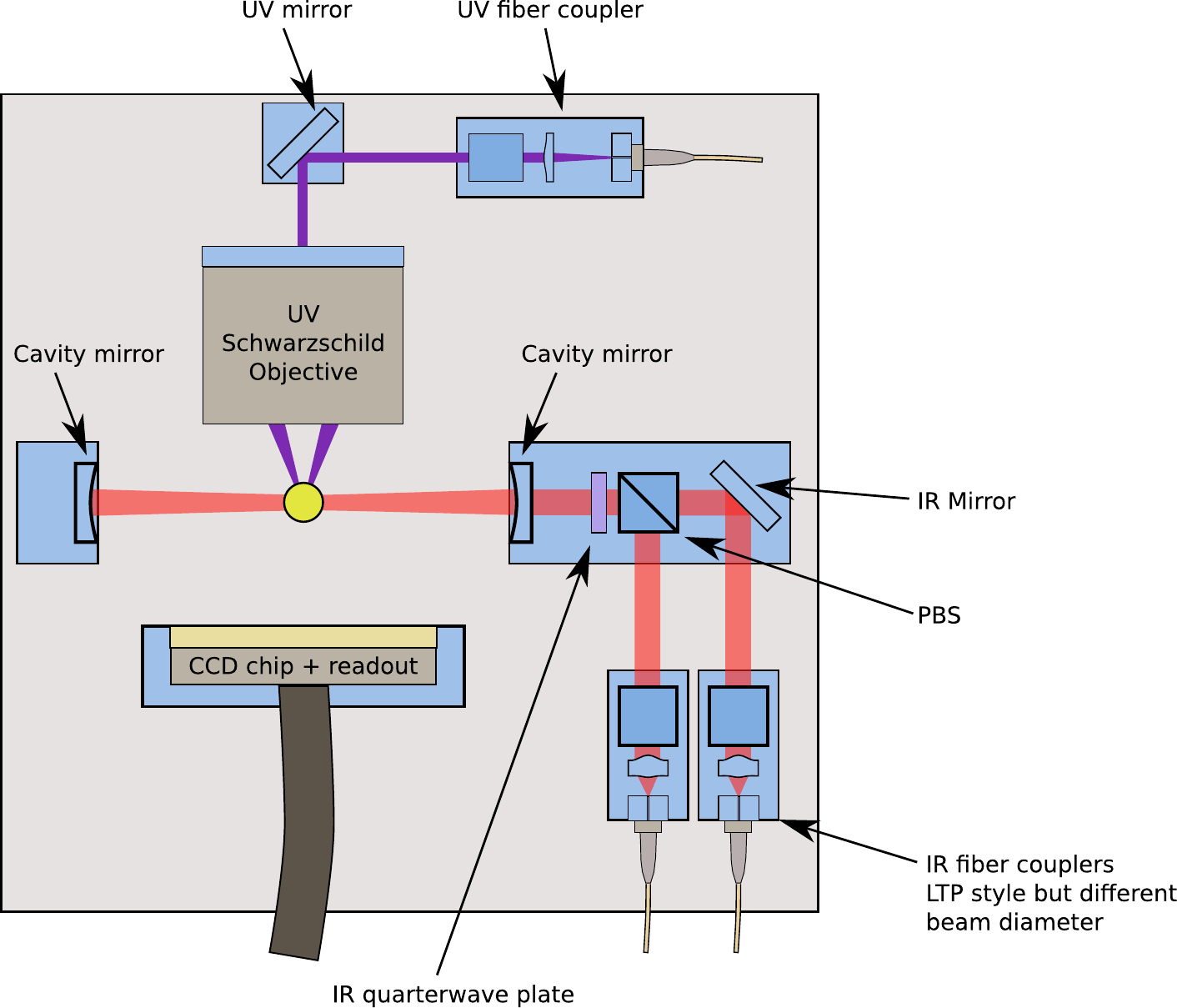}
  \includegraphics[width=0.44\linewidth]{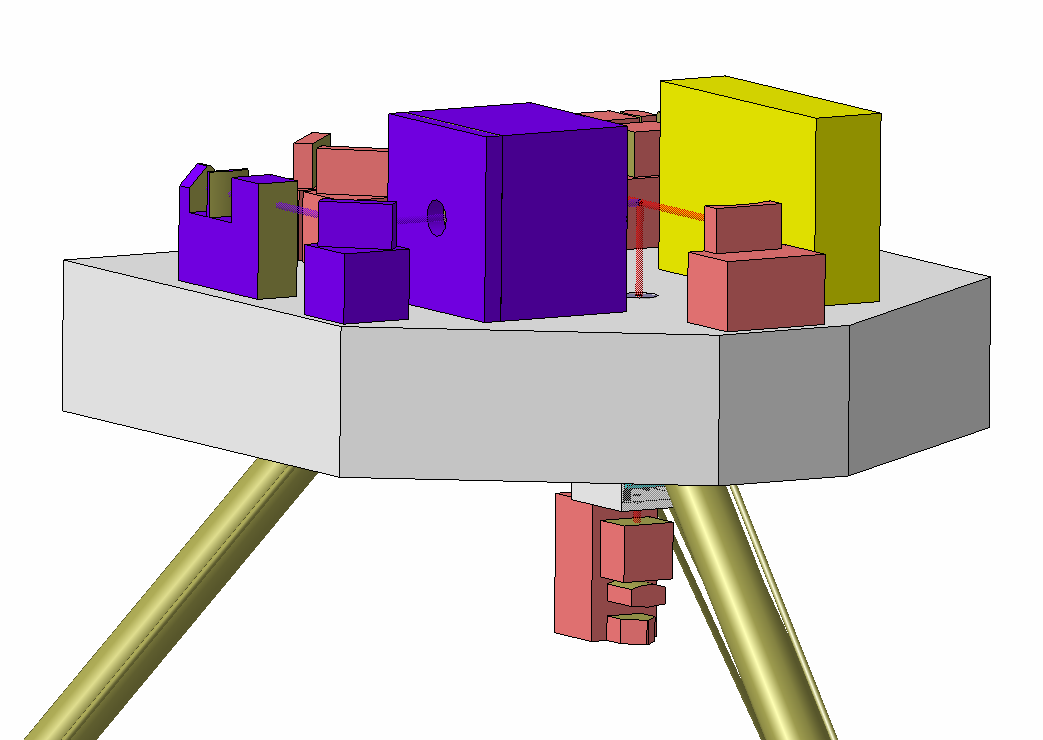}
   \caption{\textbf{Optical bench for DECIDE.} (left) Top-down view of the optical bench. (right) Corresponding 3D representation.
  \label{fig::benchDECIDE}}
 \end{center}
\end{figure}

\begin{itemize}
 \item A confocal cavity with finesse $\sim 10^4$ for trapping, cooling, manipulating and high-precision position readout of dielectric nanospheres. The mirrors have a curvature radius of $5\,$cm, are separated by $10\,$cm, and have reflectivities of $99.95$\% and $99.99$\% for the input and end mirror, respectively. The beam waist within the cavity is $10\,\mu$m.
 \item A high-numerical-aperture reflective objective for focusing a UV beam for particle manipulation. Focal length: $13\,$mm, numerical aperture: $0.4$, working distance: $24\,$mm.
 \item Polarization optics to separate the beam reflected from the cavity (for signal readout and Pound-Drever-Hall cavity locking) from the input laser beam. These modes are shared by two frequency-shifted beams, one for optical trapping, another for back-action cooling. The laser ($1064\,$nm), detectors etc. are placed on the laser module within the spacecraft (not shown). The laser beams are directed onto the exterior platform via single-mode fibers.
 \item A CCD camera (or alternatively a CMOS camera) for the observation of UV light scattered from trapped particles and for detecting particle positions for calibration purposes. A UV lens (not shown) with a focal length of $5\,$cm is used to image the experimental region onto the camera. The position of the camera in the figures is meant to be illustrative but in the actual setup the camera should be mounted such that the UV beam does not hit it. A possible position would be above the reflective objective and angled down toward the experimental region.
 \item UV single-mode fiber coupler for $350\,$nm light. Similar to the NIR single-mode fiber couplers, this coupler will be used to collimate the UV laser beam supplied through a single-mode fiber that is connected to a low power ($\sim 10\,$mW) UV laser within the spacecraft.
 \item A nano-particle loading mechanism is mounted below the optical bench to load the cavity with nanospheres whenever needed. This loading mechanism is described in detail in section \ref{subsub::loading}.
 \item A quadrant-diode (not shown) to measure NIR light scattered from the trapped particle and a lens with a focal length of $3 -5\,$cm to image the light onto the diode.
\end{itemize}

\begin{figure}[t]
 \begin{center}
  \includegraphics[width=0.8\linewidth]{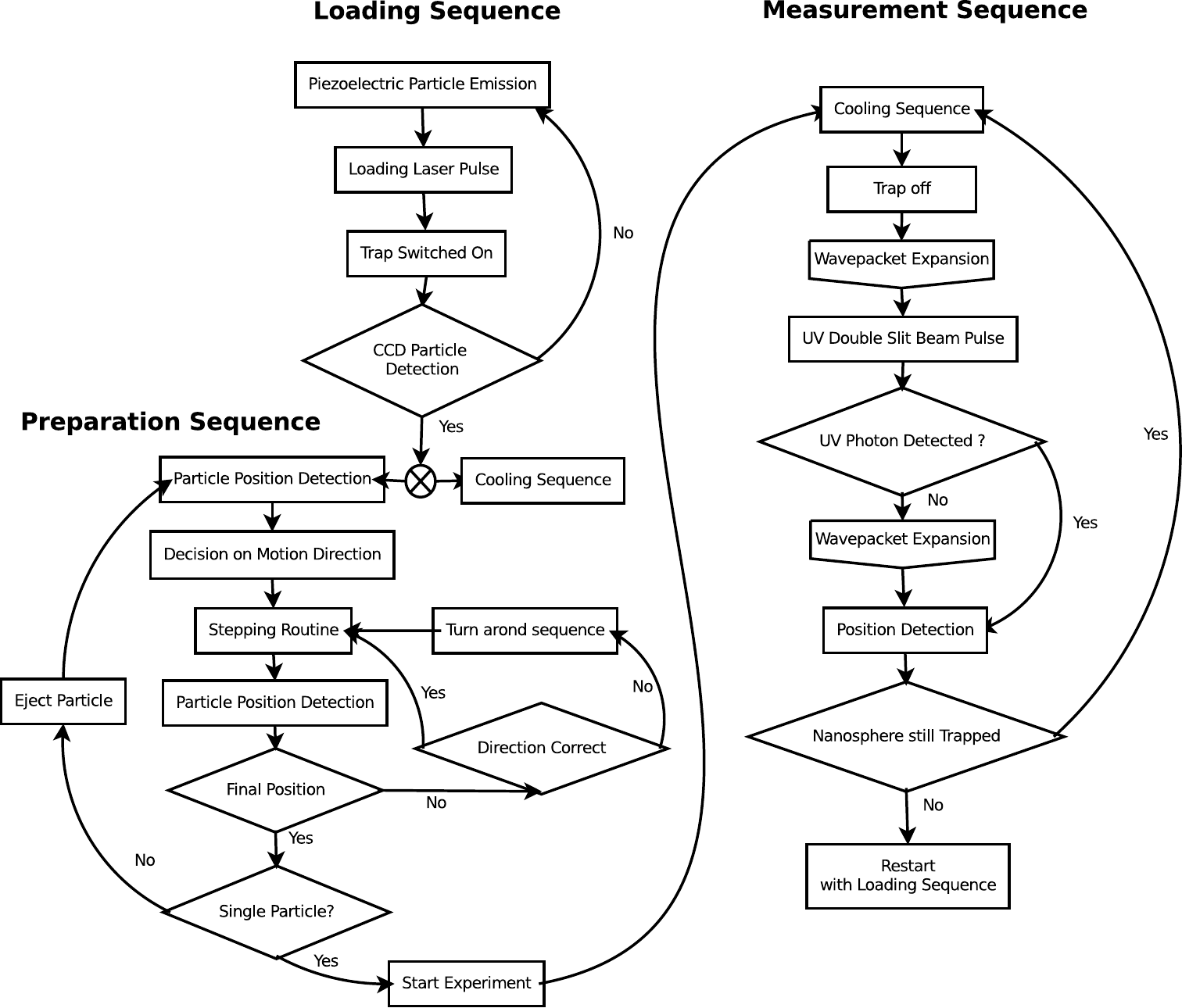}
   \caption{\textbf{Operating procedure for DECIDE.}
  \label{fig::flowchart}}
 \end{center}
\end{figure}

\subsection{Operations and measurement technique}
\label{sub::operations}
The overall flow of operations for DECIDE is described in the experimental flow chart in Fig.~\ref{fig::flowchart}. It can be divided into three distinct sequences. They will be described in the following.

\subsubsection{Loading and manipulating single nanospheres}
\label{subsub::loading}
DECIDE as well as CASE need reliable mechanisms to load and manipulate nano- and microspheres, respectively. We will concentrate on the procedure as envisioned for DECIDE. It is adaptable to CASE, and some aspects will even be significantly simpler; for example, the deterministic piezoelectric ejection is easier for micro- than for nanospheres because of the lower Casimir forces.

The nano-particles can be ejected probabilistically from piezoelectric elements. Techniques to achieve high enough accelerations to overcome the Casimir force need to be developed. Possible approaches are to micro-structure the surface of the piezoelectric element or to use surface acoustic waves to generate the necessary forces. A short laser pulse is used to velocity select the released nanospheres and to weakly accelerate them towards the trapping beam.

Once particles are optically trapped within the standing wave formed by the trapping beam within the cavity, another, frequency-shifted beam can be used to move the particle between possible trap positions \cite{Schrader2001a}. If there is more than one particle trapped simultaneously, spurious particles can be ejected via the radiation pressure from a UV pulse that is focused onto a specific trap. As soon as only a single particle is trapped within the cavity, experimental runs can begin. In principle, the loaded particle can be kept within the optical trap indefinitely and can be used for repeated experiments. After each experimental run, one can manipulate the particle and cool it to the ground state again.

\subsubsection{Preparation and detection}
\label{subsub::exprun}
For an overview over a typical experimental run, also see section \ref{subsec::interference}. Once the nanosphere's center-of-mass motion (CM) has been cooled down close to the quantum ground state, the trapping and cooling beams are switched off, and the wavefunction will expand freely \cite{RomeroIsart2011a,RomeroIsart2011b,RomeroIsart2011c}. After a time $t_1$, when the wavefunction is wider than the focus of the UV objective, a pulse of UV light is sent through the center of the wave packet. If no photons are scattered off the particle, the nanosphere must have been either left or right of the position of the UV beam. If light is scattered, the wavefunction will decohere, and the nanosphere will be well localized. 

Because we cannot determine for every UV photon whether it has been scattered, the density matrix of the nanosphere will be a statistical mixture of a decohered part and a coherent superposition similar to a Schr\"odinger cat state. The distribution of particle positions after many repetions of the experiment will then exhibit interference fringes on top of a broad Gaussian distribution from the decohered part of the density matrix.

After waiting for a time, $t_2$, during which interference fringes form, the cavity beam is switched on in order to measure the position of the nanosphere. Repeating this procedure yields a histogram of measured positions. According to quantum theory, this histogram should form an interference pattern. Such histograms and whether they exhibit interference fringes will be the main result of DECIDE. Note that recent, yet unpublished results \cite{Kaltenbaek2012a} indicate that $t_2$ has to be significantly longer than $t_1$ in order for the two parts of the wavefunction to overlap again after the preparation of the double slit (see also sections \ref{sub::DECcomp} and \ref{subsec::CaseForSpace}).

\subsection{The thermal shield}
\label{sub::thermal}
\subsubsection{General shield design}
The DECIDE experiment requires an experimental volume of very low temperature (we assume $32$ K) that cannot be achieved anywhere inside the spacecraft. For this purpose we propose an external heat shield with an onion-type design structure that uses heritage from the Darwin Proposal \cite{Leger2007} and the Gaia mission \cite{GAIA2011}. A schematic of the MAQRO heat shield is given in Figure \ref{figure_thermal_shield_3d} below:
\begin{figure}[th]
 \begin{center}
 \includegraphics[width=0.49 \linewidth]{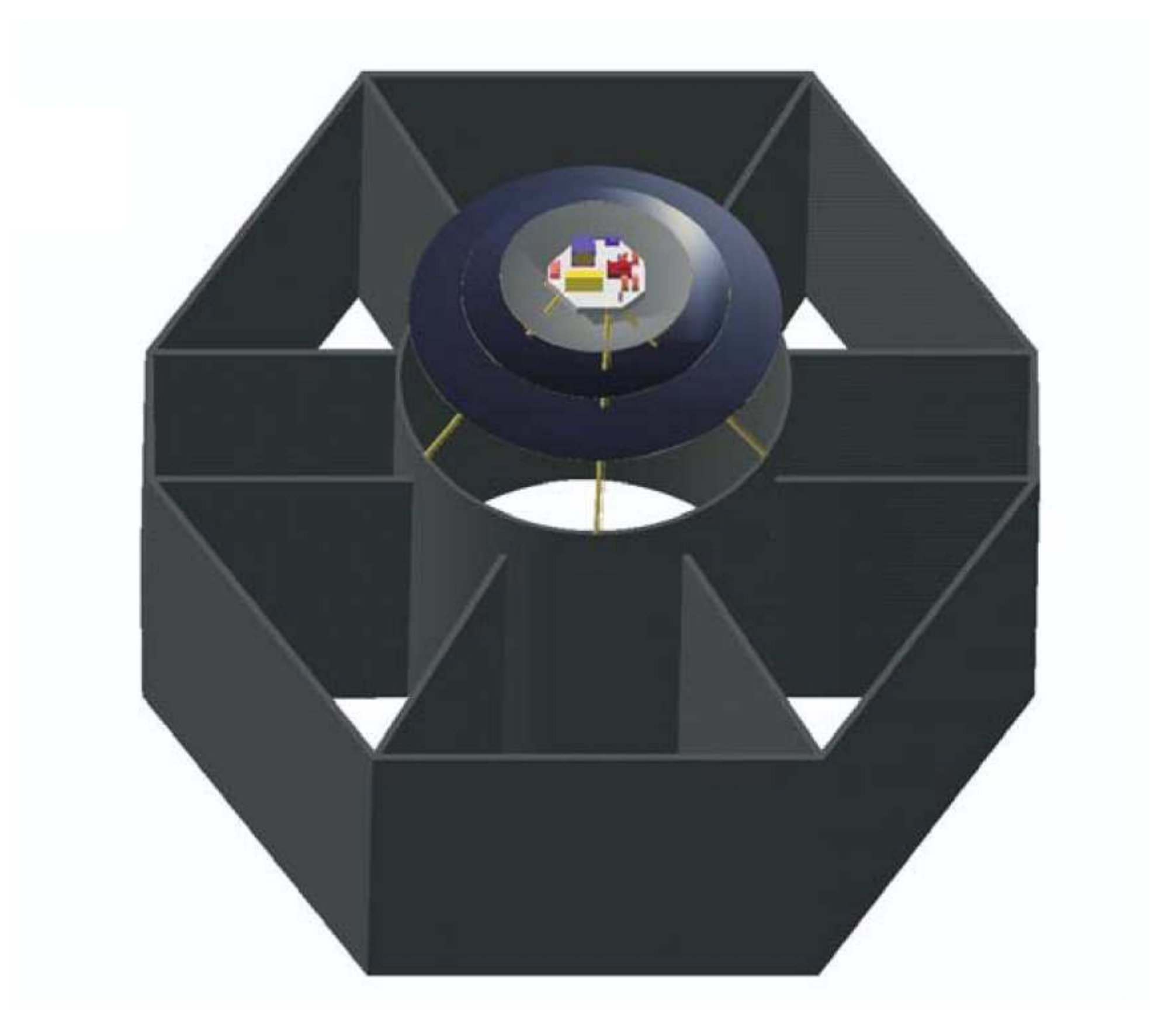}
 \includegraphics[width=0.49 \linewidth]{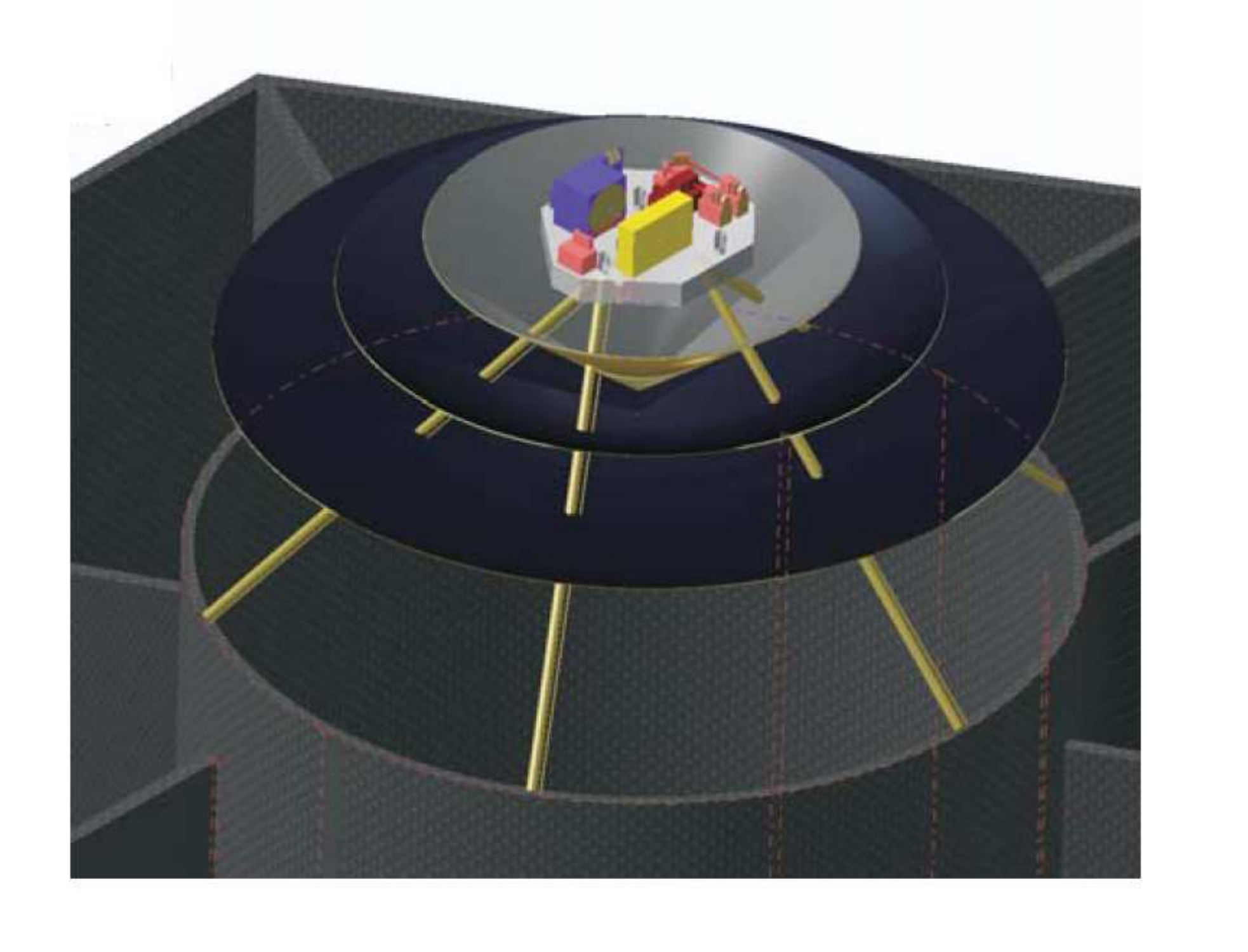}
 \caption{Left: The heat shield is attached to the spacecraft cylinder of the LISA Pathfinder science platform. Right: A close-up of the 3-layered conical heat shield design and the optical bench with various optical elements. }
 \label{figure_thermal_shield_3d}
 \end{center}
\end{figure}
The shield consists of three layers in the shape of either three cones or optionally, for ease of manufacturing, three pyramids in nested configuration. The vertex of the outermost cone is  $\approx12~cm$ distant from the spacecraft surface and there is $\approx5~cm$ spacing between the vertices of the individual pyramids. The angle  between the pyramid edge and the spacecraft plane gradually increases from $12^{\circ}$ for the outer pyramid, $24^{\circ}$ for the middle pyramid, to $36^{\circ}$ for the inner pyramid. Such a design with varying opening angles obviously improves the radiative cooling effect of the individual shields compared to a design with identical opening angles, giving each shield a greater solid angle for interaction and photon emission to deep space.
The shield structure is placed on the cold side of the spacecraft opposite to the solar panel. It is designed to fit well into the inner spacecraft cylinder to which the separable propulsion module is attached. The shields are gold coated on the side facing the spacecraft and have high emissivity (black coating) on the side facing deep space. The temperature of the spacecraft outer surface, i.e., the outermost layer of the multi-layer insulation (MLI) is assumed to be 150 K-170 K, the surface temperature of the outer shield is $\approx120~K$, of the middle shield $\approx70~K$, and of the inner shield $\approx30-40~K$. Note that the width of all shields is chosen sufficiently large so that no part of the ``hot'' spacecraft surface or outer shield layer is in direct line of sight with any optical bench component harbored by the innermost (coldest) shield. The shield is mechanically attached to the 3 pairs of rods of a tripod which are fixed at the inner cylinder of the spacecraft.
The thick, stable rods support the structure during the launch phase whereas the thin fragile rods which are drawn in parallel to the former ones support the structure after commissioning. Although the thick rods are built from a material of very low thermal conductivity (e.g. CFRP) their comparatively large cross-section - required for reasons of mechanical stability - is still conductive to heat transfer and limits the achievable temperature. To reach temperatures as low as 30 K it is therefore necessary to break the material prior to commissioning to interrupt the thermal flow through it. This can be achieved by controlled release of a spring-mechanism, or by a solution based on pyro nuts as applied in the GAIA mission\cite{GAIA2011}. To minimize remaining effects of thermal photon emissivity, which additionally deteriorate the thermal balance, the thick rods are covered by MLI with a low-emissive finish. Note that the harness lines leading to/from the experiment (not drawn in the figure) on the optical bench (2 IR glass fibers, 1 UV glass fiber, and 1 CCD sensor readout line) are either attached to one of the supporting rods or guided through one of the rods which is designed for low thermal conductivity. More detailed analysis for the optimal concept is required.

\subsubsection{Protective shield cover and bake-out}
The surface area of the shield (all layers) is approximately 1${\rm m^2}$. For a first conservative estimate, we assume a specific density of ${\rm \rho_{Al}\approx 2.7g/cm^3}$ (aluminum) and a thickness of 1 mm with a mass of ${\rm m\approx3~kg}$. Using a chemically inert and minimally outgassing refractory metal for the innermost shield, e.g. Niobium with ${\rm \rho_{Nb}=9~g/cm^3}$, would increase the mass to $\approx4$ kg. The hollow struts are made from CFRP of very low thermal conductivity and expansion as well as good mechanical stability. They are 40 cm long, 2 cm in diameter and have a wall thickness of 1.6 mm, giving a combined mass of less than one 1 kg: ${\rm m_{struts}\approx 0.6 kg}$.  The struts are fitted to the bushings inserted into the base-plate of the optical bench. Each of the three inserts has a mass of $\approx200$ g. The load-carrying struts are fixed to the spacecraft inner cylinder by launch lock mechanisms. Each of the 6 mechanisms has a mass of about 300 g.
The total mass of the shield with inserts and launch lock mechanism plus harness is approximately ${\rm m_{tot}\approx 7kg}$.
\\
During launch and before commissioning, the thermal shield is covered by an additional protective cover. The mass of the cover is estimated to be $\approx5$ kg. In an alternative configurational concept to the fixed and static shield depicted in Figure \ref{figure_thermal_shield_3d}, the whole shield and experiment assembly could be harbored inside the central structural cylinder and deployed by a dedicated mechanism once in space.
\\
Vacuum quality and low outgassing are key requirements for DECIDE. From our analysis, we found that outgassing is practically completely frozen out at temperatures as low as $\approx30$ K. Nevertheless, mainly as a means of risk mitigation for as yet unaccounted effects, it would be very useful to consider bake-out of the thermal shield and the exterior optical bench before commissioning. For that purpose heaters could be attached to the outermost shield and the optical bench. Considering that the  total radiative surface of inner shield plus optical bench is approximately ${\rm 0.23 m^2}$, we require a heating power of ${\rm P = 105\,W}$ if we bake-out at 300 K, and a heating power of ${\rm P = 330\,W}$ if we bake-out at 400 K. Providing 100 to 150 W for that purpose during commissioning phase while many units are inactive would - as an example - pose no problem for LPF resources.

\subsubsection{Single-mode fibers at cryogenic temperatures}
From previous studies \cite{Flatscher2011} we find that single-mode glass fibers can in principle be operated at temperatures as low as 10 K without structural damage to the core. The study described in \cite{Flatscher2011} deals with the design, manufacturing, and extensive testing of single-mode waveguides in the mid-infrared for a typical Darwin \cite{Leger2007} application. The environmental tests comprised a vacuum test at ambient temperature, a cryogenic test at 10 K, proton radiation test, and gamma radiation tests. All performance tests were done at CO-laser wavelength of 5.6 micron and at CO2-laser wavelength of 10.6 micron (representative of the wavelength used in MAQRO) within a Darwin-representative interferometer. A major conclusion from the study is that the low temperatures cause no problem for the fibers themselves but the connector design has to cope with the variable expansion coefficients of the materials used (from bake-out to commissioning).

\subsection{Vacuum Requirements}
\label{subsub::vacuum}
Missions that also deal with rather stringent requirements on vacuum in payload elements are LISA Pathfinder \cite{Armano2009} and LISA \cite{Bell2008}. In the former, the vacuum is maintained inside a vacuum enclosure (inertial sensor) and the required pressure is $10^{-7}$ mbar. In the latter, a choice has been made not to use a vacuum enclosure but to vent to space in order to achieve the required pressure of $10^{-8}$ mbar. This can be conveniently achieved after venting to space for a duration of several weeks, as shown in \cite{Hammesfahr2005}.
Based on the latter study for an alternative vacuum concept for LTP \cite{Hammesfahr2005}, we shall also use ``venting to space'' to achieve a good vacuum of $10^{-9}$ mbar for the interior experiment (CASE) of MAQRO. Two aluminum tubes of 10 cm diameter and 1.2 m length vent the molecular gases on the ``cold side'' of the spacecraft, where the pipes pass the exterior heat shield and therefore avoid contaminating the DECIDE experiment. The feasibility of such a concept from a vacuum as well as a thermal balancing and stability point of view has also been demonstrated in \cite{Hammesfahr2005}.

For the DECIDE experiment, the vacuum requirements are much more stringent
(lower than $10^{-14}$ mbar). While such or even lower pressures are 
achievable in a lab environment \cite{Gabrielse1990a}, it is not so straight 
forward to achieve such pressures while still allowing for full optical access and the scattering of not insignificant amounts of light. In our proposal, we achieve the necessary vacuum conditions by using a 
platform outside the spacecraft taking advantage of the vacuum and low 
temperature environment provided by space itself as long as our experimental 
apparatus is shielded well enough from the rest of the spacecraft (see section \ref{sub::thermal}). In particular, the experimental setup of DECIDE is 
shielded from solar wind and S/C areas and takes advantage of very low 
outgassing materials for elements adjacent to the sensitive zone. 
Furthermore, at such low temperatures the outgassing is practically frozen 
out. We shall briefly discuss some key vacuum aspects applicable to MAQRO:

\subsubsection{Outgassing from a plane}
Quite generally, the ougassing rate $D_{out}[kg/s]$ is given by the following expression \cite{Hammesfahr2005},\cite{Cho2004}.
\begin{equation}
\label{outgassing_rate}
    D_{out}=m_{tot}\times\sum_{species\hspace{3pt} i}\frac{TML_{i}(\%)}{100}\frac{e^{-t/\tau_i}}{\tau_i}
\end{equation}
where $\tau_i$ is the outgassing time constant of molecular species $i$ for a certain material of mass $m_{tot}$  and $TML_{i}(\%)$ is the total mass loss through outgassing of species $i$ in percent. In \cite{Hammesfahr2005},\cite{Cho2004} an outgassing analysis for the Kompsat-2 mission was performed from which outgassing rates for certain materials were deduced. Typically every material outgasses various molecular species with different outgassing time constants and total mass loss ratios. Those molecular species which outgas with very short time constants (on the order of a few hours up to some hundred hours) are not considered anymore as they are negligible on a mission timescale, in particular after bake-out.
From the remaining molecular species the dominant ones are listed for three different materials in table \ref{table_outgassing_parameters}:
\begin{table}
\begin{tabular}{lrr}
  Material  & \multicolumn{1}{c}{TML} & \multicolumn{1}{c}{$\tau_i$} \\\hline
  Adhesive (EC2216) & 0.558 & 1.20E3 \\
  CFRP & 0.207 & 2.00E3 \\
  Kapton & 0.0311 & 1.00E4 \\
\end{tabular}
\caption{Outgassing rates at 300 K of the dominant (on the mission timescale) molecular species for three different synthetics commonly used on a spacecraft. TML: Total mass loss; $\tau_i$: Outgassing time constant.\label{table_outgassing_parameters}}
\end{table}
We shall define the particle emission rate $\gamma_0$ as the number of particles that are outgassed per time and per unit area from the surface of the plane. We deduce the emission rate from $D_{out}$ by dividing through the outgassing area $A_{out}$ and the molecular mass  of the outgassed species $i$:
\begin{equation}
\label{equation_gamma_zero}
    \gamma_0=\frac{D_{out}}{m_iA_{out}}
\end{equation}
Assuming an infinite outgassing plane, it is apriori plausible to assume (based on the ideal gas law) that the steady state density $n=N/V$, where $N$ is the number of particles in the volume $V$, and the pressure $P$  are given by
\begin{equation}
\label{equation_P_outgassing}
    P=\gamma_0 m_i v_{rms}\hspace{2cm}n_0=\frac{\gamma_0}{v_{rms}},
\end{equation}
Here the root-mean-square velocity of the gas molecules is denoted by $v_{rms}$.
The product of mean density and mean velocity is given by the particle emission rate $\gamma_0$. The mean collision rate $\Gamma_{coll}$ is given by
\begin{equation}
\label{equation_collision_rate}
    \Gamma_{coll}=n v_{rms}\sigma=\gamma_{0}\sigma,
\end{equation}
where $\sigma=\pi R_s^2$ is the interaction cross section and $R_s$ is the radius of the sphere.
Taking the data from table \ref{table_outgassing_parameters} and considering the typical surface area of the components for which the outgassing was measured, we construct table \ref{table_summary_table}:
\begin{table}
\begin{tabular}{lrrrrrrr}
  Material & \multicolumn{1}{c}{$D_\mathrm{out}$} & \multicolumn{1}{c}{$m_\mathrm{gas}$} & \multicolumn{1}{c}{$\tau_i$} & \multicolumn{1}{c}{$\gamma$} & \multicolumn{1}{c}{$P$} & \multicolumn{1}{c}{$n$} & \multicolumn{1}{c}{$\Gamma_\mathrm{coll}$} \\
   & \multicolumn{1}{c}{${\rm [kg\, s^{-1}]}$} & \multicolumn{1}{c}{${\rm [m_u]}$} & \multicolumn{1}{c}{${\rm [h]}$} & \multicolumn{1}{c}{${\rm [m^{-2}s^{-1}]}$} & \multicolumn{1}{c}{${\rm [mbar]}$} & \multicolumn{1}{c}{${\rm [m^{-3}]}$} & \multicolumn{1}{c}{${\rm [s^{-1}]}$} \\
  \hline
 CFRP & 5E-9 & 30 & 2E3 & 48E14 & 7.1E-10 & 17E12 & 603\\
 Kapton & 4E-12 & 30 & 10E3 & 9E14 & 1.3E-10 & 3E12 & 113\\
 Adhesives & 9E-12 & 30 & 12E3 & 310E14 & 44E-10 & 108E12 & 3896\\
\end{tabular}
\caption{Outgassing properties at 300 K for CFRP, Kapton and composite resins. The calculated collision rate (column 6) assumes a sphere of radius 200 nm.\label{table_summary_table}}
\end{table}
The outgassing rates in table \ref{table_summary_table} are applicable for an infinite surface at 300 K and are greatly reduced by two effects: geometric dilution and decreasing rates at lower temperatures.

\subsubsection{Geometric dilution}
\begin{figure}[th]
 \begin{center}
 \includegraphics[width=1.0\linewidth]{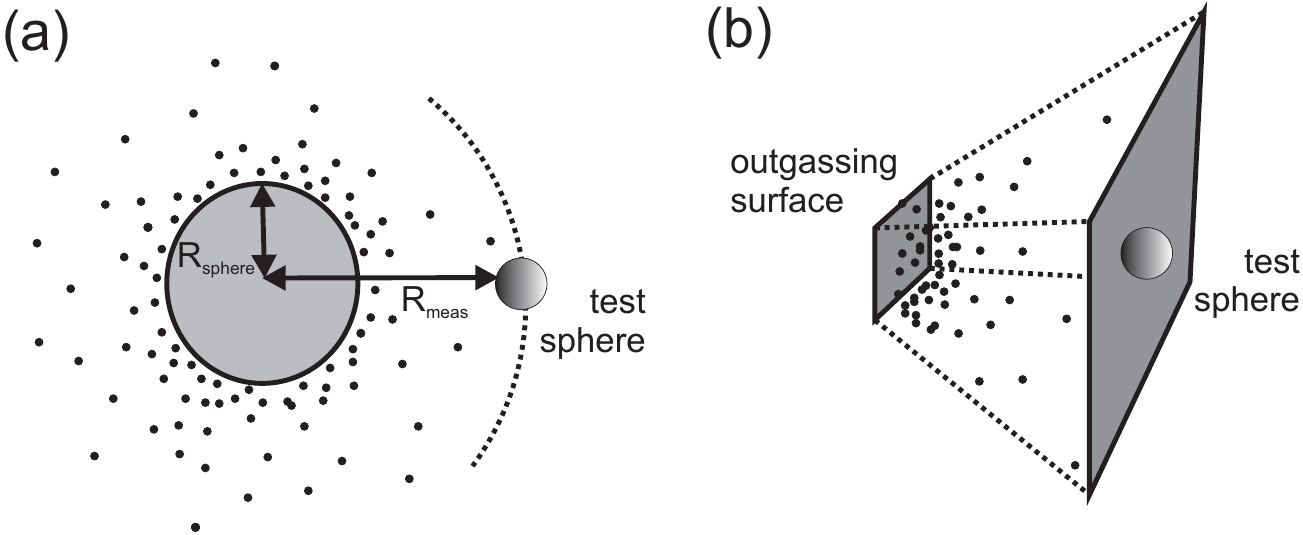}
 \caption{Outgassing from a sphere (a) and a quadratic surface element (b). In both cases the density of the outgassed molecules scales inversely proportional to the square distance.}
 \label{figure_outgassing_geometry}
 \end{center}
\end{figure}

As an example, we look at the outgassing from a sphere of radius $R_s$ and an outgassing rate of $\gamma_0$  at its surface and at the outgassing of a small surface element of area $A_{out}$. The two scenarios are depicted in Figure \ref{figure_outgassing_geometry}a and b, respectively.
The particle density $n$ at a distance $x_s$ from the center of the outgassing sphere and at a distance $x_d$ from the center of a surface element are given by equation \ref{equation_density_sphere} (left) and (right), respectively. Here we assumed that the distance between the outgassing surface element and the measurement sphere is much larger than the width of the surface element.
\begin{equation}
    \label{equation_density_sphere}
    n(x_s)=n_0\frac{R_s^2}{x_s^2}=\frac{\gamma_0}{v_{rms}}\frac{R_s^2}{x_s^2} \hspace{2cm}
    n(x_d)\approx n_0\frac{A_{out}}{4x_d^2}2=n_0\frac{A_{out}}{x_d^2}\frac{1}{2}
\end{equation}
Based on equation \ref{equation_density_sphere} we find that for a typical scenario (e.g. Kapton fiber head, 1 mm diameter, 10 cm distant from microsphere) the densities/collision rates are suppressed by a factor of $3\times 10^{-5}$, which greatly mitigates the outgassing problem.

\subsubsection{Reducing the temperature}
The temperature dependence of the outgassing time constants $\tau(T)$, also referred to as \emph{residence times}, is generally given by the Arrhenius law (see e.g. \cite{Hammesfahr2005},\cite{Cho2004}):
\begin{equation}
\label{equation_Arrhenius_law}
    \tau(T)=\tau_0 e^{\frac{E_{A}}{R_s T}}
\end{equation}
where $T[K]$ is the temperature, $R_s[J K^{-1} mol^{-1}]$ is the universal gas constant, and $E_A[J mol^{-1}]$ is the activation energy. We see that by increasing the temperature the outgassing process can be exponentially accelerated and decreasing the temperature it is dramatically reduced.
Dynamic outgassing tests are performed at ESA/ESTEC with the purpose of quantifying outgassing and condensation of materials as function of temperature and time, to support mathematical models used for the prediction of molecular contaminant generation, migration, and deposition.
Typically a Vacuum Balance Quartz Crystal (VBQC) is used in a standard program with 5 steps of 25 degrees to determine acceleration factors, temperature dependence of the residence time, and activation energy.
VBQC outgassing kinetic tests at ESA/ESTEC usually show acceleration factors of between 3 and 10 for each $25^{\circ}$C temperature step.

The equation for the Arrhenius law \ref{equation_Arrhenius_law} is combined with the equation for the outgassing rate \ref{outgassing_rate} and the pressure \ref{equation_P_outgassing} to yield:
\begin{equation}
    \label{equation_temperature_dependence}
    P=\frac{D_{out}}{m_i A_{out}}m_i v_{rms}=m_{tot}\frac{TML_i}{100}\frac{1}{\tau_i}\frac{v_{rms}}{A_{out}}=Const\cdot\sqrt{T}\cdot e^{-\frac{E_A}{R_s T}}
\end{equation}
This equation gives the dependence of the pressure on temperature and can also be used to extrapolate the vapor pressure once the activation energy $E_A$ is known.
The acceleration factors between 3 and 10 can then be used to calculate the activation energies $\tilde{E_A}$ per particle and we find $10~E_{room} < \tilde{E_A} <30~E_{room}$, where where $E_{room}$ is the energy associated with room temperature. From these typical activation energies of composite materials, we obtain the attenuation factors $F_a$ for the outgassing rates (and therefore for the vapor pressure) when the temperature is reduced from 300 K to 30 K. We find that $F_a<10^{40}$, indicating that even materials which strongly outgas at room temperature have practically no outgassing at temperatures as low as 30 K (the temperature of the experimental volume in DECIDE). From fits of equation \ref{equation_temperature_dependence} to the data tables for vapor pressure provided in \cite{VacAero2010} we extracted the activation energies of various chemically inert refractory metals and found good agreement with those values found from field emission microscopy: $E_A=140,148,128\hspace{5pt} E_{room}$, for Tungsten, Tantalum, and Niobium, respectively. The activation energies were used to extrapolate the pressures to very low temperatures. For these refractory elements the outgassing suppression is practically infinite, and -being chemically inert- they are ideally suited for the thermal shield of DECIDE.

\subsection{Requirements on position readout}
\label{scireq::readout}
A central experimental result that will have to be determined is the visibility of interference fringes. That visibility will allow to study the influence of various decoherence mechanisms, the dependence of the visibility on particle properties and the possible influence of physical collapse mechanisms as predicted by macrorealistic models (see subsections \ref{subsec::interference} and \ref{subsec::limits}).

In order to determine the visibility of the interference fringes, the minimum requirement is that the resolution of the position measurements of the nanospheres must be better than the expected fringe spacing. For $t_1=t_2$, this spacing is typically $5-10\,$pm. In order to resolve these fringes, a position readout with an accuracy of $1\,$pm or better would be required. It is possible to achieve that level of accuracy by using the cavity that is also used for trapping and cooling the nanospheres. As we have mentioned in sections \ref{sub::DECcomp} and \ref{subsec::CaseForSpace}, recent results indicate that $t_2$ has to be significantly longer than $t_1$. For these higher values of $t_2$, the resulting fringe spacing is significantly higher, i.e., on the order of $20-40\,$nm \cite{Kaltenbaek2012a}. In this case, a position resolution better than $\sim 5\,$nm would be sufficient to resolve the interference pattern.

\subsection{Requirements on the micro-propulsion system}
\label{scireq::dragfree}
The main purpose of the micro-propulsion system is to to (1) counter-act all non-gravitational fores acting on the spacecraft when the experiment DECIDE is run, and (2) to allow for drag-free control when running CASE. We will now discuss the requirements on the micro-propulsion system for these two cases.

In our original proposal, for each data points the nanosphere has to freely evolve over a time of up to $10\,$s. The fringe spacing of the interference pattern in the original proposal, where $t_2=t_1$, is $5 - 10\,$pm. Resolving this pattern requires a position stability of the spacecraft of $< 3\,\mbox{pm}$ over the course of $10\,$s.

The force noise of state-of-the-art micro thrusters as they are intended to be used for the LISA Pathfinder mission and for LISA is on the order of $10^{-8}\,\mathrm{N/\sqrt{Hz}}$. If we assume the spacecraft to have a mass of $2000\,$kg, this results in an acceleration noise of $5\times 10^{-10} \mathrm{m s^{-2}/\sqrt{Hz}}$. Over a time of $10\,$s, this leads to a position inaccuracy of less than $40\,$pm. That means, our original proposal would have required a significantly better thruster system than LISA.

However, as we have described earlier, new results show that we need $t_2\gg t_1$ for an interference pattern to form. To violate all macrorealistic models considered here, one needs a total free-fall time of $t_{tot} = t_1+t_2\approx 200\,$s. For these parameters, the fringe spacing is $20-40\,$nm. Given state-of-the-art micro thrusters as discussed above, the position inaccuracy over $t_{tot}$ is less than $1\,$nm. This should allow to clearly resolve the interference pattern.

For the CASE inertial sensors, we assume a position sensitivity on the order of picometers per $\mathrm{\sqrt{Hz}}$. For a micro-thruster acceleration noise of $5\times 10^{-10} \mathrm{m s^{-2}/\sqrt{Hz}}$, the position inaccuracy due to the thrusters will be $4\,\mathrm{pm/\sqrt{Hz}}$. The force noise of the thrusters should therefore not seriously impede the sensitivity of the test of the equivalence principle. 

Further studies will be required in order to find an optimized design for the drag-free control to allow for the necessary positioning accuracy of the spacecraft with respect to a free-falling test mass.

\subsection{Vacuum and Temperature Requirements}
\label{scireq::vactemp}
DECIDE has very high vacuum requirements (pressures below $10^{-12}$Pa) because a single collision of the nanosphere with a gas particle will lead to the decoherence of the wavefunction. In addition, it is necessary to have a low environmental temperature to reduce the detrimental effects of scattering blackbody radiation. For lower temperatures the blackbody wavelengths are longer on average and localize the quantum system less accurately, resulting in a smaller decoherence parameter. The \textit{standard parameters} we assume for the DECIDE experiment, are a pressure of $10^{-12}$Pa or less, and a temperature of $32$K. In the future, we will aim at adapting the thermal-shield design to lower the temperature even further, ideally below $20\,$K.

\subsection{Pointing stability}
\label{scireq::pointing}
The most critical element in terms of pointing stability will be the UV assembly to prepare the quantum superposition states. This assembly focuses a UV beam on a spot with a radius $\le 350$nm. The position of this has to move less than half the fringe spacing of the interference pattern, i.e., $\sim 10\,$nm. This corresponds to a pointing accuracy better than $10^{-2}$mrad. If we can measure the position of the UV spot accurately enough during each experimental shot, then the pointing stability is not critical. If we cannot, then the pointing has to be stable to that accuracy over the duration of a whole experimental run, which would render this a critical issue.

Since the cavity used in DECIDE is confocal, the pointing stability of the cavity mirrors is non-critical. However, the CCD assembly and the loading mechanism have to be stable with respect to the cavity system. If we assume a distance of 5cm between the imaging lens and the CCD, and if we want to monitor movements of the nanospheres with an accuracy of $< 1\,\mu\mathrm{m}$, this requires a pointing stability of $< 0.1\,\mbox{mrad}$ over the time of the manipulation sequence, i.e., $1$s to $10$s. The requirements on the loading mechanism are non-critical because any inaccuracies will only lead to a reduction of the probability of a successful loading event. This does not seriously influence the overall performance of the setup.

\begin{table}[t]
 \begin{center}
  \begin{tabular}{lc}
   \multicolumn{2}{c}{\textbf{DECIDE}} \\
   \multicolumn{2}{c}{\textbf{Scientific Requirements}} \\\hline
   \textbf{Parameter} & \textbf{Requirement} \\\hline
   Rate of collisions with gas particle & $< 0.01\,$Hz \\\hline
   Ambient temperature & $<35\,$K \\\hline
   Internal temperature of nanosphere & $< 100\,$K \\\hline
   Position readout accuracy & $< 5\,$nm \\\hline
   Spacecraft position stability & \textsc{critical} \\
   \hspace{1em} Along cavity axis & $< 5\,$nm over $200\,$s \\
   \hspace{1em} Perpendicular to cavity axis & $< 1\,\mu$m over $200\,$s \\\hline
   Pointing stability & \\
   \hspace{1em} Cavity mirrors & non-critical (confocal cavity) \\
   \hspace{1em} UV beam (relative to cavity) & \textsc{critical} \\
   & if beam position is continuously measured: \\
   & $< 10^{-2}$mrad over $200\,$s\\
   & if beam position is not measured: \\
   & $< 10^{-2}$mrad over several days\\
   \hspace{1em} CCD assembly & $< 1\,$mrad over 1s\\
   \hspace{1em} IR fibers & non-critical (cavity defines reference) \\
   \hspace{1em} Particle loading mechanism & non-critical (only reduces loading prob.) \\\hline
   Laser stability & \\
   \hspace{1em} IR laser & LTP stability more than sufficient \\
   \hspace{1em} UV laser & non-critical (only coarse adjustment necessary) \\\hline
   Nanosphere material absorption & \textsc{critical}\\
   \hspace{1em} at $1064\,$nm & lowest possible \\
   \hspace{1em} over blackbody spectrum & lowest possible \\\hline
   Nanosphere mass density & \textsc{critical}: as high as possible\\\hline
   Nanosphere size & non-critical\\[0.3ex]
   & (whole run measured with the same sphere)\\[0.3ex]\hline
   Nanosphere shape & to be determined\\[0.3ex]\hline
  \end{tabular}
  \caption{Overview of the various scientific requirements for DECIDE.\label{tab::DESIREreq}}
 \end{center}
\end{table}

\subsection{Laser stability}
\label{scireq::laser}
Laser noise can be a problem if one intends to use side-band cooling to cool a mechanical resonator to its ground state. According to \cite{Rabl2009a}, ground-state cooling in the presence of laser phase noise with a power spectrum $S_{\dot{\phi}}(\omega_m)$ is possible if $S_{\dot{\phi}}(\omega_m) < \frac{g^2_0}{\Gamma_m}$. Here,
$\Gamma_m = k_B T / \hbar Q$ is the thermalization rate, $Q$ is the mechanical quality factor of the resonator, $k_B$ is the Boltzmann constant, $T$ is the temperature of the environment, $\omega_m$ is the frequency of the mechanical resonator, and $g_0$ is its single-photon coupling strength.

With the experimental parameters we propose for DECIDE, this amounts to $S_{\dot{\phi}}(\omega_m) < 10^{19}\,\mbox{Hz}$ \cite{Rabl2009a}, a condition that should easily be fulfilled. For comparison, take the results presented in \cite{Camatel2006a}, where a model is fitted to the phase noise measured in a laser that shows poorer performance than the narrow line-width laser used on the LTP module, which we propose to also use for MAQRO. The measured data in \cite{Camatel2006a} agree well with the suggested model that includes white noise, flicker and random-walk noise contributions. Using that laser, we would expect $S_{\dot{\phi}}(\omega_m) = 10^{-8}$Hz, easily fulfilling the stability requirements for ground-state cooling.

Intensity fluctuations will change the trap frequency and could thus change the interference pattern. However, the fringe spacing in the interferogram is proportional to the square root of the laser power, and to significantly change the fringe spacing, immense fluctuations would be necessary. Thus, intensity fluctuations are non-critical.

The exact power of the UV laser beam is irrelevant. It only has to be switchable between completely off and a laser power of several mW for preparing the quantum superposition by scattering and for ejecting spurious particles from the sphere.

\subsection{Critical issues}
\label{subsub::crit}
Several techniques used and requirements needed are critical for the mission and have to be further investigated in technical studies. In particular, some of the techniques required for the implementation of the proposed experiments do not yet have the required technological readiness level for space experiments. Yet, recent technological progress in the various fields in question has been rapid, and we are confident that the necessary technological readiness level will be reached within a few years. The critical issues that have to be addressed are:
\begin{itemize}
 \item \textbf{Position stability of the spacecraft}:
       The position of the cavity with respect to the freely propagating nanosphere has to be kept stable with an accuracy better than $5\,\mbox{nm}$ over $t_1 + t_2 \approx 200\,$s. This should be achievable using state-of-the-art micro-propulsion systems.
 \item \textbf{Position readout}:
       The readout sensitivity for the position measurement for each data point has to be significantly better than the fringe spacing, i.e., on the order of $<5\,$nm. That the nanosphere after $t_2$ can be anywhere over a range of several wavelengths of the cavity field may negatively influence the position sensitivity and has to be studied in detail.
 \item \textbf{Loading mechanism}:\\
       The release of particles via ultrasonic vibrations from a glass plate has so far only been demonstrated with microspheres of several $\mu\mbox{m}$ in diameter \cite{Ashkin1977a,Li2011a}. The nanospheres to be used in DECIDE have a radius of $\sim 100\,$nm or smaller. We are currently working on a loading mechanism for spheres of that size.
  \item \textbf{Ground-state cooling}:\\
       This has recently been demonstrated for various architectures \cite{OConnell2010a,Teufel2011a,Chan2011a}. While the mechanical structures in all these experiments have had high mechanical frequencies (GHz), we expect that similar results will soon be achievable for mechanical systems with lower mechanical frequencies \cite{Groeblacher2009a,Riviere2011a}. However, while feedback cooling of optically trapped dielectric spheres has been demonstrated \cite{Ashkin1977a,Li2011a}, cavity cooling and, in particular, cooling to the ground state of motion has yet to be shown.
  \item \textbf{Cavities in space}:\\
       The proposed mission requires a cavity with a high finesse of $> 10000$. So far, such high-finesse cavities have not been demonstrated in space missions but several proposed missions rely on this technique because high-finesse cavities are promising candidates for high-precision frequency standards (see, e.g., \cite{Jiang2011a}). We are confident that the ongoing development \cite{Folkner2010a} effort will soon provide feasible venues for a technological realization of this central element of our experiment.
  \item \textbf{CCD cameras}:\\
       While CCD cameras in the IR and deep IR have been developed for use in space missions (see, e.g., \cite{Fazio2004a}), CCD cameras working in the NIR range and in the UV will have to be developed. In particular, the camera will have to operate at very low temperatures ($< 35$K) and under extreme vacuum conditions (interplanetary vacuum level). Recently developed CMOS cameras might provide a feasible alternative \cite{Bai2008a,Moehle2010a}.
  \item \textbf{Low-absorption dielectric materials}:\\
  		With typical state-of-the-art dielectric materials, optical trapping leads to a internal temperature high compared to the environment temperature. As a result, the decoherence rate due to the emission of blackbody radiation limits the performance of DECIDE. The development of dielectric materials with lower absorption coefficients at $1064\,$nm is therefore essential.
  \item \textbf{Influence of magnetic fields and charging}:\\
        This will have to be studied in detail in the future and will be addressed by an ESA study performed by some of the authors \cite{Kaltenbaek2012a}.
  \item \textbf{Gravitational field of the spacecraft}:\\
        Because of the very long free expansion times \cite{Kaltenbaek2012a}, the gravitational attraction of the nanosphere towards the spacecraft is critical. Compensating masses on the platform of DECIDE may solve this issue but further studies are necessary.
  \item \textbf{Transverse expansion of the wavefunction}:\\
        While the position of the nanosphere transverse to the cavity mode is not critical in itself, one has to assure that the particle does not leave the cavity mode. We will have to study the prerequisites for the particle to stay within the cavity mode during the free expansion of the wavepacket.
\end{itemize}

\subsection{State-of-the-art optomechanical experiments}
The field of optomechanics has seen tremendous progress over the last few years. In the wake of the first demonstrations of back-action cooling of nanomechanical systems in 2006 \cite{Naik2006a,Gigan2006a}, a race towards preparing mechanical systems in the ground-state of motion led towards the recent achievement of this goal in various architectures \cite{OConnell2010a,Teufel2011a,Chan2011a}. A limiting factor in quantum optomechanical experiments is the coupling to the environment. Several proposals have been put forward to realize levitated mechanical resonators in order to minimize dissipation to the environment \cite{Chang2009a,RomeroIsart2010a}. Recently, it has been demonstrated that the motion of optically trapped dielectric spheres can be cooled using feed-back cooling \cite{Li2011a}. Several groups are attempting to achieve the ground-state of motion for this type of mechanical system. We are confident that this necessary prerequisite for DECIDE will soon be achieved.

\subsection{Case for space}
\label{subsec::CaseForSpace}
In the following, we will give a series of arguments that require the proposed experiments to be performed in space. 

\subsubsection{CASE}
In CASE, microspheres are optically trapped, and the shift of their center-of-mass position due to accelerations is measured. The field trapping the microsphere necessarily has to be weak in order to achieve high sensitivity and to reduce the effect of heating of the center-of-mass motion by the trapping laser. Such a weak field can only trap the microsphere in a micro-gravity environment. Micro gravity can, in principle, be achieved in Earth-bound experiments by using a drop tower or parabola flights. In drop towers, the free-fall time is limited to a few seconds, severely limiting the integration time and, therefore, the sensitivity of the experiment. An additional drawback of drop-tower experiments is the residual gravitational acceleration. This advantage of a space environment is even more pronounced when compared to parabola flights.

\subsubsection{DECIDE}
The coherent-expansion-time (CET) for the K model as well as for the Di\'osi-Penrose model is on the order of seconds. In order to conclusively test these models, the experimental parameters have to be chosen such that the CET predicted by quantum theory is significantly longer than that timescale. Such free-fall times are, in principle, possible in drop towers but they typically allow only for a few runs per day because of the time it takes to evacuate the tower. In order to resolve the interference fringes in DECIDE, one needs at the very least thousands of data points up to $10^6$, depending on the choice of $t_2$ and the macrorealistic models to be tested. This rules out drop-tower experiments.

The case for space for DECIDE becomes even more pronounced if one takes into account more recent results where $t_2 \gg t_1$ \cite{Kaltenbaek2012a}. In order to rule out all macrorealistic models considered here, $t_2$ will have to be $\approx 200\,$s. Such free-fall times are not possible in Earth-bound experiments. 

Using our method or alternative methods \cite{RomeroIsart2011b,RomeroIsart2011c}, it might be possible to test the CSL model for a range of parameters $\lambda$ and possibly even the QG model on Earth. To test more demanding macrorealistic models, the same considerations apply as above, i.e., such experiments would have to be performed in space. For space experiments, our approach is better suited because the nanosphere can remain in one cavity instead of propagating through three separate ones. Moreover, it is not clear whether the method of Refs.~\cite{RomeroIsart2011b,RomeroIsart2011c} works for the large displacements necessary for violating, e.g., the K model or the DP model.

\section{Mission profile and spacecraft design}

\subsection{Orbit requirements}
\begin{figure}[th]
 \begin{center}
 \includegraphics[width=0.99\linewidth]{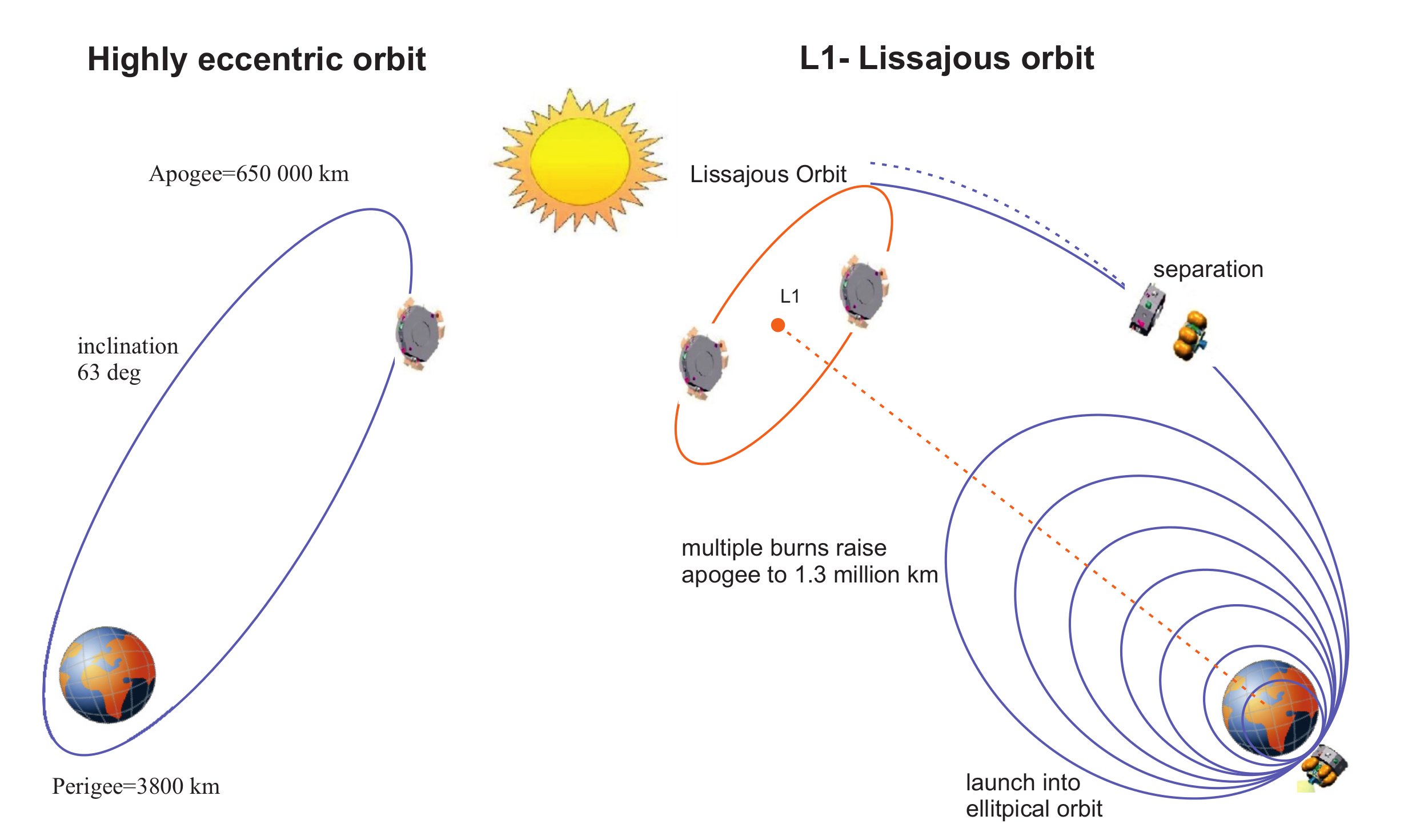}
 \caption{Left: A highly eccentric orbit is the baseline for MAQRO. It serves the needs of both experiments, DECIDE and CASE. Right: The transfer and final Lissajous orbit around L1 for the LISA Pathfinder mission would be ideal for MAQRO's DECIDE experiment. Image based on a similar plot by ESA, \cite{LPFgraphics}.}
 \label{figure_alternative_orbits}
 \end{center}
\end{figure}

\emph{A highly eccentric orbit (HEO)}: The science requirements indicate that an extremely good vacuum, very low temperatures and experimental measurement times of several seconds are required for DECIDE. On the other hand, CASE requires a medium-quality vacuum, room temperature and very long experimental measurement times in a high-gravity environment with sufficiently small drag forces. A mission to the sun/earth Lagrange points L1 or L2 (figure \ref{figure_alternative_orbits} right) would be ideally suited for DECIDE but does not offer the high gravitational field gradients necessary for tests of the equivalence principle. To satisfy the needs of both experiments, DECIDE and CASE, and also to improve the possibility of combining MAQRO with other fundamental science missions, we suggest using a highly eccentric orbit (figure \ref{figure_alternative_orbits} left). Considering an ellipse of $63^{\circ}$ inclination and apogee/perigee of $650000\,$km / $3800\,$km, the orbital period is $\approx 22$ days, from which $\approx 2$ weeks are spent around the apogee. There the conditions are similar to those at L1, which is suitable for science experiments with DECIDE. The proposed HEO is a sun-synchronous orbit (it rotates together with the earth around the sun) which guarantees that the sun is always incident perpendicular to the body-mounted solar array. 
\\\\
\emph{Perigee passage for equivalence principle measurements}: For measurements of the equivalence principle (CASE) we use the large gravitional acceleration at the perigee ($\approx0.4$ times the gravitational acceleration on Earth), which allows us to perform a measurement of the external gravitational acceleration with fractional accuracy of ${\rm 2.5\times 10^{-13}ms^{-2}/\sqrt{ Hz}}$. Considering that the time spent at the perigee (spacecraft within 3800 km-4500 km height) is 20.2 minutes, the integrated measurement time yields a total fractional accuracy of $\approx 5\times10^{-15}$. Longer experimental integration times might in principle be feasible but require different orbits and more detailed analysis. The calibration of the residual spacecraft gravitational field gradients could be performed during the two weeks while the spacecraft is at the apogee, together with operating the DECIDE experiment.
\\\\
\emph{Radiation doses}: The impact of heavy radiation doses when crossing the Van-Allen belt and large thermal gradients and strains when approaching earth must be further investigated before a final judgment can be made on the suitability of a HEO orbit. However, considering that because of a boost motor failure the Hipparcos satellite \cite{Hipparcos2011} remained in a geostationary transfer orbit and therefore unintentionally exposed to heavy radiation during its successful 3 year mission lifetime, it seems feasible to operate MAQRO in a similar HEO during its much smaller mission lifetime of only 6 months (amounting to 8 orbital periods).
\\\\
\emph{An alternative L1/L2 orbit}: Based on the mission scenario for LISA Pathfinder, the MAQRO spacecraft is injected into a halo orbit round the sun/earth Lagrange point L1 (L2 also feasible) at 1.5 million km distance from earth, following the initial injection into elliptical earth orbit and 8 apogee raising orbits.
Shortly before reaching the final on-station orbit around L1, the Propulsion Module (PRM) is separated from the Science Module (SCM). The nominal attitude profile is maintained using the micro-propulsion subsystems. In contrast to LISA Pathfinder, MAQRO will only use Cesium-slit FEEP thrusters (specified for $>2000$ Ns firing) and no additional colloidal thrusters of the disturbance reduction system (DRS), which shall be removed. Asides from a considerable simplification, this effectively decreases the spacecraft mass by $\approx37$ kg.

\subsection{Other mission Parameters}

\emph{Communication and data storage}:
Communication for MAQRO will be on X-band using low gain hemispherical (HEO orbit) or medium gain horn antennas (for L1 orbit). A communication bandwidth of $60$kbps fulfills the down-link bandwidth requirements for MAQRO. Therefore $\approx6$W of transmitted RF-power are sufficient to establish the required downlink rate for on-station nominal operation. As in Pathfinder, it is suggested to use the 35 m antenna of the ground station Cebreros in Spain.
A communication window of $>8$ hours per day is sufficient to transfer science data to ground. Data are received by the 35 m antenna and transferred to ESOC for further processing. If a highly-eccentric orbit is chosen, there will be an interruption of ground communication for several hours during passage through the perigee, which implies that there is never any ground station contact during the CASE experiment and all steps of the experiment have to be uploaded to the on-board computer for automated execution.\\\\

\emph{Spacecraft thermal Design}: The standard thermal control tasks are to keep the overall S/C and its external and internal units and equipment within the allowable temperature ranges by a proper thermal balance between isolating and radiating outer surfaces, supported by active control elements such as heaters. For the MAQRO mission the thermal design has to focus on a good thermal stability within the S/C (for CASE) and a proper thermal interface design from the warm S/C to the extremely cold external payload of DECIDE. Optimal thermal stability for the DECIDE experiment is achieved by further de-coupling from an already very stable S/C and by good coupling to the ultra-stable 4 K environment of deep space.
In order to obtain a good thermal stability for the CASE experiment the MAQRO S/C internal dissipation fluctuations are minimized and the S/C interior is isolated from the solar array which inherently introduces solar fluctuations into the S/C. For the DECIDE experiment the (warm) mechanical interface is designed as cold as possible, e.g. 270 K, and the S/C surfaces facing towards the external payload are covered by a high-efficient multi-layer insulation (20 layers), where the outermost layer has a high emissivity $> 0.8$. These measures optimize the radiative pre-cooling of the outer thermal shield of the payload which facilitates reaching the required 30 K environment at the experimental volume behind the inner shield. 

An overview over the orbit- and other mission parameters is given in table \ref{table_mission_parameters}.

\begin{table}
\begin{tabular}{|l|l|}
  \textbf{Mission requirement} & \textbf{Proposed choice} \\
  \hline
  Launcher & Rockot/Vega \\
  \hline
  science platform & LISA Pathfinder + external platform \\
  \hline
  Orbit & Highly elliptical ($650\times 10^3$ km / $3.8\times10^3$ km) at $63^{\circ}$ incl.\\
    & Alternative: L1 / L2 Lissajous\\
  \hline
  Mission Lifetime & 6 months total, 1-2 months to reach final orbit \\
  \hline
  Communication &  X-Band, 60 kb/s, $>8$ h  daily coverage,\\
  \hline
  &hemi-spherical antenna (HEO), horn antenna (L1/L2)\\
  \hline
  Ground segment & Cerebros, Spain (35 m)
\end{tabular}
\caption{Summary of mission and orbit requirements.\label{table_mission_parameters}}
\end{table}

\subsection{Scientific payload mass and power allocations}
The MAQRO mass budget in table \ref{table_mass_budget_total} is closely based upon the one of LISA Pathfinder.
The spacecraft platform of MAQRO is identical to the one of LPF and the MAQRO payload is similar to LTP for many units.
Note that in LPF the payload module and service module are combined in the "Science Module". The latter is attached to the propulsion module which is ejected after the final burn, if a HEO is chosen.

Replacing the heavy inertial sensor of LTP by an ONERA accelerometer (see e.g. \cite{Onera2008}) reduces the payload mass of MAQRO by approx. 70 kg. The UV laser assembly is assumed to be similar in mass to the IR laser assembly without modulators. The heat shield and external optical bench only add $\approx 20$ kg to the total payload mass.
If "venting to space" \cite{Hammesfahr2005} is implemented to achieve a good interior vacuum,
the use of venting ducts increases the spacecraft mass by approximately 3 kg.

To obtain the dry total mass of the spacecraft we add the mass of the science spacecraft and of the propulsion module to the payload mass. Note that the mass of the science spacecraft for MAQRO is considerably reduced (by 37 kg) with respect to LPF because the disturbance reduction system (DRS) has been removed.
We conservatively add the same amount of propellant as for the heavier Pathfinder spacecraft.
\begin{table}[t]
 \begin{center}
  \begin{tabular}{lll}
  \textbf{Launch composite} & \textbf{LPF mass [kg]} & \textbf{MAQRO mass [kg]}\\
  \hline
  Payload (LTP) & 144 & 97(+7)\\
  \hline
  Science Spacecraft & 274 & 237\\
  \hline
  Propulsion Module & 210 & 210\\
  \hline\hline
  Launch composite dry total & 628 & 544(+7)\\
  \hline
  Consumables & 1110 & 1110\\
  \hline
  Launch composite wet total & 1738 & 1654(+7)\\
  \hline
  \end{tabular}
  \caption{The total mass budget of MAQRO in comparison to LPF.
  The optional weight arising from the shield extension and bake-out mechanisms is given in brackets.
  \label{table_mass_budget_total}}
 \end{center}
\end{table}

In table \ref{table_power_budget} the total power budget of the MAQRO payload
is compared to LPF to demonstrate that the power requirements are also very similar.
One can conclude that the Pathfinder solar array of $\approx680$ W is sufficient for the needs of MAQRO.
Note that a bake out mechanism for the outermost heat shield (+optical bench) can be optionally included for MAQRO.
The heater requires 105 W of power for bake-out at 300 K. Before commissioning, LTP and likewise MAQRO only requires 30 W of power so that the power difference of 110 W with respect to the LTP science mode are sufficient for bake-out. During bake-out only the experimental data management unit (DMU) and the shield heaters must be active.

\begin{table}[t]
 \begin{center}
  \begin{tabular}{lll}
  \textbf{Payload Operating Mode} & \textbf{Req. Power LPF} & \textbf{Req. Power MAQRO}\\ \hline
  Science mode & 141 & 141\\
  \hline
  Maximal Power & 163  & 141\\
  \hline
  Minimal power & 30  & 30 \\
  \hline
  Optional heater for shields && 105\\
  \hline
  Minimal power\&heater && 135\\
  \hline
  \end{tabular}
  \caption{The power requirements of the MAQRO payload (right side) are compared to those of LTP (left side).\label{table_power_budget}}
 \end{center}
\end{table}

\section{Conclusion}
We have presented the proposal for a medium-sized space mission, MAQRO, consisting of two experiments, DECIDE and CASE. Both of these experiments are essentially independent and feature a light-weight, modular design such that each of the experiments could separately be added to other missions that have similar orbit and micro-gravity requirements. The main scientific objective of the mission is addressed by the experiment DECIDE, which aims at testing quantum theory in an interference experiment with macroscopic resonators. We have designed a thermal shield that allows to perform DECIDE on a separate platform outside the spacecraft in order to fulfill the strict temperature and vacuum requirements of DECIDE. Our analysis has shown that it should, in principle, be possible to perform such interference experiments where quantum theory predicts reasonable interference visibility while alternative, macrorealistic theories predict that no interference should occur. For the second experiment, CASE, a novel inertial sensor based on the position detection of optically trapped microspheres has been presented. This new inertial sensor is to be compared against a state-of-the-art capacitive sensor, and a combined operation of the two sensors allows for a test of the principle of universal free fall.

\section{Outlook}
We aim at further developing the technological readiness level of all the central techniques for the proposed experiments. In particular, we are currently working on a study for the European Space Agency (ESA) in order to investigate the possibility of quantum experiments with macroscopic resonators in more detail \cite{Kaltenbaek2012a}. Moreover, we work on the developments of several techniques that are necessary for the implementation of DECIDE. For example, we currently work on the implementation of a novel loading mechanism for nanospheres into an optical trap in vacuum. Other goals within the next months and years will be to achieve ground-state cooling for optically trapped nanospheres and a proof-of-principle demonstration of the double-slit preparation via a UV pulse as it is proposed to be used for DECIDE. Recent theoretical results \cite{Kaltenbaek2012a} will be analyzed in more detail, and they will allow us to give a significantly more detailed design of a future space mission, confirming the feasibility of the concepts proposed in the present work.

\begin{acknowledgments}
We thank S. Hofer, G. Cole, K. Hammerer, A. Pflanzer and J. I. Cirac for valuable discussions, Johannes Burkhard for his help in designing the heat shield, T. Ziegler for his help with the 1-d DFACS for the CASE experiment, N. Brandt for his help and advice regarding the platform and experimental design, and we thank Jens Burkhard for the 3D graphics of the heat shield and the optical setups.
R. K. acknowledges support from the Austrian Program for Advanced  Research and Technology (APART) of the Austrian Academy of Sciences and support from the European Commission (Marie Curie, FP7-PEOPLE-2010-RG). O. R.-I. and N. K. acknowledge funding by the Alexander von Humboldt foundation, and M. A. acknowledges funding by the Austrian Science Fund FWF (START, FOQUS), the European Research Council (ERC StG QOM), and the European Commission (FP7 STREP MINOS, Q-ESSENCE).
\end{acknowledgments}

\bibliography{references}  

\begin{thebibliography}{86}
\expandafter\ifx\csname natexlab\endcsname\relax\def\natexlab#1{#1}\fi
\expandafter\ifx\csname bibnamefont\endcsname\relax
  \def\bibnamefont#1{#1}\fi
\expandafter\ifx\csname bibfnamefont\endcsname\relax
  \def\bibfnamefont#1{#1}\fi
\expandafter\ifx\csname citenamefont\endcsname\relax
  \def\citenamefont#1{#1}\fi
\expandafter\ifx\csname url\endcsname\relax
  \def\url#1{\texttt{#1}}\fi
\expandafter\ifx\csname urlprefix\endcsname\relax\def\urlprefix{URL }\fi
\providecommand{\bibinfo}[2]{#2}
\providecommand{\eprint}[2][]{\url{#2}}

\bibitem[{\citenamefont{Armano et~al.}(2009)\citenamefont{Armano, Benedetti,
  Bogenstahl, Bortoluzzi, Bosetti, Brandt, Cavalleri, Ciani, Cristofolini,
  Cruise et~al.}}]{Armano2009}
\bibinfo{author}{\bibfnamefont{M.}~\bibnamefont{Armano}},
  \bibinfo{author}{\bibfnamefont{M.}~\bibnamefont{Benedetti}},
  \bibinfo{author}{\bibfnamefont{J.}~\bibnamefont{Bogenstahl}},
  \bibinfo{author}{\bibfnamefont{D.}~\bibnamefont{Bortoluzzi}},
  \bibinfo{author}{\bibfnamefont{P.}~\bibnamefont{Bosetti}},
  \bibinfo{author}{\bibfnamefont{N.}~\bibnamefont{Brandt}},
  \bibinfo{author}{\bibfnamefont{A.}~\bibnamefont{Cavalleri}},
  \bibinfo{author}{\bibfnamefont{G.}~\bibnamefont{Ciani}},
  \bibinfo{author}{\bibfnamefont{I.}~\bibnamefont{Cristofolini}},
  \bibinfo{author}{\bibfnamefont{A.~M.} \bibnamefont{Cruise}},
  \bibnamefont{et~al.}, \bibinfo{journal}{Class. Quantum Grav.}
  \textbf{\bibinfo{volume}{26}}, \bibinfo{pages}{094001}
  (\bibinfo{year}{2009}).

\bibitem[{\citenamefont{Schr{\" o}dinger}(1935)}]{Schroedinger1935a}
\bibinfo{author}{\bibfnamefont{E.}~\bibnamefont{Schr{\" o}dinger}},
  \bibinfo{journal}{{D}ie {N}aturwissenschaften} \textbf{\bibinfo{volume}{23}},
  \bibinfo{pages}{807} (\bibinfo{year}{1935}).

\bibitem[{\citenamefont{Nielsen and Chuang}(2000)}]{Nielsen2000a}
\bibinfo{author}{\bibfnamefont{M.}~\bibnamefont{Nielsen}} \bibnamefont{and}
  \bibinfo{author}{\bibfnamefont{I.}~\bibnamefont{Chuang}},
  \emph{\bibinfo{title}{Quantum Computation and Quantum Information Theory}}
  (\bibinfo{publisher}{Cambridge Univ. Press}, \bibinfo{address}{Cambridge},
  \bibinfo{year}{2000}).

\bibitem[{\citenamefont{Bohr}(1969)}]{Bohr1949a}
\bibinfo{author}{\bibfnamefont{N.}~\bibnamefont{Bohr}}, in
  \emph{\bibinfo{booktitle}{{Albert Einstein. Philosopher-Scientist.}}}, edited
  by \bibinfo{editor}{\bibfnamefont{P.~A.} \bibnamefont{Schlipp}}
  (\bibinfo{publisher}{New York: MJF Books}, \bibinfo{year}{1969}),
  \bibinfo{edition}{3rd} ed.

\bibitem[{\citenamefont{Zeilinger et~al.}(1982)\citenamefont{Zeilinger,
  Gähler, Shull, and Treimer}}]{Zeilinger1981a}
\bibinfo{author}{\bibfnamefont{A.}~\bibnamefont{Zeilinger}},
  \bibinfo{author}{\bibfnamefont{R.}~\bibnamefont{Gähler}},
  \bibinfo{author}{\bibfnamefont{C.~G.} \bibnamefont{Shull}}, \bibnamefont{and}
  \bibinfo{author}{\bibfnamefont{W.}~\bibnamefont{Treimer}}, in
  \emph{\bibinfo{booktitle}{Proceedings on the Symposium on Neutron Scattering,
  Argonne 1981}}, edited by
  \bibinfo{editor}{\bibfnamefont{J.}~\bibnamefont{Faber}}
  (\bibinfo{publisher}{AIP New York}, \bibinfo{year}{1982}),
  vol.~\bibinfo{volume}{89}, pp. \bibinfo{pages}{93--99}.

\bibitem[{\citenamefont{Tsuchiya et~al.}(1986)\citenamefont{Tsuchiya, Inuzuka,
  Kurono, and Hosoda}}]{Tsuchiya1986a}
\bibinfo{author}{\bibfnamefont{Y.}~\bibnamefont{Tsuchiya}},
  \bibinfo{author}{\bibfnamefont{E.}~\bibnamefont{Inuzuka}},
  \bibinfo{author}{\bibfnamefont{T.}~\bibnamefont{Kurono}}, \bibnamefont{and}
  \bibinfo{author}{\bibfnamefont{M.}~\bibnamefont{Hosoda}}, in
  \emph{\bibinfo{booktitle}{Photo-Electronic Image Devices Proceedings of the
  Eighth Symposium}}, edited by \bibinfo{editor}{\bibfnamefont{B.~L.}
  \bibnamefont{Morgan}} (\bibinfo{publisher}{Academic Press},
  \bibinfo{year}{1986}), vol. \bibinfo{volume}{64, Part A} of
  \emph{\bibinfo{series}{Advances in Electronics and Electron Physics}}, pp.
  \bibinfo{pages}{21 -- 31}.

\bibitem[{\citenamefont{Tonomura et~al.}(1989)\citenamefont{Tonomura, Endo,
  Matsuda, Kawasaki, and Ezawa}}]{Tonomura1989a}
\bibinfo{author}{\bibfnamefont{A.}~\bibnamefont{Tonomura}},
  \bibinfo{author}{\bibfnamefont{J.}~\bibnamefont{Endo}},
  \bibinfo{author}{\bibfnamefont{T.}~\bibnamefont{Matsuda}},
  \bibinfo{author}{\bibfnamefont{T.}~\bibnamefont{Kawasaki}}, \bibnamefont{and}
  \bibinfo{author}{\bibfnamefont{H.}~\bibnamefont{Ezawa}},
  \bibinfo{journal}{American Journal of Physics} \textbf{\bibinfo{volume}{57}},
  \bibinfo{pages}{117} (\bibinfo{year}{1989}), ISSN \bibinfo{issn}{00029505}.

\bibitem[{\citenamefont{Einstein et~al.}(1935)\citenamefont{Einstein, Podolsky,
  and Rosen}}]{Einstein1935a}
\bibinfo{author}{\bibfnamefont{A.}~\bibnamefont{Einstein}},
  \bibinfo{author}{\bibfnamefont{B.}~\bibnamefont{Podolsky}}, \bibnamefont{and}
  \bibinfo{author}{\bibfnamefont{N.}~\bibnamefont{Rosen}},
  \bibinfo{journal}{Phys. Rev.} \textbf{\bibinfo{volume}{47}},
  \bibinfo{pages}{777} (\bibinfo{year}{1935}).

\bibitem[{\citenamefont{Bell}(1964)}]{Bell1964a}
\bibinfo{author}{\bibfnamefont{J.~S.} \bibnamefont{Bell}},
  \bibinfo{journal}{Physics} \textbf{\bibinfo{volume}{1}}, \bibinfo{pages}{195}
  (\bibinfo{year}{1964}).

\bibitem[{\citenamefont{Marton et~al.}(1953)\citenamefont{Marton, Simpson, and
  Suddeth}}]{Marton1953a}
\bibinfo{author}{\bibfnamefont{L.}~\bibnamefont{Marton}},
  \bibinfo{author}{\bibfnamefont{J.~A.} \bibnamefont{Simpson}},
  \bibnamefont{and} \bibinfo{author}{\bibfnamefont{J.~A.}
  \bibnamefont{Suddeth}}, \bibinfo{journal}{Phys. Rev.}
  \textbf{\bibinfo{volume}{90}}, \bibinfo{pages}{490} (\bibinfo{year}{1953}).

\bibitem[{\citenamefont{Marton et~al.}(1954)\citenamefont{Marton, Simpson, and
  Suddeth}}]{Marton1954a}
\bibinfo{author}{\bibfnamefont{L.}~\bibnamefont{Marton}},
  \bibinfo{author}{\bibfnamefont{J.~A.} \bibnamefont{Simpson}},
  \bibnamefont{and} \bibinfo{author}{\bibfnamefont{J.~A.}
  \bibnamefont{Suddeth}}, \bibinfo{journal}{Review of Scientific Instruments}
  \textbf{\bibinfo{volume}{25}}, \bibinfo{pages}{1099} (\bibinfo{year}{1954}).

\bibitem[{\citenamefont{Rauch et~al.}(1974)\citenamefont{Rauch, Treimer, and
  Bonse}}]{Rauch1974a}
\bibinfo{author}{\bibfnamefont{H.}~\bibnamefont{Rauch}},
  \bibinfo{author}{\bibfnamefont{W.}~\bibnamefont{Treimer}}, \bibnamefont{and}
  \bibinfo{author}{\bibfnamefont{U.}~\bibnamefont{Bonse}},
  \bibinfo{journal}{Phys. Lett.~A} \textbf{\bibinfo{volume}{47}},
  \bibinfo{pages}{369} (\bibinfo{year}{1974}).

\bibitem[{\citenamefont{Rauch and Werner}(2000)}]{Rauch2000a}
\bibinfo{author}{\bibfnamefont{H.}~\bibnamefont{Rauch}} \bibnamefont{and}
  \bibinfo{author}{\bibfnamefont{S.~A.} \bibnamefont{Werner}},
  \emph{\bibinfo{title}{{N}eutron {I}nterferometry: {L}essons in {E}xperimental
  {Q}uantum {M}echanics}} (\bibinfo{publisher}{Oxford University Press},
  \bibinfo{address}{Oxford, New York}, \bibinfo{year}{2000}).

\bibitem[{\citenamefont{Cronin et~al.}(2009)\citenamefont{Cronin, Schmiedmayer,
  and Pritchard}}]{Cronin2009a}
\bibinfo{author}{\bibfnamefont{A.~D.} \bibnamefont{Cronin}},
  \bibinfo{author}{\bibfnamefont{J.}~\bibnamefont{Schmiedmayer}},
  \bibnamefont{and} \bibinfo{author}{\bibfnamefont{D.~E.}
  \bibnamefont{Pritchard}}, \bibinfo{journal}{Rev. Mod. Phys.}
  \textbf{\bibinfo{volume}{81}}, \bibinfo{pages}{1051} (\bibinfo{year}{2009}).

\bibitem[{\citenamefont{Arndt et~al.}(1999)\citenamefont{Arndt, Nairz,
  Voss-Andreae, Keller, Van~der Zouw, and Zeilinger}}]{Arndt1999a}
\bibinfo{author}{\bibfnamefont{M.}~\bibnamefont{Arndt}},
  \bibinfo{author}{\bibfnamefont{O.}~\bibnamefont{Nairz}},
  \bibinfo{author}{\bibfnamefont{J.}~\bibnamefont{Voss-Andreae}},
  \bibinfo{author}{\bibfnamefont{C.}~\bibnamefont{Keller}},
  \bibinfo{author}{\bibfnamefont{G.}~\bibnamefont{Van~der Zouw}},
  \bibnamefont{and}
  \bibinfo{author}{\bibfnamefont{A.}~\bibnamefont{Zeilinger}},
  \bibinfo{journal}{Nature} \textbf{\bibinfo{volume}{401}},
  \bibinfo{pages}{680} (\bibinfo{year}{1999}).

\bibitem[{\citenamefont{Hackerm{\"u}ller
  et~al.}(2004)\citenamefont{Hackerm{\"u}ller, Hornberger, Brezger, Zeilinger,
  and Arndt}}]{Hackermueller2004a}
\bibinfo{author}{\bibfnamefont{L.}~\bibnamefont{Hackerm{\"u}ller}},
  \bibinfo{author}{\bibfnamefont{K.}~\bibnamefont{Hornberger}},
  \bibinfo{author}{\bibfnamefont{B.}~\bibnamefont{Brezger}},
  \bibinfo{author}{\bibfnamefont{A.}~\bibnamefont{Zeilinger}},
  \bibnamefont{and} \bibinfo{author}{\bibfnamefont{M.}~\bibnamefont{Arndt}},
  \bibinfo{journal}{Nature} \textbf{\bibinfo{volume}{427}},
  \bibinfo{pages}{711} (\bibinfo{year}{2004}).

\bibitem[{\citenamefont{Gerlich et~al.}(2011)\citenamefont{Gerlich,
  Eibenberger, Tomandl, Nimmrichter, Hornberger, Fagan, T\"{u}xen, Mayor, and
  Arndt}}]{Gerlich2011a}
\bibinfo{author}{\bibfnamefont{S.}~\bibnamefont{Gerlich}},
  \bibinfo{author}{\bibfnamefont{S.}~\bibnamefont{Eibenberger}},
  \bibinfo{author}{\bibfnamefont{M.}~\bibnamefont{Tomandl}},
  \bibinfo{author}{\bibfnamefont{S.}~\bibnamefont{Nimmrichter}},
  \bibinfo{author}{\bibfnamefont{K.}~\bibnamefont{Hornberger}},
  \bibinfo{author}{\bibfnamefont{P.~J.} \bibnamefont{Fagan}},
  \bibinfo{author}{\bibfnamefont{J.}~\bibnamefont{T\"{u}xen}},
  \bibinfo{author}{\bibfnamefont{M.}~\bibnamefont{Mayor}}, \bibnamefont{and}
  \bibinfo{author}{\bibfnamefont{M.}~\bibnamefont{Arndt}},
  \bibinfo{journal}{Nature communications} \textbf{\bibinfo{volume}{2}},
  \bibinfo{pages}{263} (\bibinfo{year}{2011}).

\bibitem[{\citenamefont{Julsgaard et~al.}(2001)\citenamefont{Julsgaard,
  Kozhekin, and Polzik}}]{Julsgaard2001a}
\bibinfo{author}{\bibfnamefont{B.}~\bibnamefont{Julsgaard}},
  \bibinfo{author}{\bibfnamefont{A.}~\bibnamefont{Kozhekin}}, \bibnamefont{and}
  \bibinfo{author}{\bibfnamefont{E.~S.} \bibnamefont{Polzik}},
  \bibinfo{journal}{Nature} \textbf{\bibinfo{volume}{413}},
  \bibinfo{pages}{400} (\bibinfo{year}{2001}).

\bibitem[{\citenamefont{Hagley et~al.}(1999)\citenamefont{Hagley, Deng, Kozuma,
  Wen, Helmerson, Rolston, and Phillips}}]{Hagley1999a}
\bibinfo{author}{\bibfnamefont{E.~W.} \bibnamefont{Hagley}},
  \bibinfo{author}{\bibfnamefont{L.}~\bibnamefont{Deng}},
  \bibinfo{author}{\bibfnamefont{M.}~\bibnamefont{Kozuma}},
  \bibinfo{author}{\bibfnamefont{J.}~\bibnamefont{Wen}},
  \bibinfo{author}{\bibfnamefont{K.}~\bibnamefont{Helmerson}},
  \bibinfo{author}{\bibfnamefont{S.~L.} \bibnamefont{Rolston}},
  \bibnamefont{and} \bibinfo{author}{\bibfnamefont{W.~D.}
  \bibnamefont{Phillips}}, \bibinfo{journal}{Science}
  \textbf{\bibinfo{volume}{283}}, \bibinfo{pages}{1706} (\bibinfo{year}{1999}),
  ISSN \bibinfo{issn}{00368075}.

\bibitem[{\citenamefont{Bloch et~al.}(1999)\citenamefont{Bloch, H\"{a}nsch, and
  Esslinger}}]{Bloch1999a}
\bibinfo{author}{\bibfnamefont{I.}~\bibnamefont{Bloch}},
  \bibinfo{author}{\bibfnamefont{T.}~\bibnamefont{H\"{a}nsch}},
  \bibnamefont{and}
  \bibinfo{author}{\bibfnamefont{T.}~\bibnamefont{Esslinger}},
  \bibinfo{journal}{Phys. Rev. Lett.} \textbf{\bibinfo{volume}{82}},
  \bibinfo{pages}{3008} (\bibinfo{year}{1999}), ISSN \bibinfo{issn}{0031-9007}.

\bibitem[{\citenamefont{Schwab and Roukes}(2005)}]{Schwab2005a}
\bibinfo{author}{\bibfnamefont{K.~C.} \bibnamefont{Schwab}} \bibnamefont{and}
  \bibinfo{author}{\bibfnamefont{M.~L.} \bibnamefont{Roukes}},
  \bibinfo{journal}{Physics Today} \textbf{\bibinfo{volume}{58}},
  \bibinfo{pages}{36} (\bibinfo{year}{2005}).

\bibitem[{\citenamefont{Kippenberg and Vahala}(2008)}]{Kippenberg2008a}
\bibinfo{author}{\bibfnamefont{T.~J.} \bibnamefont{Kippenberg}}
  \bibnamefont{and} \bibinfo{author}{\bibfnamefont{K.~J.}
  \bibnamefont{Vahala}}, \bibinfo{journal}{Science}
  \textbf{\bibinfo{volume}{321}}, \bibinfo{pages}{1172} (\bibinfo{year}{2008}).

\bibitem[{\citenamefont{Aspelmeyer}(2010)}]{Aspelmeyer2010a}
\bibinfo{author}{\bibfnamefont{M.}~\bibnamefont{Aspelmeyer}},
  \bibinfo{journal}{Nature} \textbf{\bibinfo{volume}{464}},
  \bibinfo{pages}{685} (\bibinfo{year}{2010}).

\bibitem[{\citenamefont{Chang et~al.}(2009)\citenamefont{Chang, Regal, Papp,
  Wilson, Y, Painter, Kimble, and Zoller}}]{Chang2009a}
\bibinfo{author}{\bibfnamefont{D.~E.} \bibnamefont{Chang}},
  \bibinfo{author}{\bibfnamefont{C.~A.} \bibnamefont{Regal}},
  \bibinfo{author}{\bibfnamefont{S.~B.} \bibnamefont{Papp}},
  \bibinfo{author}{\bibfnamefont{D.~J.} \bibnamefont{Wilson}},
  \bibinfo{author}{\bibfnamefont{J.}~\bibnamefont{Y}},
  \bibinfo{author}{\bibfnamefont{O.}~\bibnamefont{Painter}},
  \bibinfo{author}{\bibfnamefont{H.~J.} \bibnamefont{Kimble}},
  \bibnamefont{and} \bibinfo{author}{\bibfnamefont{P.}~\bibnamefont{Zoller}},
  \bibinfo{journal}{Proc. Natl. Acad. Sci. USA} \textbf{\bibinfo{volume}{107}},
  \bibinfo{pages}{1005} (\bibinfo{year}{2009}).

\bibitem[{\citenamefont{Romero-Isart et~al.}(2010)\citenamefont{Romero-Isart,
  Juan, Quidant, and Cirac}}]{RomeroIsart2010a}
\bibinfo{author}{\bibfnamefont{O.}~\bibnamefont{Romero-Isart}},
  \bibinfo{author}{\bibfnamefont{M.~L.} \bibnamefont{Juan}},
  \bibinfo{author}{\bibfnamefont{R.}~\bibnamefont{Quidant}}, \bibnamefont{and}
  \bibinfo{author}{\bibfnamefont{J.~I.} \bibnamefont{Cirac}},
  \bibinfo{journal}{New J. Phys.} \textbf{\bibinfo{volume}{12}},
  \bibinfo{pages}{033015} (\bibinfo{year}{2010}).

\bibitem[{\citenamefont{Romero-Isart
  et~al.}(2011{\natexlab{a}})\citenamefont{Romero-Isart, Pflanzer, Juan,
  Quidant, Kiesel, Aspelmeyer, and Cirac}}]{RomeroIsart2011a}
\bibinfo{author}{\bibfnamefont{O.}~\bibnamefont{Romero-Isart}},
  \bibinfo{author}{\bibfnamefont{A.~C.} \bibnamefont{Pflanzer}},
  \bibinfo{author}{\bibfnamefont{M.~L.} \bibnamefont{Juan}},
  \bibinfo{author}{\bibfnamefont{R.}~\bibnamefont{Quidant}},
  \bibinfo{author}{\bibfnamefont{N.}~\bibnamefont{Kiesel}},
  \bibinfo{author}{\bibfnamefont{M.}~\bibnamefont{Aspelmeyer}},
  \bibnamefont{and} \bibinfo{author}{\bibfnamefont{J.~I.} \bibnamefont{Cirac}},
  \bibinfo{journal}{Phys. Rev. A} \textbf{\bibinfo{volume}{83}},
  \bibinfo{pages}{013803} (\bibinfo{year}{2011}{\natexlab{a}}).

\bibitem[{\citenamefont{Romero-Isart
  et~al.}(2011{\natexlab{b}})\citenamefont{Romero-Isart, Pflanzer, Blaser,
  Kaltenbaek, Kiesel, Aspelmeyer, and Cirac}}]{RomeroIsart2011b}
\bibinfo{author}{\bibfnamefont{O.}~\bibnamefont{Romero-Isart}},
  \bibinfo{author}{\bibfnamefont{A.~C.} \bibnamefont{Pflanzer}},
  \bibinfo{author}{\bibfnamefont{F.}~\bibnamefont{Blaser}},
  \bibinfo{author}{\bibfnamefont{R.}~\bibnamefont{Kaltenbaek}},
  \bibinfo{author}{\bibfnamefont{N.}~\bibnamefont{Kiesel}},
  \bibinfo{author}{\bibfnamefont{M.}~\bibnamefont{Aspelmeyer}},
  \bibnamefont{and} \bibinfo{author}{\bibfnamefont{J.~I.} \bibnamefont{Cirac}},
  \bibinfo{journal}{Phys. Rev. Lett.} \textbf{\bibinfo{volume}{107}},
  \bibinfo{pages}{020405} (\bibinfo{year}{2011}{\natexlab{b}}).

\bibitem[{\citenamefont{Wilson-Rae et~al.}(2008)\citenamefont{Wilson-Rae,
  Nooshi, Dobrindt, Kippenberg, and Zwerger}}]{WilsonRae2008a}
\bibinfo{author}{\bibfnamefont{I.}~\bibnamefont{Wilson-Rae}},
  \bibinfo{author}{\bibfnamefont{N.}~\bibnamefont{Nooshi}},
  \bibinfo{author}{\bibfnamefont{J.}~\bibnamefont{Dobrindt}},
  \bibinfo{author}{\bibfnamefont{T.~J.} \bibnamefont{Kippenberg}},
  \bibnamefont{and} \bibinfo{author}{\bibfnamefont{W.}~\bibnamefont{Zwerger}},
  \bibinfo{journal}{New J. Phys.} \textbf{\bibinfo{volume}{10}},
  \bibinfo{pages}{095007} (\bibinfo{year}{2008}).

\bibitem[{\citenamefont{Gr{\"o}blacher}(2009)}]{Groeblacher2009a}
\bibinfo{author}{\bibfnamefont{S.}~\bibnamefont{Gr{\"o}blacher}},
  \bibinfo{journal}{Nature Physics} \textbf{\bibinfo{volume}{5}},
  \bibinfo{pages}{485} (\bibinfo{year}{2009}).

\bibitem[{\citenamefont{O'Connell et~al.}(2010)\citenamefont{O'Connell,
  Hofheinz, Ansmann, Bialczak, Lenander, Lucero, Neeley, Sank, Wang, Weides
  et~al.}}]{OConnell2010a}
\bibinfo{author}{\bibfnamefont{A.~D.} \bibnamefont{O'Connell}},
  \bibinfo{author}{\bibfnamefont{M.}~\bibnamefont{Hofheinz}},
  \bibinfo{author}{\bibfnamefont{M.}~\bibnamefont{Ansmann}},
  \bibinfo{author}{\bibfnamefont{R.~C.} \bibnamefont{Bialczak}},
  \bibinfo{author}{\bibfnamefont{M.}~\bibnamefont{Lenander}},
  \bibinfo{author}{\bibfnamefont{E.}~\bibnamefont{Lucero}},
  \bibinfo{author}{\bibfnamefont{M.}~\bibnamefont{Neeley}},
  \bibinfo{author}{\bibfnamefont{D.}~\bibnamefont{Sank}},
  \bibinfo{author}{\bibfnamefont{H.}~\bibnamefont{Wang}},
  \bibinfo{author}{\bibfnamefont{M.}~\bibnamefont{Weides}},
  \bibnamefont{et~al.}, \bibinfo{journal}{Nature}
  \textbf{\bibinfo{volume}{464}}, \bibinfo{pages}{697} (\bibinfo{year}{2010}).

\bibitem[{\citenamefont{Ghirardi et~al.}(1990)\citenamefont{Ghirardi, Pearle,
  and Rimini}}]{Ghirardi1990a}
\bibinfo{author}{\bibfnamefont{G.~C.} \bibnamefont{Ghirardi}},
  \bibinfo{author}{\bibfnamefont{P.}~\bibnamefont{Pearle}}, \bibnamefont{and}
  \bibinfo{author}{\bibfnamefont{A.}~\bibnamefont{Rimini}},
  \bibinfo{journal}{Phys. Rev. A} \textbf{\bibinfo{volume}{42}},
  \bibinfo{pages}{78} (\bibinfo{year}{1990}).

\bibitem[{\citenamefont{Collett and Pearle}(2003)}]{Collett2003a}
\bibinfo{author}{\bibfnamefont{B.}~\bibnamefont{Collett}} \bibnamefont{and}
  \bibinfo{author}{\bibfnamefont{P.}~\bibnamefont{Pearle}},
  \bibinfo{journal}{Foundations of Physics} \textbf{\bibinfo{volume}{33}},
  \bibinfo{pages}{1495} (\bibinfo{year}{2003}), ISSN \bibinfo{issn}{0015-9018},
  \bibinfo{note}{10.1023/A:1026048530567}.

\bibitem[{\citenamefont{Ghirardi et~al.}(1986)\citenamefont{Ghirardi, Rimini,
  and Weber}}]{Ghirardi1986a}
\bibinfo{author}{\bibfnamefont{G.~C.} \bibnamefont{Ghirardi}},
  \bibinfo{author}{\bibfnamefont{A.}~\bibnamefont{Rimini}}, \bibnamefont{and}
  \bibinfo{author}{\bibfnamefont{T.}~\bibnamefont{Weber}},
  \bibinfo{journal}{Phys. Rev.~D} \textbf{\bibinfo{volume}{34}},
  \bibinfo{pages}{470} (\bibinfo{year}{1986}).

\bibitem[{\citenamefont{Pearle}(1976)}]{Pearle1976a}
\bibinfo{author}{\bibfnamefont{P.}~\bibnamefont{Pearle}},
  \bibinfo{journal}{Phys. Rev. D} \textbf{\bibinfo{volume}{13}},
  \bibinfo{pages}{857} (\bibinfo{year}{1976}).

\bibitem[{\citenamefont{Pearle}(1989)}]{Pearle1989a}
\bibinfo{author}{\bibfnamefont{P.}~\bibnamefont{Pearle}},
  \bibinfo{journal}{Phys. Rev.~A} \textbf{\bibinfo{volume}{39}},
  \bibinfo{pages}{2277} (\bibinfo{year}{1989}).

\bibitem[{\citenamefont{Gisin}(1989)}]{Gisin1989a}
\bibinfo{author}{\bibfnamefont{N.}~\bibnamefont{Gisin}},
  \bibinfo{journal}{Helv. Phys. Acta} \textbf{\bibinfo{volume}{62}},
  \bibinfo{pages}{363} (\bibinfo{year}{1989}).

\bibitem[{\citenamefont{Ellis et~al.}(1989)\citenamefont{Ellis, Mohanty, and
  Nanopoulos}}]{Ellis1989a}
\bibinfo{author}{\bibfnamefont{J.}~\bibnamefont{Ellis}},
  \bibinfo{author}{\bibfnamefont{S.}~\bibnamefont{Mohanty}}, \bibnamefont{and}
  \bibinfo{author}{\bibfnamefont{D.~V.} \bibnamefont{Nanopoulos}},
  \bibinfo{journal}{Phys. Lett.~B} \textbf{\bibinfo{volume}{221}},
  \bibinfo{pages}{113 } (\bibinfo{year}{1989}).

\bibitem[{\citenamefont{Ellis et~al.}(1992)\citenamefont{Ellis, Mavromatos, and
  Nanopoulos}}]{Ellis1992a}
\bibinfo{author}{\bibfnamefont{J.}~\bibnamefont{Ellis}},
  \bibinfo{author}{\bibfnamefont{N.~E.} \bibnamefont{Mavromatos}},
  \bibnamefont{and} \bibinfo{author}{\bibfnamefont{D.~V.}
  \bibnamefont{Nanopoulos}}, \bibinfo{journal}{Phys. Lett.~B}
  \textbf{\bibinfo{volume}{293}}, \bibinfo{pages}{37 } (\bibinfo{year}{1992}).

\bibitem[{\citenamefont{K\'arolyh\'azy}(1966)}]{Karolyhazy1966a}
\bibinfo{author}{\bibfnamefont{F.}~\bibnamefont{K\'arolyh\'azy}},
  \bibinfo{journal}{Nuovo Cimento A} \textbf{\bibinfo{volume}{52}},
  \bibinfo{pages}{390} (\bibinfo{year}{1966}).

\bibitem[{\citenamefont{Di{\'o}si}(2005)}]{Diosi2005a}
\bibinfo{author}{\bibfnamefont{L.}~\bibnamefont{Di{\'o}si}},
  \bibinfo{journal}{Brazilian Journal of Physics}
  \textbf{\bibinfo{volume}{35}}, \bibinfo{pages}{260} (\bibinfo{year}{2005}).

\bibitem[{\citenamefont{Penrose}(1996)}]{Penrose1996a}
\bibinfo{author}{\bibfnamefont{R.}~\bibnamefont{Penrose}},
  \bibinfo{journal}{{G}en. {R}el. {G}rav.} \textbf{\bibinfo{volume}{28}},
  \bibinfo{pages}{581} (\bibinfo{year}{1996}).

\bibitem[{\citenamefont{Adler}(2007)}]{Adler2007a}
\bibinfo{author}{\bibfnamefont{S.~L.} \bibnamefont{Adler}},
  \bibinfo{journal}{J. Phys.~A. Math. Gen.} \textbf{\bibinfo{volume}{40}},
  \bibinfo{pages}{2935} (\bibinfo{year}{2007}).

\bibitem[{\citenamefont{Adler and Bassi}(2009)}]{Adler2009a}
\bibinfo{author}{\bibfnamefont{S.~L.} \bibnamefont{Adler}} \bibnamefont{and}
  \bibinfo{author}{\bibfnamefont{A.}~\bibnamefont{Bassi}},
  \bibinfo{journal}{Science (New York, N.Y.)} \textbf{\bibinfo{volume}{325}},
  \bibinfo{pages}{275} (\bibinfo{year}{2009}).

\bibitem[{\citenamefont{Nimmrichter et~al.}(2011)\citenamefont{Nimmrichter,
  Hornberger, Haslinger, and Arndt}}]{Nimmrichter2011a}
\bibinfo{author}{\bibfnamefont{S.}~\bibnamefont{Nimmrichter}},
  \bibinfo{author}{\bibfnamefont{K.}~\bibnamefont{Hornberger}},
  \bibinfo{author}{\bibfnamefont{P.}~\bibnamefont{Haslinger}},
  \bibnamefont{and} \bibinfo{author}{\bibfnamefont{M.}~\bibnamefont{Arndt}},
  \bibinfo{journal}{Phys. Rev. A} \textbf{\bibinfo{volume}{83}},
  \bibinfo{pages}{043621} (\bibinfo{year}{2011}).

\bibitem[{\citenamefont{Gallis and Fleming}(1990)}]{Gallis1990a}
\bibinfo{author}{\bibfnamefont{M.}~\bibnamefont{Gallis}} \bibnamefont{and}
  \bibinfo{author}{\bibfnamefont{G.}~\bibnamefont{Fleming}},
  \bibinfo{journal}{Phys. Rev.~A} \textbf{\bibinfo{volume}{42}},
  \bibinfo{pages}{38} (\bibinfo{year}{1990}), ISSN \bibinfo{issn}{1050-2947}.

\bibitem[{\citenamefont{Schlosshauer}(2007)}]{Schlosshauer2007a}
\bibinfo{author}{\bibfnamefont{M.~A.} \bibnamefont{Schlosshauer}},
  \emph{\bibinfo{title}{{D}ecoherence and the {Q}uantum-to-{C}lassical
  {T}ransition}} (\bibinfo{publisher}{Springer}, \bibinfo{address}{Berlin},
  \bibinfo{year}{2007}).

\bibitem[{\citenamefont{Bohren and Huffman}(1998)}]{Bohren1998a}
\bibinfo{author}{\bibfnamefont{C.~F.} \bibnamefont{Bohren}} \bibnamefont{and}
  \bibinfo{author}{\bibfnamefont{D.~R.} \bibnamefont{Huffman}},
  \emph{\bibinfo{title}{{Absorption and Scattering of Light by Small
  Particles}}} (\bibinfo{publisher}{John Wiley \& Sons}, \bibinfo{year}{1998}).

\bibitem[{\citenamefont{Romero-Isart}(2011)}]{RomeroIsart2011c}
\bibinfo{author}{\bibfnamefont{O.}~\bibnamefont{Romero-Isart}},
  \bibinfo{journal}{Phys. Rev. A} \textbf{\bibinfo{volume}{84}},
  \bibinfo{pages}{052121} (\bibinfo{year}{2011}).

\bibitem[{\citenamefont{Frenkel}(1990)}]{Frenkel1990a}
\bibinfo{author}{\bibfnamefont{A.}~\bibnamefont{Frenkel}},
  \bibinfo{journal}{Found. Phys.} \textbf{\bibinfo{volume}{20}},
  \bibinfo{pages}{159} (\bibinfo{year}{1990}).

\bibitem[{\citenamefont{Arai et~al.}(1988)\citenamefont{Arai, Imai, Hosono,
  Abe, and Imagawa}}]{Arai1988a}
\bibinfo{author}{\bibfnamefont{K.}~\bibnamefont{Arai}},
  \bibinfo{author}{\bibfnamefont{H.}~\bibnamefont{Imai}},
  \bibinfo{author}{\bibfnamefont{H.}~\bibnamefont{Hosono}},
  \bibinfo{author}{\bibfnamefont{Y.}~\bibnamefont{Abe}}, \bibnamefont{and}
  \bibinfo{author}{\bibfnamefont{H.}~\bibnamefont{Imagawa}},
  \bibinfo{journal}{Appl. Phys. Lett.} \textbf{\bibinfo{volume}{53}},
  \bibinfo{pages}{1891} (\bibinfo{year}{1988}).

\bibitem[{\citenamefont{Tsai et~al.}(1994)\citenamefont{Tsai, Friebele,
  Rajaram, and Mukhapadhyay}}]{Tsai1994a}
\bibinfo{author}{\bibfnamefont{T.~E.} \bibnamefont{Tsai}},
  \bibinfo{author}{\bibfnamefont{E.~J.} \bibnamefont{Friebele}},
  \bibinfo{author}{\bibfnamefont{M.}~\bibnamefont{Rajaram}}, \bibnamefont{and}
  \bibinfo{author}{\bibfnamefont{S.}~\bibnamefont{Mukhapadhyay}},
  \bibinfo{journal}{Appl. Phys. Lett.} \textbf{\bibinfo{volume}{64}},
  \bibinfo{pages}{1481} (\bibinfo{year}{1994}).

\bibitem[{\citenamefont{Bagratashvili et~al.}(1996)\citenamefont{Bagratashvili,
  Tsypina, Chernov, Rybaltovskii, Zavorotny, Alimpiev, Simanovskii, Dong, and
  Russel}}]{Bagratashvili1996a}
\bibinfo{author}{\bibfnamefont{V.~N.} \bibnamefont{Bagratashvili}},
  \bibinfo{author}{\bibfnamefont{S.~I.} \bibnamefont{Tsypina}},
  \bibinfo{author}{\bibfnamefont{P.~V.} \bibnamefont{Chernov}},
  \bibinfo{author}{\bibfnamefont{A.~O.} \bibnamefont{Rybaltovskii}},
  \bibinfo{author}{\bibfnamefont{Y.~S.} \bibnamefont{Zavorotny}},
  \bibinfo{author}{\bibfnamefont{S.~S.} \bibnamefont{Alimpiev}},
  \bibinfo{author}{\bibfnamefont{Y.~O.} \bibnamefont{Simanovskii}},
  \bibinfo{author}{\bibfnamefont{L.}~\bibnamefont{Dong}}, \bibnamefont{and}
  \bibinfo{author}{\bibfnamefont{P.~S.~J.} \bibnamefont{Russel}},
  \bibinfo{journal}{Appl. Phys. Lett.} \textbf{\bibinfo{volume}{68}},
  \bibinfo{pages}{1616} (\bibinfo{year}{1996}).

\bibitem[{\citenamefont{Heinbuch et~al.}(2005)\citenamefont{Heinbuch, Grisham,
  Martz, and Rocca}}]{Heinbuch2005a}
\bibinfo{author}{\bibfnamefont{S.}~\bibnamefont{Heinbuch}},
  \bibinfo{author}{\bibfnamefont{M.}~\bibnamefont{Grisham}},
  \bibinfo{author}{\bibfnamefont{D.}~\bibnamefont{Martz}}, \bibnamefont{and}
  \bibinfo{author}{\bibfnamefont{J.~J.} \bibnamefont{Rocca}},
  \bibinfo{journal}{Optics Express} \textbf{\bibinfo{volume}{13}},
  \bibinfo{pages}{4050} (\bibinfo{year}{2005}).

\bibitem[{\citenamefont{Kaltenbaek et~al.}(2012)\citenamefont{Kaltenbaek,
  Hechenblaikner, Kiesel, Wieczorek, Hofer, Gr{\"o}blacher, Vanner, Blaser,
  Johann, and Aspelmeyer}}]{Kaltenbaek2012a}
\bibinfo{author}{\bibfnamefont{R.}~\bibnamefont{Kaltenbaek}},
  \bibinfo{author}{\bibfnamefont{G.}~\bibnamefont{Hechenblaikner}},
  \bibinfo{author}{\bibfnamefont{N.}~\bibnamefont{Kiesel}},
  \bibinfo{author}{\bibfnamefont{W.}~\bibnamefont{Wieczorek}},
  \bibinfo{author}{\bibfnamefont{S.}~\bibnamefont{Hofer}},
  \bibinfo{author}{\bibfnamefont{S.}~\bibnamefont{Gr{\"o}blacher}},
  \bibinfo{author}{\bibfnamefont{M.~R.} \bibnamefont{Vanner}},
  \bibinfo{author}{\bibfnamefont{F.}~\bibnamefont{Blaser}},
  \bibinfo{author}{\bibfnamefont{U.}~\bibnamefont{Johann}}, \bibnamefont{and}
  \bibinfo{author}{\bibfnamefont{M.}~\bibnamefont{Aspelmeyer}},
  \bibinfo{type}{Tech. Rep.}, \bibinfo{institution}{Study conducted under
  contract with the European Space Agency} (\bibinfo{year}{2012}).

\bibitem[{\citenamefont{Marque et~al.}(2008)\citenamefont{Marque, Christophe,
  Liourzou, Bodoville, Foulon, Guerard, and Lebat}}]{Onera2008}
\bibinfo{author}{\bibfnamefont{J.-P.} \bibnamefont{Marque}},
  \bibinfo{author}{\bibfnamefont{B.}~\bibnamefont{Christophe}},
  \bibinfo{author}{\bibfnamefont{F.}~\bibnamefont{Liourzou}},
  \bibinfo{author}{\bibfnamefont{G.}~\bibnamefont{Bodoville}},
  \bibinfo{author}{\bibfnamefont{B.}~\bibnamefont{Foulon}},
  \bibinfo{author}{\bibfnamefont{J.}~\bibnamefont{Guerard}}, \bibnamefont{and}
  \bibinfo{author}{\bibfnamefont{V.}~\bibnamefont{Lebat}}, \bibinfo{journal}{59
  th International Astronautical Congress (IAC-08-B1.3.7)} pp.
  \bibinfo{pages}{TP 2008--137} (\bibinfo{year}{2008}).

\bibitem[{\citenamefont{Touboul et~al.}(2001)\citenamefont{Touboul, Rodrigues,
  M{\'e}tris, and Tatry}}]{Touboul2001a}
\bibinfo{author}{\bibfnamefont{P.}~\bibnamefont{Touboul}},
  \bibinfo{author}{\bibfnamefont{M.}~\bibnamefont{Rodrigues}},
  \bibinfo{author}{\bibfnamefont{G.}~\bibnamefont{M{\'e}tris}},
  \bibnamefont{and} \bibinfo{author}{\bibfnamefont{B.}~\bibnamefont{Tatry}},
  \bibinfo{journal}{Comptes Rendus de l'Académie des Sciences - Series IV -
  Physics} \textbf{\bibinfo{volume}{2}}, \bibinfo{pages}{1271 }
  (\bibinfo{year}{2001}), ISSN \bibinfo{issn}{1296-2147}.

\bibitem[{\citenamefont{Ashkin and Dziedzic}(1971)}]{Ashkin1971a}
\bibinfo{author}{\bibfnamefont{A.}~\bibnamefont{Ashkin}} \bibnamefont{and}
  \bibinfo{author}{\bibfnamefont{J.~M.} \bibnamefont{Dziedzic}},
  \bibinfo{journal}{Appl. Phys. Lett.} \textbf{\bibinfo{volume}{19}},
  \bibinfo{pages}{283} (\bibinfo{year}{1971}).

\bibitem[{\citenamefont{Volpe et~al.}(2007)\citenamefont{Volpe, Kozyreff, and
  Petrov}}]{Volpe2007a}
\bibinfo{author}{\bibfnamefont{G.}~\bibnamefont{Volpe}},
  \bibinfo{author}{\bibfnamefont{G.}~\bibnamefont{Kozyreff}}, \bibnamefont{and}
  \bibinfo{author}{\bibfnamefont{D.}~\bibnamefont{Petrov}},
  \bibinfo{journal}{J. Appl. Phys.} \textbf{\bibinfo{volume}{102}},
  \bibinfo{pages}{084701} (\bibinfo{year}{2007}), ISSN
  \bibinfo{issn}{00218979}.

\bibitem[{\citenamefont{Damour}(1996)}]{Damour1996a}
\bibinfo{author}{\bibfnamefont{T.}~\bibnamefont{Damour}},
  \bibinfo{journal}{Classical and Quantum Gravity}
  \textbf{\bibinfo{volume}{13}}, \bibinfo{pages}{A33} (\bibinfo{year}{1996}).

\bibitem[{\citenamefont{Adelberger et~al.}(2009)\citenamefont{Adelberger,
  Gundlach, Heckel, Hoedl, and Schlamminger}}]{Adelberger2009a}
\bibinfo{author}{\bibfnamefont{E.}~\bibnamefont{Adelberger}},
  \bibinfo{author}{\bibfnamefont{J.}~\bibnamefont{Gundlach}},
  \bibinfo{author}{\bibfnamefont{B.}~\bibnamefont{Heckel}},
  \bibinfo{author}{\bibfnamefont{S.}~\bibnamefont{Hoedl}}, \bibnamefont{and}
  \bibinfo{author}{\bibfnamefont{S.}~\bibnamefont{Schlamminger}},
  \bibinfo{journal}{Prog. Part. Nucl. Phys.} \textbf{\bibinfo{volume}{62}},
  \bibinfo{pages}{102 } (\bibinfo{year}{2009}), ISSN \bibinfo{issn}{0146-6410}.

\bibitem[{\citenamefont{{ESA}}()}]{LPFgraphics}
\bibinfo{author}{\bibnamefont{{ESA}}},
  \urlprefix\url{http://sci.esa.int/lisapf}.

\bibitem[{\citenamefont{Fichter et~al.}(2005)\citenamefont{Fichter, Gath,
  Vitale, and Bortoluzzi}}]{Fichter2005a}
\bibinfo{author}{\bibfnamefont{W.}~\bibnamefont{Fichter}},
  \bibinfo{author}{\bibfnamefont{P.}~\bibnamefont{Gath}},
  \bibinfo{author}{\bibfnamefont{S.}~\bibnamefont{Vitale}}, \bibnamefont{and}
  \bibinfo{author}{\bibfnamefont{D.}~\bibnamefont{Bortoluzzi}},
  \bibinfo{journal}{Classical and Quantum Gravity}
  \textbf{\bibinfo{volume}{22}}, \bibinfo{pages}{S139} (\bibinfo{year}{2005}).

\bibitem[{\citenamefont{Schrader et~al.}(2001)\citenamefont{Schrader, Kuhr,
  Alt, M{\"u}ller, Gomer, and Meschede}}]{Schrader2001a}
\bibinfo{author}{\bibfnamefont{D.}~\bibnamefont{Schrader}},
  \bibinfo{author}{\bibfnamefont{S.}~\bibnamefont{Kuhr}},
  \bibinfo{author}{\bibfnamefont{W.}~\bibnamefont{Alt}},
  \bibinfo{author}{\bibfnamefont{M.}~\bibnamefont{M{\"u}ller}},
  \bibinfo{author}{\bibfnamefont{V.}~\bibnamefont{Gomer}}, \bibnamefont{and}
  \bibinfo{author}{\bibfnamefont{D.}~\bibnamefont{Meschede}},
  \bibinfo{journal}{Appl. Phys.~B} \textbf{\bibinfo{volume}{73}},
  \bibinfo{pages}{819} (\bibinfo{year}{2001}).

\bibitem[{\citenamefont{Leger}(2007)}]{Leger2007}
\bibinfo{author}{\bibfnamefont{A.}~\bibnamefont{Leger}},
  \bibinfo{journal}{arXiv:0707.3385v1 [astro-ph]}  (\bibinfo{year}{2007}).

\bibitem[{\citenamefont{{ESA Science and
  Technology}}(2011{\natexlab{a}})}]{GAIA2011}
\bibinfo{author}{\bibnamefont{{ESA Science and Technology}}}
  (\bibinfo{year}{2011}{\natexlab{a}}),
  \urlprefix\url{http://sci.esa.int/GAIA}.

\bibitem[{\citenamefont{Flatscher}(2011)}]{Flatscher2011}
\bibinfo{author}{\bibfnamefont{R.}~\bibnamefont{Flatscher}}
  (\bibinfo{year}{2011}), \bibinfo{note}{doc.No. SMW-TDP, issue 2}.

\bibitem[{\citenamefont{Bell}(2008)}]{Bell2008}
\bibinfo{author}{\bibfnamefont{T.}~\bibnamefont{Bell}},
  \bibinfo{journal}{Nature} \textbf{\bibinfo{volume}{452}}, \bibinfo{pages}{18}
  (\bibinfo{year}{2008}).

\bibitem[{\citenamefont{Hammesfahr}(2005)}]{Hammesfahr2005}
\bibinfo{author}{\bibfnamefont{A.}~\bibnamefont{Hammesfahr}},
  \bibinfo{journal}{Article S2-ASD-RP-3016 for the Project LISA Technology
  Package (LTP)}  (\bibinfo{year}{2005}), \bibinfo{note}{issue 1}.

\bibitem[{\citenamefont{Gabrielse et~al.}(1990)\citenamefont{Gabrielse, Fei,
  Orozco, Tjoelker, Haas, Kalinowsky, Trainor, and Kells}}]{Gabrielse1990a}
\bibinfo{author}{\bibfnamefont{G.}~\bibnamefont{Gabrielse}},
  \bibinfo{author}{\bibfnamefont{X.}~\bibnamefont{Fei}},
  \bibinfo{author}{\bibfnamefont{L.~A.} \bibnamefont{Orozco}},
  \bibinfo{author}{\bibfnamefont{R.~L.} \bibnamefont{Tjoelker}},
  \bibinfo{author}{\bibfnamefont{J.}~\bibnamefont{Haas}},
  \bibinfo{author}{\bibfnamefont{H.}~\bibnamefont{Kalinowsky}},
  \bibinfo{author}{\bibfnamefont{T.~A.} \bibnamefont{Trainor}},
  \bibnamefont{and} \bibinfo{author}{\bibfnamefont{W.}~\bibnamefont{Kells}},
  \bibinfo{journal}{Phys. Rev. Lett.} \textbf{\bibinfo{volume}{65}},
  \bibinfo{pages}{1317} (\bibinfo{year}{1990}).

\bibitem[{\citenamefont{Cho}(2004)}]{Cho2004}
\bibinfo{author}{\bibfnamefont{H.}~\bibnamefont{Cho}},
  \bibinfo{journal}{Proceedings of the $5^{th}$ International Symposium on
  Environmental Testing for Space Programs}  (\bibinfo{year}{2004}).

\bibitem[{\citenamefont{International}()}]{VacAero2010}
\bibinfo{author}{\bibfnamefont{V.}~\bibnamefont{International}},
  \emph{\bibinfo{title}{Various data tables and documents retrieved from
  website http://vacaero.com}}.

\bibitem[{\citenamefont{Rabl et~al.}(2009)\citenamefont{Rabl, Genes, Hammerer,
  and Aspelmeyer}}]{Rabl2009a}
\bibinfo{author}{\bibfnamefont{P.}~\bibnamefont{Rabl}},
  \bibinfo{author}{\bibfnamefont{C.}~\bibnamefont{Genes}},
  \bibinfo{author}{\bibfnamefont{K.}~\bibnamefont{Hammerer}}, \bibnamefont{and}
  \bibinfo{author}{\bibfnamefont{M.}~\bibnamefont{Aspelmeyer}},
  \bibinfo{journal}{Phys. Rev. A} \textbf{\bibinfo{volume}{80}},
  \bibinfo{pages}{063819} (\bibinfo{year}{2009}).

\bibitem[{\citenamefont{Camatel and Ferrero}(2006)}]{Camatel2006a}
\bibinfo{author}{\bibfnamefont{S.}~\bibnamefont{Camatel}} \bibnamefont{and}
  \bibinfo{author}{\bibfnamefont{V.}~\bibnamefont{Ferrero}},
  \bibinfo{journal}{Phot. Tech. Lett., IEEE} \textbf{\bibinfo{volume}{18}},
  \bibinfo{pages}{2529 } (\bibinfo{year}{2006}), ISSN
  \bibinfo{issn}{1041-1135}.

\bibitem[{\citenamefont{Ashkin and Dziedzic}(1977)}]{Ashkin1977a}
\bibinfo{author}{\bibfnamefont{A.}~\bibnamefont{Ashkin}} \bibnamefont{and}
  \bibinfo{author}{\bibfnamefont{J.~M.} \bibnamefont{Dziedzic}},
  \bibinfo{journal}{Appl. Phys. Lett.} \textbf{\bibinfo{volume}{30}},
  \bibinfo{pages}{202} (\bibinfo{year}{1977}), ISSN \bibinfo{issn}{00036951}.

\bibitem[{\citenamefont{Li et~al.}(2011)\citenamefont{Li, Kheifets, and
  Raizen}}]{Li2011a}
\bibinfo{author}{\bibfnamefont{T.}~\bibnamefont{Li}},
  \bibinfo{author}{\bibfnamefont{S.}~\bibnamefont{Kheifets}}, \bibnamefont{and}
  \bibinfo{author}{\bibfnamefont{M.~G.} \bibnamefont{Raizen}},
  \bibinfo{journal}{Nature Physics} \textbf{\bibinfo{volume}{7}},
  \bibinfo{pages}{527} (\bibinfo{year}{2011}).

\bibitem[{\citenamefont{Teufel et~al.}(2011)\citenamefont{Teufel, Donner, Li,
  Harlow, Allman, Cicak, Sirois, Whittaker, Lehnert, and
  Simmonds}}]{Teufel2011a}
\bibinfo{author}{\bibfnamefont{J.~D.} \bibnamefont{Teufel}},
  \bibinfo{author}{\bibfnamefont{T.}~\bibnamefont{Donner}},
  \bibinfo{author}{\bibfnamefont{D.}~\bibnamefont{Li}},
  \bibinfo{author}{\bibfnamefont{J.~W.} \bibnamefont{Harlow}},
  \bibinfo{author}{\bibfnamefont{M.~S.} \bibnamefont{Allman}},
  \bibinfo{author}{\bibfnamefont{K.}~\bibnamefont{Cicak}},
  \bibinfo{author}{\bibfnamefont{A.~J.} \bibnamefont{Sirois}},
  \bibinfo{author}{\bibfnamefont{J.~D.} \bibnamefont{Whittaker}},
  \bibinfo{author}{\bibfnamefont{K.~W.} \bibnamefont{Lehnert}},
  \bibnamefont{and} \bibinfo{author}{\bibfnamefont{R.~W.}
  \bibnamefont{Simmonds}}, \bibinfo{journal}{Nature}
  \textbf{\bibinfo{volume}{475}}, \bibinfo{pages}{359} (\bibinfo{year}{2011}).

\bibitem[{\citenamefont{Chan et~al.}(2011)\citenamefont{Chan, Alegre,
  Safavi-Naeini, Hill, Krause, Gr\"{o}blacher, Aspelmeyer, and
  Painter}}]{Chan2011a}
\bibinfo{author}{\bibfnamefont{J.}~\bibnamefont{Chan}},
  \bibinfo{author}{\bibfnamefont{T.~P.~M.} \bibnamefont{Alegre}},
  \bibinfo{author}{\bibfnamefont{A.~H.} \bibnamefont{Safavi-Naeini}},
  \bibinfo{author}{\bibfnamefont{J.~T.} \bibnamefont{Hill}},
  \bibinfo{author}{\bibfnamefont{A.}~\bibnamefont{Krause}},
  \bibinfo{author}{\bibfnamefont{S.}~\bibnamefont{Gr\"{o}blacher}},
  \bibinfo{author}{\bibfnamefont{M.}~\bibnamefont{Aspelmeyer}},
  \bibnamefont{and} \bibinfo{author}{\bibfnamefont{O.}~\bibnamefont{Painter}},
  \bibinfo{journal}{Nature} \textbf{\bibinfo{volume}{478}}, \bibinfo{pages}{89}
  (\bibinfo{year}{2011}).

\bibitem[{\citenamefont{Rivi\`ere et~al.}(2011)\citenamefont{Rivi\`ere,
  Del\'eglise, Weis, Gavartin, Arcizet, Schliesser, and
  Kippenberg}}]{Riviere2011a}
\bibinfo{author}{\bibfnamefont{R.}~\bibnamefont{Rivi\`ere}},
  \bibinfo{author}{\bibfnamefont{S.}~\bibnamefont{Del\'eglise}},
  \bibinfo{author}{\bibfnamefont{S.}~\bibnamefont{Weis}},
  \bibinfo{author}{\bibfnamefont{E.}~\bibnamefont{Gavartin}},
  \bibinfo{author}{\bibfnamefont{O.}~\bibnamefont{Arcizet}},
  \bibinfo{author}{\bibfnamefont{A.}~\bibnamefont{Schliesser}},
  \bibnamefont{and} \bibinfo{author}{\bibfnamefont{T.~J.}
  \bibnamefont{Kippenberg}}, \bibinfo{journal}{Phys. Rev. A}
  \textbf{\bibinfo{volume}{83}}, \bibinfo{pages}{063835}
  (\bibinfo{year}{2011}).

\bibitem[{\citenamefont{Jiang et~al.}(2011)\citenamefont{Jiang, Ludlow, Lemke,
  Fox, Sherman, Ma, and Oates}}]{Jiang2011a}
\bibinfo{author}{\bibfnamefont{Y.~Y.} \bibnamefont{Jiang}},
  \bibinfo{author}{\bibfnamefont{A.~D.} \bibnamefont{Ludlow}},
  \bibinfo{author}{\bibfnamefont{N.~D.} \bibnamefont{Lemke}},
  \bibinfo{author}{\bibfnamefont{R.~W.} \bibnamefont{Fox}},
  \bibinfo{author}{\bibfnamefont{J.~A.} \bibnamefont{Sherman}},
  \bibinfo{author}{\bibfnamefont{L.-S.} \bibnamefont{Ma}}, \bibnamefont{and}
  \bibinfo{author}{\bibfnamefont{C.~W.} \bibnamefont{Oates}},
  \bibinfo{journal}{Nature Photonics} \textbf{\bibinfo{volume}{5}},
  \bibinfo{pages}{158} (\bibinfo{year}{2011}).

\bibitem[{\citenamefont{Folkner et~al.}(2010)\citenamefont{Folkner, deVine,
  Klipstein, McKenzie, Shaddock, Spero, Thompson, Wuchenich, Yu, Stephens
  et~al.}}]{Folkner2010a}
\bibinfo{author}{\bibfnamefont{W.~M.} \bibnamefont{Folkner}},
  \bibinfo{author}{\bibfnamefont{G.}~\bibnamefont{deVine}},
  \bibinfo{author}{\bibfnamefont{W.~M.} \bibnamefont{Klipstein}},
  \bibinfo{author}{\bibfnamefont{K.}~\bibnamefont{McKenzie}},
  \bibinfo{author}{\bibfnamefont{D.}~\bibnamefont{Shaddock}},
  \bibinfo{author}{\bibfnamefont{R.}~\bibnamefont{Spero}},
  \bibinfo{author}{\bibfnamefont{R.}~\bibnamefont{Thompson}},
  \bibinfo{author}{\bibfnamefont{D.}~\bibnamefont{Wuchenich}},
  \bibinfo{author}{\bibfnamefont{N.}~\bibnamefont{Yu}},
  \bibinfo{author}{\bibfnamefont{M.}~\bibnamefont{Stephens}},
  \bibnamefont{et~al.}, in \emph{\bibinfo{booktitle}{Earth Science Technology
  Forum, Arlington, Virginia}} (\bibinfo{year}{2010}),
  \urlprefix\url{http://hdl.handle.net/2014/41635}.

\bibitem[{\citenamefont{Fazio et~al.}(2004)\citenamefont{Fazio, Hora, Allen,
  Ashby, Barmby, Deutsch, Huang, Kleiner, Marengo, Megeath
  et~al.}}]{Fazio2004a}
\bibinfo{author}{\bibfnamefont{G.~G.} \bibnamefont{Fazio}},
  \bibinfo{author}{\bibfnamefont{J.~L.} \bibnamefont{Hora}},
  \bibinfo{author}{\bibfnamefont{L.~E.} \bibnamefont{Allen}},
  \bibinfo{author}{\bibfnamefont{M.~L.~N.} \bibnamefont{Ashby}},
  \bibinfo{author}{\bibfnamefont{P.}~\bibnamefont{Barmby}},
  \bibinfo{author}{\bibfnamefont{L.~K.} \bibnamefont{Deutsch}},
  \bibinfo{author}{\bibfnamefont{J.}~\bibnamefont{Huang}},
  \bibinfo{author}{\bibfnamefont{S.}~\bibnamefont{Kleiner}},
  \bibinfo{author}{\bibfnamefont{M.}~\bibnamefont{Marengo}},
  \bibinfo{author}{\bibfnamefont{S.~T.} \bibnamefont{Megeath}},
  \bibnamefont{et~al.}, \bibinfo{journal}{Astr. J. Supp. Series}
  \textbf{\bibinfo{volume}{154}}, \bibinfo{pages}{10} (\bibinfo{year}{2004}),
  ISSN \bibinfo{issn}{0067-0049}.

\bibitem[{\citenamefont{Bai et~al.}(2008)\citenamefont{Bai, Bajaj, Beletic,
  Farris, Joshi, Lauxtermann, Petersen, and Williams}}]{Bai2008a}
\bibinfo{author}{\bibfnamefont{Y.}~\bibnamefont{Bai}},
  \bibinfo{author}{\bibfnamefont{J.}~\bibnamefont{Bajaj}},
  \bibinfo{author}{\bibfnamefont{J.~W.} \bibnamefont{Beletic}},
  \bibinfo{author}{\bibfnamefont{M.~C.} \bibnamefont{Farris}},
  \bibinfo{author}{\bibfnamefont{A.}~\bibnamefont{Joshi}},
  \bibinfo{author}{\bibfnamefont{S.}~\bibnamefont{Lauxtermann}},
  \bibinfo{author}{\bibfnamefont{A.}~\bibnamefont{Petersen}}, \bibnamefont{and}
  \bibinfo{author}{\bibfnamefont{G.}~\bibnamefont{Williams}}, in
  \emph{\bibinfo{booktitle}{{High Energy, Optical, and Infrared Detectors for
  Astronomy III}}} (\bibinfo{organization}{Proceedings of SPIE},
  \bibinfo{year}{2008}), vol. \bibinfo{volume}{7021}, p.
  \bibinfo{pages}{702102}.

\bibitem[{\citenamefont{M{\"o}hle et~al.}(2010)\citenamefont{M{\"o}hle,
  D{\"o}ringshoff, Nagel, Kovalchuk, and Peters}}]{Moehle2010a}
\bibinfo{author}{\bibfnamefont{K.}~\bibnamefont{M{\"o}hle}},
  \bibinfo{author}{\bibfnamefont{K.}~\bibnamefont{D{\"o}ringshoff}},
  \bibinfo{author}{\bibfnamefont{M.}~\bibnamefont{Nagel}},
  \bibinfo{author}{\bibfnamefont{E.}~\bibnamefont{Kovalchuk}},
  \bibnamefont{and} \bibinfo{author}{\bibfnamefont{A.}~\bibnamefont{Peters}},
  in \emph{\bibinfo{booktitle}{{Proceedings of the 24th European Frequency and
  Time Forum}}} (\bibinfo{year}{2010}).

\bibitem[{\citenamefont{Naik et~al.}(2006)\citenamefont{Naik, Buu, LaHaye,
  Armour, Clerk, Blencowe, and Schwab}}]{Naik2006a}
\bibinfo{author}{\bibfnamefont{A.}~\bibnamefont{Naik}},
  \bibinfo{author}{\bibfnamefont{O.}~\bibnamefont{Buu}},
  \bibinfo{author}{\bibfnamefont{M.~D.} \bibnamefont{LaHaye}},
  \bibinfo{author}{\bibfnamefont{A.~D.} \bibnamefont{Armour}},
  \bibinfo{author}{\bibfnamefont{A.~A.} \bibnamefont{Clerk}},
  \bibinfo{author}{\bibfnamefont{M.~P.} \bibnamefont{Blencowe}},
  \bibnamefont{and} \bibinfo{author}{\bibfnamefont{K.~C.}
  \bibnamefont{Schwab}}, \bibinfo{journal}{Nature}
  \textbf{\bibinfo{volume}{443}}, \bibinfo{pages}{193} (\bibinfo{year}{2006}).

\bibitem[{\citenamefont{Gigan et~al.}(2006)\citenamefont{Gigan, B�hm,
  Paternostro, Blaser, Langer, Hertzberg, Schwab, B�uerle, Aspelmeyer, and
  Zeilinger}}]{Gigan2006a}
\bibinfo{author}{\bibfnamefont{S.}~\bibnamefont{Gigan}},
  \bibinfo{author}{\bibfnamefont{H.~R.} \bibnamefont{B�hm}},
  \bibinfo{author}{\bibfnamefont{M.}~\bibnamefont{Paternostro}},
  \bibinfo{author}{\bibfnamefont{F.}~\bibnamefont{Blaser}},
  \bibinfo{author}{\bibfnamefont{G.}~\bibnamefont{Langer}},
  \bibinfo{author}{\bibfnamefont{J.~B.} \bibnamefont{Hertzberg}},
  \bibinfo{author}{\bibfnamefont{K.~C.} \bibnamefont{Schwab}},
  \bibinfo{author}{\bibfnamefont{D.}~\bibnamefont{B�uerle}},
  \bibinfo{author}{\bibfnamefont{M.}~\bibnamefont{Aspelmeyer}},
  \bibnamefont{and}
  \bibinfo{author}{\bibfnamefont{A.}~\bibnamefont{Zeilinger}},
  \bibinfo{journal}{Nature} \textbf{\bibinfo{volume}{444}}, \bibinfo{pages}{67}
  (\bibinfo{year}{2006}).

\bibitem[{\citenamefont{{ESA Science and
  Technology}}(2011{\natexlab{b}})}]{Hipparcos2011}
\bibinfo{author}{\bibnamefont{{ESA Science and Technology}}}
  (\bibinfo{year}{2011}{\natexlab{b}}),
  \urlprefix\url{http://sci.esa.int/hipparcos}.

\end{thebibliography}

\end{document}